\documentclass[iop,twocolumn,twocolappendix,numberedappendix]{openjournal}
\usepackage{hyperref} % Gives \url
\hypersetup{colorlinks=true,linkcolor=blue,citecolor=blue,filecolor=blue,urlcolor=blue}
\usepackage{graphicx}	% Including figure files
\usepackage{amsmath}	% Advanced maths 
\usepackage{orcidlink}
\usepackage{amssymb}
\usepackage{bm}
\usepackage{xcolor}
\usepackage[T1]{fontenc}
\usepackage{ae,aecompl}
\usepackage{newtxtext,newtxmath}
\usepackage{lineno}

\newcommand{\mycomment}[1]{} % Multi-line commenting (pseudo-function)
\def\galfit{{\tt GALFIT}}

\begin{document}
\begin{nolinenumbers}
\vspace*{-\headsep}\vspace*{\headheight}
\footnotesize \hfill  FERMILAB-PUB-26-0147-PPD\\
\vspace*{-\headsep}\vspace*{\headheight}
\footnotesize \hfill DES-2025-0908
\end{nolinenumbers}

%%%%%%%%%%%%%%%%%%% TITLE PAGE %%%%%%%%%%%%%%%%%%%
\title{\vspace{-0.8cm}Brightest Cluster Galaxy ellipticity as proxy for halo shape: Orientation bias, assembly bias, and potential selection effects in SZ-selected clusters\vspace{-1.5cm}}

\author{
Radhakrishnan Srinivasan$^{1}\star$\orcidlink{0000-0002-4508-4581},
Tae-hyeon Shin$^{1}\dagger$\orcidlink{0000-0002-6389-5409},
Anja von der Linden$^{1}\ddagger$\orcidlink{0000-0002-3881-7724},
Ricardo Herbonnet$^{1}$, 
Matthias Klein$^{2}$,
Tamas N. Varga$^{2}$, 
Antonio Frigo$^{1}$, 
Lindsey E. Bleem$^{3,4}$, 
Hao-Yi Wu$^{5}$, 
Zhuowen Zhang$^{6}$, 
Benjamin Levine$^{1}$, 
%
% Second Tier - Starts here
%
Alex Alarcon$^{7}$\orcidlink{0000-0001-8505-1269}, 
Alexandra Amon$^{8}$\orcidlink{0000-0002-6445-0559},
Matthew B. Bayliss$^{9}$, 
Keith Bechtol$^{10}$\orcidlink{0000-0001-8156-0429},
Matthew Becker$^{11}$\orcidlink{0000-0001-7774-2246},
Gary Bernstein$^{12}$\orcidlink{0000-0002-8613-8259},
Sebastian Bocquet$^{2}$, 
Andresa Campos$^{13,14}$,
Aurelio Carnero Rosell$^{15,16,17}$\orcidlink{0000-0003-3044-5150},
Matias Carrasco Kind$^{18,19}$\orcidlink{0000-0002-4802-3194},
Chihway Chang$^{4,20}$\orcidlink{0000-0002-7887-0896},
Rebecca Chen$^{21}$,
Ami Choi$^{22}$\orcidlink{0000-0002-5636-233X},
Juan De Vicente$^{23}$\orcidlink{0000-0001-8318-6813},
Joseph DeRose$^{24}$\orcidlink{0000-0002-0728-0960},
Scott Dodelson$^{4,20,25}$\orcidlink{0000-0002-8446-3859},
Cyrille Doux$^{12,26}$\orcidlink{0000-0003-4480-0096},
Alex Drlica-Wagner$^{4,20,25}$\orcidlink{0000-0001-8251-933X},
Jack Elvin-Poole$^{27}$\orcidlink{0000-0001-5148-9203},	
Spencer Everett$^{28}$,	
Agnès Ferté$^{29}$\orcidlink{0000-0003-3065-9941},
Marco Gatti$^{4,7}$\orcidlink{0000-0001-6134-8797},
Raven Gassis$^{9}$, 
Michael D. Gladders$^{4,20}$, 
Sebastian Grandis$^{30}$, 
Daniel Gruen$^{2}$\orcidlink{0000-0003-3270-7644},
Robert Gruendl$^{18,19}$\orcidlink{0000-0002-4588-6517},
Ian Harrison$^{31}$\orcidlink{0000-0002-4437-0770},
Mike Jarvis$^{12}$\orcidlink{0000-0002-4179-5175},
Niall MacCrann$^{32}$\orcidlink{0000-0002-8998-3909},
Jamie McCullough$^{2,8,29,33}$\orcidlink{0000-0002-4475-3456},
Michael A. McDonald$^{34}$, 
Justin Myles$^{8}$\orcidlink{0000-0001-6145-5859},
Andres Navarro Alsina$^{35}$,
Shivam Pandey$^{12}$,
Judit Prat$^{36,37}$\orcidlink{0000-0002-5933-5150},
Marco Raveri$^{38}$\orcidlink{0000-0002-7354-3802},
Christian L. Reichardt$^{39}$, 
Richard Rollins$^{40}$,
Eli Rykoff$^{29,33}$\orcidlink{0000-0001-9376-3135},
Carles Sanchez$^{41,42}$\orcidlink{0000-0002-2744-4934},
Arnab Sarkar$^{34,43}$,
Lucas F. Secco$^{4}$,
Ignacio Sevilla$^{23}$\orcidlink{0000-0002-1831-1953},
Erin Sheldon$^{44}$\orcidlink{0000-0001-9194-0441},
Taweewat Somboonpanyakul$^{45}$, 
Brian Stalder$^{46,47}$, 
Anthony A. Stark$^{48}$, 
Michael A. Troxel$^{21}$\orcidlink{0000-0002-5622-5212},
Isaac Tutusaus$^{49}$\orcidlink{0000-0002-3199-0399},
Brian Yanny$^{25}$\orcidlink{0000-0002-9541-2678},
Boyan Yin$^{21}$\orcidlink{0000-0001-7005-8820},
%
% Third Tier - Starts here
%
Michel Aguena$^{16,50}$\orcidlink{0000-0001-5679-6747},
Sahar Allam$^{25}$\orcidlink{0000-0002-7069-7857},
Felipe Andrade-Oliveira$^{51}$\orcidlink{0000-0003-0171-6900},
David Bacon$^{52}$,
Jonathan Blazek$^{53}$\orcidlink{0000-0002-4687-4657},
David Brooks$^{54}$\orcidlink{0000-0002-8458-5047},
David Burke$^{29,33}$\orcidlink{0000-0003-1866-1950},
Ryan Camilleri$^{55}$,
Jorge Carretero$^{42}$\orcidlink{0000-0002-3130-0204},
Matteo Costanzi$^{50,56,57}$\orcidlink{0000-0001-8158-1449},
Luiz da Costa$^{16}$\orcidlink{0000-0002-7731-277X},
Maria Elidaiana da Silva Pereira$^{57}$,
Shantanu Desai$^{59}$\orcidlink{0000-0002-0466-3288},
H. Thomas Diehl$^{25}$\orcidlink{0000-0002-8357-7467},
Juan Garcia-Bellido$^{60}$\orcidlink{0000-0002-9370-8360},
Gaston Gutierrez$^{25}$\orcidlink{0000-0003-0825-0517},
Samuel Hinton$^{55}$\orcidlink{0000-0003-2071-9349},
Devon L. Hollowood$^{61}$\orcidlink{0000-0002-9369-4157},
Sujeong Lee$^{62}$\orcidlink{0000-0002-8289-740X},	
Jennifer Marshall$^{63}$\orcidlink{0000-0003-0710-9474},
Juan Mena-Fernández$^{26,64}$\orcidlink{0000-0001-9497-7266},
Felipe Menanteau$^{18,19}$\orcidlink{0000-0002-1372-2534},
Ramon Miquel$^{42,65}$\orcidlink{0000-0002-6610-4836},
Andrés Plazas Malagón$^{29,33}$\orcidlink{0000-0002-2598-0514},
Ricardo Ogando$^{66}$\orcidlink{0000-0003-2120-1154},
Kathy Romer$^{67}$\orcidlink{0000-0002-9328-879X},
Aaron Roodman$^{29,33}$\orcidlink{0000-0001-5326-3486},
Eusebio Sanchez$^{23}$\orcidlink{0000-0002-9646-8198},
David Sanchez Cid$^{23,51}$\orcidlink{0000-0003-3054-7907},
Eric Suchyta$^{68}$\orcidlink{0000-0002-7047-9358},
Molly Swanson$^{18}$,
Noah Weaverdyck$^{24,69}$\orcidlink{0000-0001-9382-5199},
Jochen Weller$^{2,70}$\orcidlink{0000-0002-8282-2010}
\newline
\small\underline{\underline{\textit{Affiliations are listed at the end of the paper}}}
}

% These dates will be filled out by the publisher
%\date{Last updated TBD; in original form TBD}
% Enter the current year, for the copyright statements etc.
%\pubyear{2026}

\shorttitle{Orientation and Assembly bias in optical observables}
\shortauthors{R. Srinivasan et al.}
\thanks{$^{\star}$E-mail:rk29121996@gmail.com}
\thanks{$^{\dagger}$E-mail:taehyeos@andrew.cmu.edu}
\thanks{$^{\ddagger}$E-mail:anja.vonderlinden@stonybrook.edu}

\begin{abstract}
The orientation of triaxial galaxy clusters with respect to the line-of-sight is expected to be one of the prime sources of scatter and potential bias in optical observables (e.g., richness and weak-lensing signal) of galaxy clusters. In this work, we use the observed shape of the central Brightest Cluster Galaxy (BCG) as proxy for the orientation along the line-of-sight for clusters selected via the Sunyaev-Zel'dovich (SZ) effect from the South Pole Telescope (SPT) and Atacama Cosmology Telescope (ACT) surveys, matched to optically selected clusters from the Dark Energy Survey Year 3 (DES). 
We construct two samples of clusters that are designed to be identical in SZ mass estimate and redshift but with the roundest vs. the most elliptical BCGs, which we expect to correspond to BCGs (and clusters) with major axes aligned along the line-of-sight vs. in the plane of the sky, respectively.  We find that the optical richness of round-BCG clusters is $\sim 10$\% larger than that of elliptical-BCG clusters, in agreement with the expectation from projection effects
and presenting the first such detection in data. 
The density profiles, however, are not in agreement with the expectation from projection effects:  the 1-halo term (below $6~h^{-1}\rm{Mpc}$) of both the weak-lensing and galaxy density profiles are the same for the subsamples, contrary to previous studies based on X-ray selected clusters.  In the 2-halo regime (above $6~h^{-1}\rm{Mpc}$), we find a significant excess of the elliptical-BCG cluster profiles compared to the round-BCG cluster profiles, which is the opposite of the expectation from numerical simulations.   
We hypothesize that the intrinsic shape of the BCG reflects not just the orientation angle, but also intrinsic properties of the cluster which can affect both the SZ signal and the amplitude of the 2-halo term.  
Indeed, we find that the round-BCG sample has a significantly higher (projected) concentration than the elliptical-BCG sample, which would both boost the SZ signal and suppress the amplitude of the 2-halo term; however, the difference in concentration alone cannot explain the stark difference in profile amplitudes. We also find a lower blue galaxy fraction in round-BCG clusters, suggesting that they are older than the elliptical-BCG clusters. 
Future multi-wavelength datasets and simulations will be needed to resolve the cause of the different bias amplitudes, but for now we conclude that the observed BCG shape appears to reflect both halo orientation and assembly bias.
\end{abstract}

\keywords{gravitational lensing: weak -- galaxies: clusters: general -- cosmology: observations}

\section{Introduction}
\label{Introduction}
Galaxy clusters are landmarks in the universe as the most massive matter density peaks that arise in the large scale cosmic web of the matter distribution. Therefore, keeping track of the number of these massive structures can lead to a better understanding of the composition of the universe. Counting galaxy clusters as a function of mass and comparing the observed number density with the predicted halo mass function can constrain a number of cosmological parameters, including the amplitude of the matter power spectrum ($\sigma_{8}$), the matter density ($\Omega_{\rm m}$), the dark energy equation of state ($w$), and the species-summed neutrino mass ($\sum m_{\nu}$), see e.g., \citet{Allen-S-2011:Clusters} for a review.  Indeed, clusters have already placed some of the tightest limits on $\sigma_8$ and $\Omega_{\rm m}$, $w$, and $\sum m_{\nu}$ \citep{Mantz2015, Bocquet2024b, Ghirardini-V-2024:eRASS1cosmology}.

Galaxy clusters are observed in multiple wavelengths, such as over-densities of red galaxies in optical and near-IR wavelengths \citep[e.g.,][]{Abell-G-1958:Optical,Koester-B-2007:Optical, Rykoff-E-2014:redmapper, Abbott-T-2020:Y1cluster}, as the hosts of extended X-ray emission by their hot intra-cluster gas \citep[e.g.,][]{Gioia-I-1990:Xray, Vikhlinin-A-2009:XrayClusterSurvey,Mantz2015, Klein-M-2019:Xray, Garrel-C-2022:XrayClusterSurvey, Ghirardini-V-2024:eRASS1cosmology}, 
and in millimeter waves by the inverse Compton scattering of Cosmic Microwave Background (CMB) photons, also commonly referred to as the Sunyaev-Zel'dovich (SZ) effect \citep[e.g.,][]{Sunyaev-R-1972:SZ, Staniszewski-Z-2009:SZ,Ade-P-2016:PlanckClusterCosmology, Bleem-L-2015:SPTSZ, Bleem-L-2020:SPTECS, Bocquet-S-2019:SPTSZ, Hilton-M-2021:ACT-Catalog}. 

The observables in cluster surveys (e.g., optical richness, SZ Compton-y, X-ray luminosity for cluster detection, and weak-lensing mass estimates for the absolute mass calibration) correlate with the mass of clusters and --- on average --- exhibit a power-law relation as predicted by the self-similarity model of clusters \citep{Fillmore-J-1984:ClusterSelfSimilarity, Bertschinger-E-1985:ClusterSelfSimilarity, Kaiser-N-1986:Clusters}. However, the observables also exhibit a substantial scatter with respect to this mean scaling relation. This scatter can be a result of triaxiality, projection effects from correlated large scale structures along the line-of-sight (LOS) and the dynamical state of the cluster, among others. Since the scatter can lead to significant selection biases, cluster count cosmology analyses need to self-consistently include the scatter in their modeling \citep[e.g.,][]{Mantz-A-2010:XrayScalingRelation,Mantz2015,Bocquet-S-2019:SPTSZ,Bocquet2024a, Ghirardini-V-2024:eRASS1cosmology}.

Optical observables are expected to be particularly affected by orientation bias. Both galaxy richness and the observed weak-lensing profile are expected to scatter ``up'' when the cluster's major axis is aligned with the LOS \citep[similarly, velocity dispersions are expected to be biased high, see e.g.,][]{Bayliss2011}. In simple terms, both galaxy richness and weak lensing are sensitive to the mass in a roughly cylindrical aperture, with the cylinder having a substantial extension along the LOS.  Therefore, orientation bias captures not only the triaxial shape of the cluster halo, but also the contributions of the correlated structures along the LOS.  Simulations indicate that halo triaxiality and orientation bias cause an intrinsic scatter of $\sim 20$\% in weak-lensing mass estimates \citep{Meneghetti-M-2010:simulation, Becker-M-2011:simulation, Noh-Y-2012:SimulationsOrientationBias, Osato-K-2018:Orientation}.  

Consequently, selecting clusters according to optical richness can lead to a biased sample. Since the observed richness of clusters aligned along the LOS is boosted, such clusters are preferentially included in optical samples. The weak-lensing signal of these clusters is also boosted high, leading to correlated scatter between richness and lensing mass \citep[][]{Angulo-R-2012:Orientation,Dietrich-J-2014:Orientation,Sunayama-T-2020:Projection, Zhang-Y-2022:Orientation,Wu-H-2022:Orientation,Zhang-Z-2023:Triaxiality, Payerne-C-2025:WLMassRichnessRelation}.  It is likely that selection biases and correlated scatter due to orientation are a major contributor to the puzzling result of the cluster count analysis of Dark Energy Survey (DES) Year 1 data, which found a 5$\sigma$ discrepancy from Planck in the $\sigma_8 - \Omega_{\rm m}$ plane \citep{Abbott-T-2020:Y1cluster, Wu-H-2022:Orientation, Salcedo-A-2024}.

In contrast to the optical observables, X-ray and SZ observables are expected to be much less affected by orientation bias, since both are sensitive only to the hot gas associated with massive halos. The gas follows the gravitational potential, which is inherently more spherical than the dark matter distribution \citep{Zhou-C-2024:ProjectionEffects, Balzer-F-2025:ProjectionEffects}.  Hence, the intrinsic scatter of weak-lensing masses has relatively little impact on cluster count analyses of X-ray and SZ-selected cluster samples.

The effects of orientation bias have mostly been studied in simulations so far, and should therefore be seen as a prediction of the standard cosmological model with cold dark matter and hierarchical structure (and galaxy) formation.  Here, we seek to identify the effects of orientation bias in observations, using the measured shape of the central cluster galaxy, commonly referred to as the Brightest Cluster Galaxy (BCG), as tracer for halo orientation.  Both simulations and observations indicate that cluster-scale dark matter halos are preferentially prolate \citep{Frenk-C-1988:ProlatenessSimulations, Cooray-A-2000:ProlatenessObservations, Bailin-J-2005:ProlatenessSimulations, Sereno-M-2006:ProlatenessObservations}, and that the shape of the BCG follows the overall shape of the dark matter halo \citep{Binggeli-B-1982:Orientation, Niederste-Ostholt-M-2010:Orientation, Ragone-Figueroa-C-2020:Alignment, Zhou-C-2023:IA, Rodriguez25}.
Thus, the BCG of a halo with major axis aligned along the LOS is expected to appear round on the sky, whereas the BCG of a halo with its major axis in the plane of the sky is expected to be elliptical \citep{Herbonnet-R-2022:Triaxiality}.
Several previous works have used relatively small, but well-studied, cluster samples to identify a correlation between the observed BCG shape and the ratio between individual weak-lensing mass estimates and a low-scatter mass proxy --- namely the SZ decrement \citep{Marrone-D-2012, Gruen-D-2014}, hydrostatic mass estimates from X-ray observations \citep{Mahdavi-A-2013}, or X-ray gas masses \citep{Herbonnet-R-2019:Ellipticity}.  In particular, \citet{Herbonnet-R-2019:Ellipticity} showed that the observed BCG shape can successfully predict the weak-lensing mass bias; they found that weak-lensing mass estimates are $\sim$~20\% high (low) for the 25\% clusters with the roundest (most elliptical) BCGs in the Weighing the Giants sample \citep[WtG;][]{vonderLinden-A-2014:WtG1}.  We note that the weak-lensing mass estimates in these studies were based on fitting only the so-called 1-halo term, i.e., the weak-lensing profile extending only over virialized region of the cluster (up to $\sim$~3~Mpc).

In this work, we study the effect of orientation bias on the optical observables (richness, weak-lensing, and galaxy number density profiles)
of massive SZ-selected galaxy clusters. We are particularly interested in measuring the lensing and galaxy density profiles to large radii, i.e., over the 2-halo regime.  Simulations predict that the excess surface mass density for halos aligned along the line-of-sight is present also in the 2-halo regime, at a level similar to, or exceeding, the excess in the 1-halo regime \citep{Osato-K-2018:Orientation, Zhang-Z-2023:Triaxiality}.  Quantitatively, these simulations predict that the 20\% of halos with the largest (smallest) inclination $\cos i$ between the halo major axis and the line-of-sight have 20--30\% larger (smaller) projected surface mass densities out to at least 10 to 100 $h^{-1} {\rm Mpc}$.

We match SZ cluster catalogs from the South Pole Telescope (SPT) and Atacama Cosmology Telescope (ACT) surveys to a catalog of optically selected clusters from the DES Year 3 data, and use the measured shape of the BCG as proxy for the dark matter halo orientation along the LOS.  Namely, we construct two subsamples for each SZ cluster sample, matched in both SZ mass estimate and redshift, but selected on the roundest 25\% and most elliptical 25\% of BCGs. 
We find that the stacked richness estimates are consistent with expectations from both SZ cluster samples. We find that the richnesses of round-BCG clusters are $\sim 10$\% larger than those of elliptical-BCG clusters.  For the lensing and galaxy density profiles, however, the measurements differ from simulation predictions significantly.  In the 1-halo regime, the profiles have nearly identical amplitudes, whereas in the 2-halo regime, the measured surface density profiles are lower for the round-BCG samples than the elliptical-BCG samples --- surprisingly, this is the exact opposite of the expectations.  While it is likely that SZ-selection is also subject to orientation bias to some degree, in the sense that clusters aligned along the LOS (i.e., the round-BCG clusters) are likely inherently less massive, this is not enough to explain the discrepancy. We therefore speculate that the observed BCG shape reflects not only the halo orientation, but also other halo properties which correlate with the amplitude of the large-scale bias, i.e., a form of halo assembly bias (dependence of the large scale clustering of haloes of a given mass on their assembly history) \citep{Gao05, Wechsler2006, Lin_22}.  
Depending on the cluster selection, assembly bias can have similar effects on the clustering signal of clusters as orientation bias \citep{Wu08}.
For cluster-mass halos, the halo properties that correlate most with the clustering amplitude are halo concentration and spin \citep{Mao18,Wechlser_Tinker_2018}.  While the observed lensing and galaxy density profiles are not well described by (spherical) NFW profiles, we do find that the galaxy concentrations of round-BCG clusters are indeed $\sim 30$\% higher than those of elliptical-BCG clusters, as expected in the assembly bias scenario.
We explore other observational proxies expected to relate to the formation history or the dynamical state (used interchangeably in this work) of a cluster, and find that elliptical-BCG clusters have a higher blue-galaxy fraction than round-BCG clusters, consistent with having a higher infall rate due to being located in denser environments. However, our investigations of the magnitude gap and X-ray relaxedness yield inconclusive results. 

The structure of the paper is as follows: in Section \ref{Data}, we summarize the data sets used in this work and the visual inspection process to identify correct BCGs. In Section \ref{BCGorientation}, we describe how we construct cluster sub-samples matched in SZ mass estimate and redshift, but corresponding to the 25th percentiles of the roundest / most elliptical BCGs. In Section \ref{Results}, we present the results of comparing optical observables such as richness, stacked weak lensing profiles and stacked galaxy density profiles between the cluster sub-samples. In Section \ref{sect:toy-model}, we interpret the observed profiles and we remark on the role of assembly bias. Lastly, in Section \ref{Discussion} we summarize and discuss our findings.

We have assumed a flat $\Lambda$CDM fiducial cosmology, unless explicitly specified, with $\Omega_{\rm m} = 0.3$, $\Omega_{\Lambda} = 0.7$ and $H_0$ = 100 $h$ km/s/Mpc where $h$ = 0.7. We use $M_{\rm 500c}$, the mass enclosed within a sphere whose mean density equals 500 times the critical density, as our mass definition and denote it as $M_{500}$ for brevity. 

\section{Data} \label{Data}

We use weak lensing and galaxy density profiles around the galaxy clusters from the DES Year 3 data matched to their SZ counterparts from SPT and ACT to assess the effect of orientation bias and possible assembly bias. We briefly describe each data set in the following sections.

\subsection{The Dark Energy Survey data set} \label{des-data}

We use the images and the galaxy catalogs produced by the DES to measure the weak lensing and the galaxy density profiles around the galaxy clusters and to infer the halo orientations along the LOS using the shapes of BCGs as a proxy. We also use the galaxy colors to investigate the difference in galaxy population between different cluster samples.

The DES used the 4-meter Blanco Telescope at the Cerro Tololo Inter-American Observatory (CTIO) with the 570 Megapixel, 3 $\rm deg^{2}$ field of view Dark Energy Camera (DECam) in five filters: \textit{g,r,i,z,Y} over the span of six years from 2013 to 2019 \citep{Abbott-T-2016:DES}. The data taken in the wide-field survey spans nearly 5000 $\rm deg^{2}$ and is optimized for weak lensing, galaxy clustering, and galaxy cluster cosmology with over 300 million galaxies at a depth of i-band magnitude < 23.50 . 

The first public data release (DR1) of the DES comprises single epoch images, coadded images, coadded source catalogs and other associated products developed from the images taken in the first three years of the survey, covering the entire footprint with four 90 second exposures in every filter \citep{Abbott-T-2018:DR1}. The Y3 GOLD catalog was built from the DR1 coadded catalogs, providing updated astrometric and photometric calibration over the DR1 catalogs, photometry of galaxies based on Single-Object Fitting (SOF) and Multi-object Fitting (MOF) as well as photometric redshifts derived using multiple algorithms \citep{Sevilla-Noarbe-I-2021:Y3GOLD, Abbott-T-2018:DR1}. Access to both the DES DR1 image cutouts and the Y3 GOLD catalogs were provided via web interfaces at \url{https://des.ncsa.illinois.edu/releases}. The DR1 image cutouts are used for the visual inspection process as detailed in Appendix \ref{BCG-VI}, and the Y3 GOLD catalogs are used to infer the shape of the BCGs, to measure lensing profiles, and for constructing the galaxy density profiles. The optical clusters and the corresponding member catalog are derived from the \textsc{redMaPPer} algorithm (v6.4.22) run on the Y3 GOLD data set. The \textsc{redMaPPer} algorithm searches for galaxy clusters in optical surveys by examining galaxy colors, luminosities, and spatial distributions to identify the red-sequence galaxies in a galaxy cluster. \textsc{redMaPPer} finds over 869,335 galaxy clusters in DES Y3 at a richness greater than 5.0 in the redshift range of [0.1,0.94] \citep{Rykoff-E-2014:redmapper}.

For weak-lensing analyses (Section \ref{Weak lensing method}), we use the galaxy shapes generated with the \textsc{ngmix} \citep{Sheldon-E-2015:NGMIX} and the \textsc{Metacalibration} algorithm \citep{Huff-E-2017:Metacalibration, Sheldon-E-2017:Metacalibration} applied on the DES Y3 data \citep{Gatti-M-2021:DESY3shear}. We use the photometric redshift catalog accompanying the GOLD catalog. Specifically, we utilize the redshifts estimated by the \textsc{DNF} algorithm \citep[see][for further details]{DNF}. 

When measuring the galaxy density profiles around the galaxy clusters (Section~\ref{sec:galden}), we use the full 6-year DES data in order to utilize the fainter galaxy magnitude limit than the Y3 data, which increases the number density of galaxies and therefore boosts our signal. The Y6 GOLD catalog \citep{Bechtol-K-2026:DESY6GOLD} was built upon the Y6 coadded images \citep{DES2021:DR2}, analogous to the Y3 GOLD data products but with an improved photometric calibration (DES Collaboration, in prep).
 
\subsection{The South Pole Telescope data set} \label{spt-data}
We use SZ selected galaxy clusters from the SPT to construct part of our cluster sample. Both the SPT-SZ and the SPTpol Extended Cluster Survey overlap with the DES Y3 footprint and thereby provide a common set of clusters to study. The SPT-SZ survey has a footprint of 2500 $\rm deg^{2}$ and has detected 409 galaxy clusters with signal-to-noise ratio (SNR) $\xi\geq5.0$ ($\geq95\,\%$ purity) \citep{Bleem-L-2015:SPTSZ}. The SPTpol Extended Cluster Survey (SPTpol-ECS) covers another 2770 $\rm deg^{2}$ north of the SPT-SZ survey \citep{Bleem-L-2020:SPTECS} and adds 266 clusters at $\xi\geq5.0$. 

The redshifts of SPT clusters are estimated using various follow-up optical and/or NIR imaging \citep{Bleem-L-2015:SPTSZ,Bocquet-S-2019:SPTSZ}. We remove the high-redshift clusters ($z>0.7$) from the sample owing to the low signal-to-noise ratio of the weak-lensing measurement, along with the difficulty in identifying BCGs. At low redshift ($z\leq0.15$) the primary CMB fluctuations interfere with the cluster detection; we therefore restrict the cluster sample to the redshift range of [0.15,0.7].
We further only keep clusters that are optically confirmed by the Multi-Component Matched Filter (MCMF) cluster confirmation technique \citep{Klein-M-2023:MCMFSPT}, with a low probability to be dominated by line-of-sight projections.  Specifically, the MCMF algorithm estimates the contamination factor ($f_{\rm cont}$) for each cluster candidate, quantifying the estimated contamination of the optical richness measurement.  We limit the sample to clusters with $f_{\rm cont} < 0.2$, which has been shown to give over 98\,\% purity for SNR$> 4.5$ in the SPT-SZ sample with over 96\,\% completeness \citep{Klein-M-2023:MCMFSPT}.  There are 284 clusters within the DES Y3 foot print that match these criteria. For the analyses in this paper, we use the mass estimate ($M_{500}$) derived using the SNR-to-mass relation from \citet{Bocquet-S-2019:SPTSZ}. For simplicity, we will henceforth refer to this estimate as the SZ mass estimate. 

For a subset of clusters, where available, we use the optical images from the Parallel Imager for Southern Cosmology Observations (PISCO) to validate the shape measurements for the BCGs from the DES because of its better depth due and targeted observations of SPT clusters. The PISCO is an imaging instrument set-up on the 6.5 m Magellan/Clay telescope at Las Campanas Observatory in Chile \citep{Stalder-B-2014:PISCO}. The PISCO data only covers a subset of the SPT clusters (300 clusters) but with a uniform depth of r=24.3 \citep{Bleem-L-2020:SPTECS}. We employ the deeper imaging for BCG selection and for validation of BCG shape measurements (see details in Section \ref{BCG selection}).

\subsection{The Atacama Cosmology Telescope data set}\label{ACT}

The public ACT DR5 catalog \citep{Hilton-M-2021:ACT-Catalog} spans over 13000 $\rm deg^{2}$ and contains 2212 SZ-selected galaxy clusters at signal-to-noise ratio ${\rm SNR}_{2.4} \geq 5.0$ ($\geq 95\,\%$ purity). To match the SPT sample, we select clusters within the redshift range [0.15, 0.7] and with $f_{\rm cont} < 0.2$ (provided by MCMF), which results in over 98\,\% purity and 98\,\% completeness as shown in \citet{Klein-M-2024:MCMFACT}. These selection criteria result in a sample of 536 clusters within the DES Y3 footprint.

For the analysis in this work, we use the ACT DR5 mass estimates rescaled according to a richness-based weak-lensing mass calibration and redshifts from the ACT DR5 public catalog \citep{Hilton-M-2021:ACT-Catalog}. The ACT cluster catalog adopts the \textsc{redMaPPer} redshifts for the majority of their clusters (1433) and the ``scanning'' mode redshifts for 154 clusters where the cluster SZ center is provided to \textsc{redMaPPer} as a prior \citep{Hilton-M-2021:ACT-Catalog}. Since there are clusters that are common between the SPT and ACT data sets, we remove these ``duplicates'' ($16\,\%$ of the ACT sample) to keep the two data sets independent of each other. We chose to remove these clusters in ACT, because of the larger sample size in comparison to SPT.

\subsection{The eROSITA eRASS1 X-ray cluster catalogs}

We use X-ray measurements of clusters to study the orientation dependence of the SZ observable (SZ mass estimate). We use the latest X-ray cluster data set delivered by the eROSITA All-Sky Survey, the eRASS1 cluster main catalog \citep{Bulbul-E-2024:eRASS1dataset}. This catalog consists of 12247 optically confirmed clusters in the 13116  $\rm{deg}^{2}$ region in the western Galactic hemisphere, which eROSITA surveyed in its first six months of operation. The clusters are optically confirmed with the DESI Legacy Survey (LS DR9N and LS DR10) and span a redshift range of [0, 1.32].

\section{Assigning BCG shape as proxy for halo line-of-sight orientation} \label{BCGorientation}

To use the BCG as a proxy for the halo LOS orientation, we need to first identify the optical counterpart (a redMaPPer cluster) for each SZ cluster. Also, the robustness of the BCG selection must be guaranteed.

\subsection{Matching cluster catalogs} \label{likelihood-matching}

We match the SZ clusters from SPT and ACT cluster catalogs to the DES Y3 \textsc{redMaPPer} cluster catalog in the redshift range of [0.15,0.7] using a likelihood-based matching. For an SZ cluster that has an MCMF counterpart with a contamination factor less than 0.2 and in the redshift range, we look for a \textsc{redMaPPer} cluster that is closest on the sky and along the LOS distance.

For each SZ cluster (rank-ordered by SZ mass), we evaluate the following likelihood for all \textsc{redMaPPer} clusters within 1500.0 kpc of the SZ center, and within $\Delta z=0.15$ from the MCMF redshift:
\begin{equation}
    \mathcal{L}(\Delta z, \Delta R, \lambda) = \mathcal{N}(\Delta z| \mu_{\Delta z}=0,\,\sigma_{\Delta z})\mathcal{N}(\Delta R| \mu_{\Delta R}=0,\,\sigma_{\Delta R})S(\lambda)\,,
    \label{eq:likelihood_matching}
\end{equation}
where $\Delta z = |z_{\rm RM} - z_{\rm MCMF}|$, $\sigma_{\Delta z} = 0.02$, $\Delta R$ is the sky separation between SZ centroid and the \textsc{redMaPPer} center, $\sigma_{\Delta R} = 500.0$ kpc, and the sigmoid function, S, is defined as
\begin{equation}
    S(\lambda) = 1/(1+e^{-m(\ln(\lambda)-\ln(\lambda_{p}))})\,,
    \label{eq:sigmoid}
\end{equation}
where $m$ = 3.0 and the pivot for richness ($\lambda$) is $\lambda_{p}=40.0$. The sigmoid function prioritizes higher richness matches. $\mathcal{N}(x | \mu,\sigma)$ denotes a Gaussian distribution with the mean $\mu$ and the standard deviation $\sigma$. All the above parameters for matching were determined empirically.

We pick the highest likelihood match for each cluster and remove it from the list of available clusters to prevent duplicate matches with a lower limit on the likelihood for any match being 0.003 to prevent spurious matches to unnecessarily low richness clusters. We keep the SZ clusters for which we find \textsc{redMaPPer} matches using the above prescription. We find matches for 94\,\% of clusters (266 out of 284) for SPT and 90\,\% of clusters (485 out of 536) for ACT.

We also utilize the likelihood-based matching algorithm to match eROSITA's eRASS1 clusters main catalog (12247 clusters) to the SPTxDES-Y3 and ACTxDES-Y3 data set. The components of the likelihood include the proximity between the SZ centroid to X-ray center and redshift difference with the same parameter values as above. We match SZ clusters with redshift in the range [0.15,0.4] to X-ray clusters. We match 50 out of 51 SPT-SZ clusters (98\,\% completeness) and 53 out of 57 ACT clusters (93\,\% completeness). We use these SZ-complete cluster samples to measure the contribution of orientation bias on the intrinsic scatter of SZ mass proxy and to test its robustness when compared to optical observables.

\subsection{BCG selection} \label{BCG selection}
The \textsc{redMaPPer} algorithm provides 5 central galaxy candidates per cluster along with centering probabilities based on their color, brightness, and local red galaxy densities \citep{Rykoff-E-2016:RedmapperDES}. However, ~20-25\,\% of optically identified clusters suffer from mis-centering because of various effects such as major mergers, projection, and masking \citep{Abbott-T-2016:DES, Zhang-Y-2019:Miscentering}. For our analysis, it is critical to correctly identify BCGs and verify their shape measurements, as they are the cornerstones to measuring the LOS orientation of the cluster halo. We choose to visually inspect the BCGs as described in Appendix \ref{BCG-VI}. 

We also run \galfit{} on the deeper \textit{gri} images from PISCO when available and thereby validate the Single Object Fitting (SOF) and SExtractor shape measurements of BCGs from the DES-Y3 GOLD catalog. We manually construct masks for removing contaminations such as bright stars and for exclusion of the brightest galaxy cores as they skew the model fits to the BCG's stellar envelope considerably. The procedure is detailed in \citet{Herbonnet-R-2019:Ellipticity}. The results are provided in Appendix \ref{Shape validation}. We also supplement DES images with PISCO image cutouts during the visual inspection as described in Appendix \ref{BCG-VI}.

We inspect all the SPTxDES clusters (266 clusters) and find that 175 (65\,\%) clusters clear the visual inspection process. We found that \textsc{redMaPPer} did not identify the correct BCGs for roughly $35\,\%$ (91 of 266) of the BCGs because the images were either scaled improperly due to foreground bright stars or needed to be masked heavily. These effects also affected the shape measurement of the BCG in 17\,\% (46 of 266) of the cases. In the remaining BCG sample, for only $5\,\%$ (9 of 175) of the clusters we disagreed with \textsc{redMaPPer}’s choice of the correct BCG. To summarize, \textsc{redMaPPer} misidentified the BCG primarily due to foreground bright objects and the visual inspection results are in agreement with redMaPPer for most of the cases. Therefore, additional selection effects from BCG selection most likely did not affect the SPT sample.

To scale this effort to larger data sets including the ACT sample, we employ feature engineering techniques. 
Since the shape measurements of BCGs can be affected by foreground bright objects causing blending or scaling issues, we first investigate whether an ellipse overlaid on the BCG with its measured axis ratio is visually representative of its extended diffuse envelope (see Figure \ref{fig:BCG_gallery}), by which we remove $\sim$14\,\% of the ACT clusters (416 clusters left out of 485 clusters).
Then by selecting the most important BCG features from \textsc{redMaPPer}, we construct an algorithm that identifies clusters with correct BCGs using the SPT clusters as a training sample. As a result, we filter the sample size to $\sim$65\,\%, leaving 263 clusters out of 416 clusters which meet our selection criteria (see Appendix \ref{feature-engg-details} for details). The feature engineering primarily chooses BCGs with a centering probability $P_{\rm cen}$\footnote{the probability of a BCG candidate to be the true BCG.} greater than 0.9. This was motivated by the large fraction of BCGs for which the visual inspection is in agreement with \textsc{redMaPPer}’s choice for SPT. Since we do discard as many as $35\,\%$ BCGs, we test whether the ACT data set could be affected by some selection effects by re-analyzing the ACT sample without removing these BCGs and find the results to be consistent. We show the weak-lensing profiles from this analysis in Appendix \ref{Validation-of-VI}.

We require the feature engineering process to only classify the top-ranked \textsc{redMaPPer} BCG for the ACT clusters. Accordingly, for our final results, we also constrain the SPT data set to the cases where the top-ranked \textsc{redMaPPer} BCG is the BCG of choice after visual inspection, which is found true for almost 79\,\% of the BCGs. The details are provided in Appendix \ref{feature-engg-details}. As an extended exercise, using the visual inspection results of SPT, we have trained several machine-learning classifiers (Random forest, Gaussian process, k-Nearest Neighbour, and Multi-layer Perceptron; see Appendix \ref{ML} for details), but found that $P_{\rm cen}$ is just as good a predictor as these trained classifiers. Therefore, we keep ACTxDES clusters with $P_{\rm cen}$ above 0.9 and that have passed the visual shape inspection. The final cluster samples comprises 166 SPT and 220 ACT clusters.

We note that robust BCG identification and ellipticity measurements were a significant challenge for this work.  Future, deeper imaging from Rubin/LSST and Euclid will significantly improve this step by including measurements of the stellar envelopes of BCGs \citep{Huang2022,Kwiecien2024}.

\subsection{Pair matching algorithm} 
\label{Pair matching}

\begin{figure*}
    \centering
    \includegraphics[width=1.7\columnwidth]{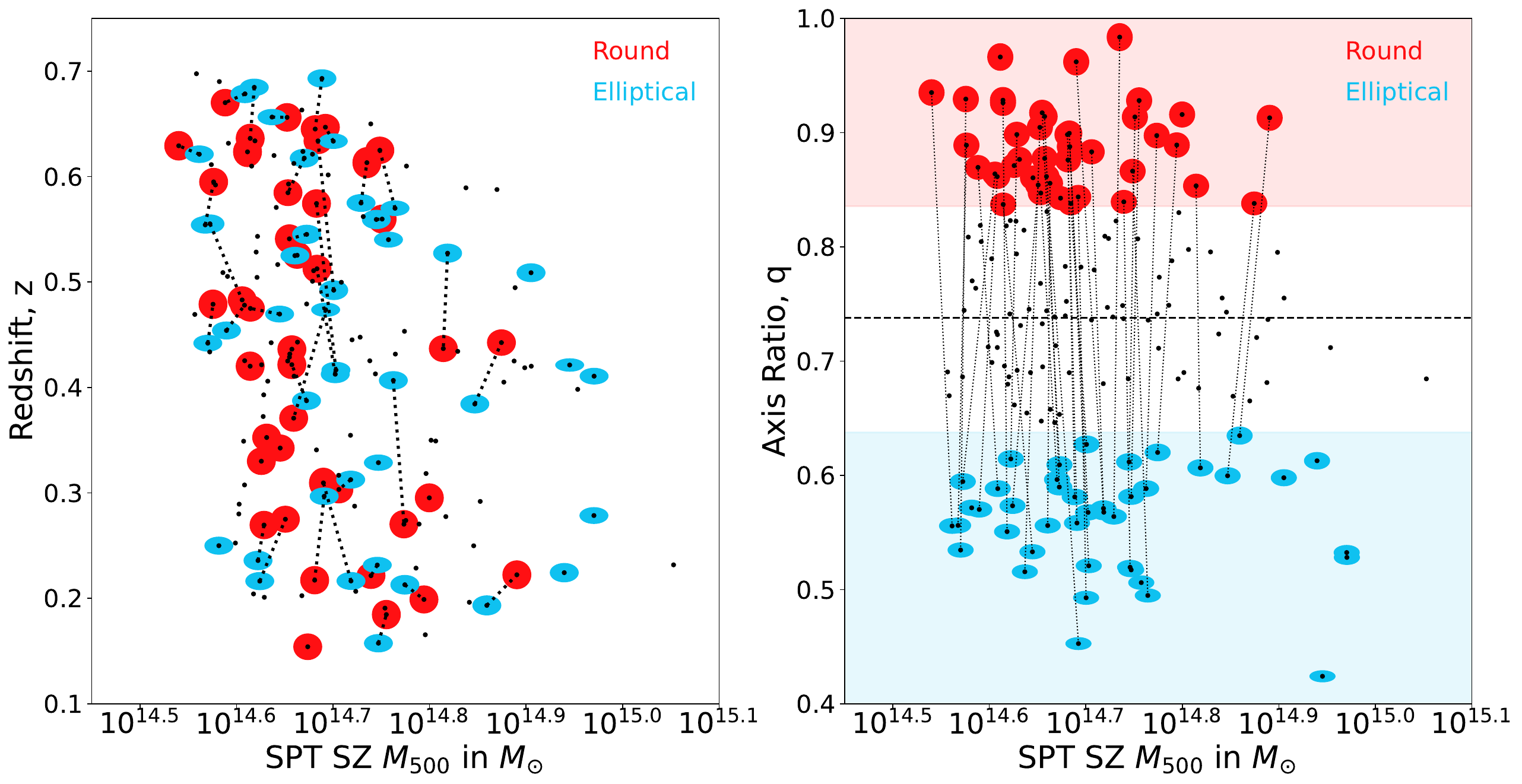}
    \caption{\textit{Left}: Matching of SPT clusters from the round and elliptical bins selected in BCG axis ratio (q). The black dots represent all the clusters in the SPT data set, the red circles represent clusters with round-BCG (q below the 25th percentile), and the blue ellipses represent clusters with elliptical BCGs (q above the 75th percentile). The pairings (dotted lines) were decided on the basis of proximity in SZ $\rm M_{500}$ mass and redshift (see Section \ref{Pair matching}). \textit{Right}: The same pairing is shown with the axis ratio on the vertical axis. The red and blue shaded regions represent the 25th and 75th percentiles in axis ratio (q) respectively. The shape of each scatter point in the round and elliptical bin is representative of the axis ratio of the respective BCG. }
    \label{fig:Matching_pairs}
\end{figure*}

For the remainder of this work, we will consider two subsets of the cluster sample, constructed to be identical for their use in SZ-based cluster cosmology (namely, in SZ-based mass estimates and in redshifts), but one representing the 25\,\% of the sample with the roundest BCGs and one representing the 25\,\% with the most elliptical BCGs. In other words, we consider the <25 and >75 percentiles in axis ratio ($q$) of the BCG. Note that, we use the following relation between axis ratio $q$ and ellipticity $e$: $q = (1-e)/(1+e)$

To construct these two sub-samples, we use a simple pair-matching algorithm in the SZ mass and redshift space. We choose cluster pairs in a window with the size of $\Delta z = 0.16$ and $\frac{1}{1.08}<M_{\rm Round}/M_{\rm Elliptical}<1.08$ in the SZ mass where $M_{\rm Round}$ is the mass of the cluster selected to be in the round-BCG sample and $M_{\rm Elliptical}$ is the mass of the cluster in the elliptical-BCG sample. The algorithm makes multiple iterations until all of the clusters that can be paired have been assigned a cluster of the opposite shape quantile to themselves. In each iteration, clusters that are isolated in this redshift-mass phase space are prioritized, thus leaving the ones with the most neighbors for the last iteration. This process maximizes the number of cluster pairs while still finding the closest matches in mass and redshift as shown in Figure \ref{fig:Matching_pairs}. There are 33 pairs of clusters for SPT and 45 pairs of clusters for ACT assigned with this algorithm.

\section{Measurement Results} \label{Results}
\subsection{Correlation of SZ mass bias and BCG axis ratio} \label{Mass_ratio}

Our analysis is based on SZ-selected clusters, as we expect the SZ selection to be close to a selection in mass due to its lower intrinsic scatter compared with other cluster observables \citep{ Carlstrom-J-2002:SZ, Motl-P-2005:SZ}. However, since the SZ signal is based on the LOS integral of the electron density, we still expect orientation bias to contribute to the intrinsic scatter similar to the optical observables, albeit to a much lower degree \citep{Battaglia-N-2012:SZProjections, Krause-E-2012:SZProjections}. Cluster gas mass $M_{\rm{gas}}$ as measured from X-rays, on the other hand, is a low-scatter mass proxy \citep{Mantz-A-2016:XrayGasMass}, and relatively immune to the orientation bias since the X-ray emission scales with electron density squared. Hence, the ratio of SZ mass estimate to $M_{\rm{gas}}$ should correlate with the cluster LOS orientation, i.e., with BCG axis ratio, $q$. We use the SPT and ACT SZ clusters matched to eRASS1 eROSITA clusters to test this hypothesis.

\begin{figure}
    \centering
    \includegraphics[height=0.85\columnwidth]{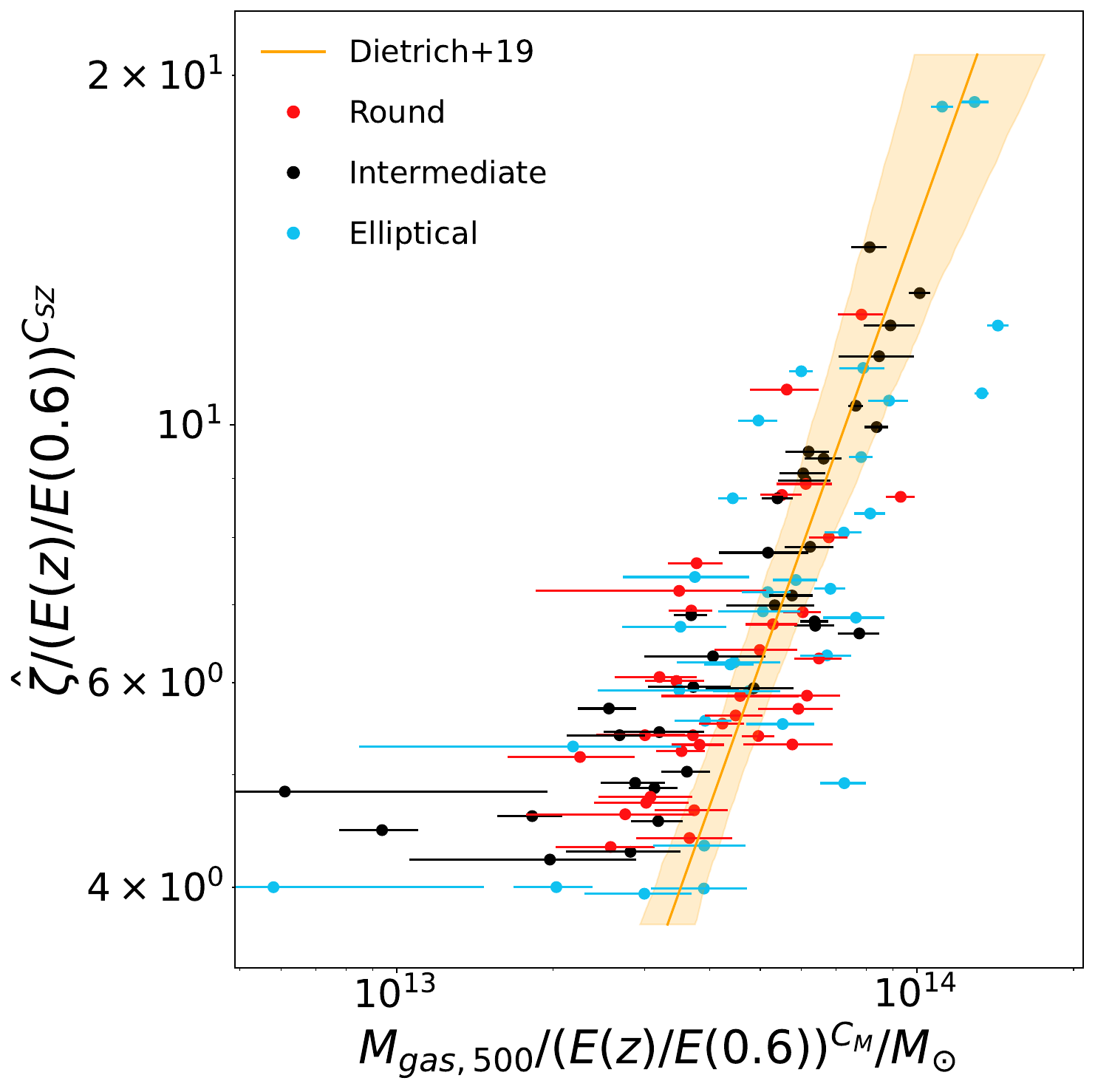}
    \caption{SZ signal-to-noise ratio, $\hat\zeta = \sqrt{\xi^{2} - 3}/\gamma$ vs X-ray gas mass scaling relation for the SPT cluster sample matched to the eRASS1 clusters. Here, the $\gamma$ factor accounts for the varying field depth across the SPT survey. The red points indicate the round-BCG cluster sample, blue indicate the elliptical-BCG cluster sample and the black points are the clusters in between. Also, the fitted scaling relation by \citet{Dietrich-J-2019:Scaling} is plotted as the orange solid line along with $1\,\sigma$ confidence shown in the shaded bands.}
    \label{fig:scaling_relation}
\end{figure}

In Figure \ref{fig:scaling_relation}, we plot the unbiased SZ detection significance $\hat\zeta = \sqrt{\xi^{2} - 3}/\gamma$ of the SPT clusters against the X-ray gas mass from eROSITA, along with the best-fit $\zeta$-mass scaling relation from \citet{Dietrich-J-2019:Scaling}, color-coding each cluster according to BCG ellipticity.  The factor $\gamma$ accounts for varying depth per field, as well as rescaling the ECS field depth to that of the SPT-SZ survey \citep{Reichardt13}, i.e.,
$\gamma_{\rm SPTSZ} = \gamma_{\rm field}$ where the values of $\gamma_{\rm field}$ are provided in Table 1 of \citet{de-Haan-T-2016:SZ} and $\gamma_{\rm SPTECS} = \gamma_{\rm ECS}\times\gamma_{\rm field}$ where the values for $\gamma_{\rm ECS}$ and $\gamma_{\rm field}$ are taken from \citet{Bleem-L-2020:SPTECS}.
If orientation bias dominates the intrinsic scatter in the SZ signal, we would expect the SZ signal of halos aligned along the line-of-sight, i.e., round-BCG clusters, to scatter above the mean relation, and elliptical-BCG cluster to scatter below the mean relation.  Figure \ref{fig:scaling_relation} shows that BCG shape alone does not predict the SZ scatter.
Therefore, Figure \ref{fig:scaling_relation} suggests that either orientation bias alone does not fully account for intrinsic scatter in the $\zeta$-mass relation, or the BCG shape is not a good tracer of line-of-sight orientation in this cluster sample.  However, we also caution that the eROSITA gas mass measurements are based on comparatively shallow data, and are likely to be subject to additional systematic scatter not accounted for in the statistical uncertainties shown here.

\begin{figure*}
    \centering
    \includegraphics[width=1.7\columnwidth]{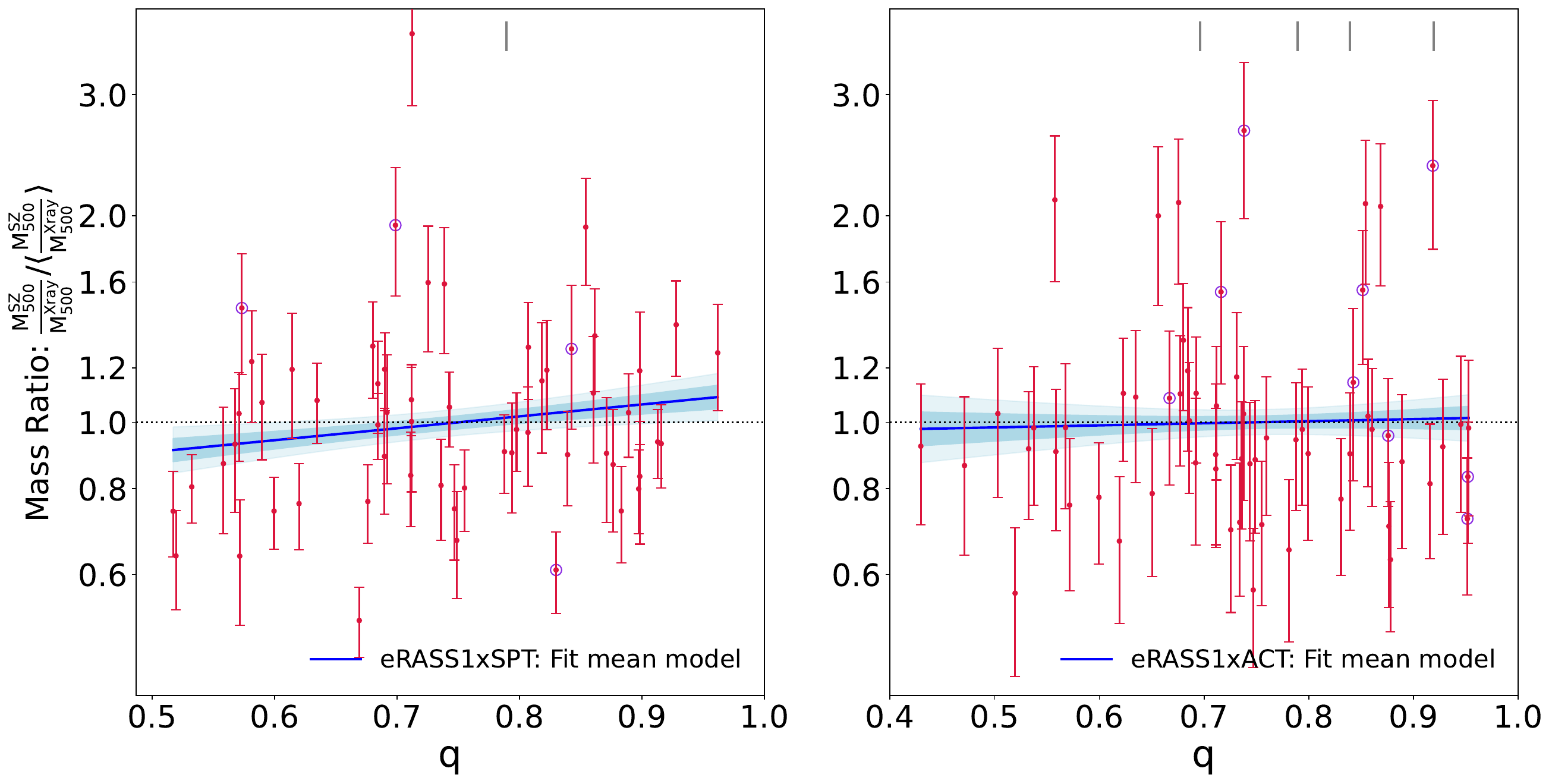}
    \caption{\textit{Left}: The ratio of SZ Mass estimate ($M^{\rm SZ}_{500}$) to X-ray gas mass (${ M}^{\rm Xray}_{500}$) normalized with its mean, against the axis ratio, $q$. We plot the eRASS1$\times$SPT data set in red with $1\,\sigma$ errorbars and fit a log-linear relation using MCMC method (blue). \textit{Right}: similar as the left panel, but for eRASS1$\times$ACT data set. The best-fit ranges are shown with the $1\,\sigma$ and the $2\,\sigma$ confidence bands in light blue. The short vertical gray lines at the top show the values of axis ratio for SZ clusters that were not matched to the X-ray dataset. The blue rings represent the faintest BCGs in the dataset defined as z-band absolute magnitude $M_z> -24.0$  (see Section~\ref{Magnitude gap} for detailed discussion).}
    \label{fig:Mass_ratio_vs_q}
\end{figure*}

\begin{figure}
    \centering
    \includegraphics[height=0.85\columnwidth]{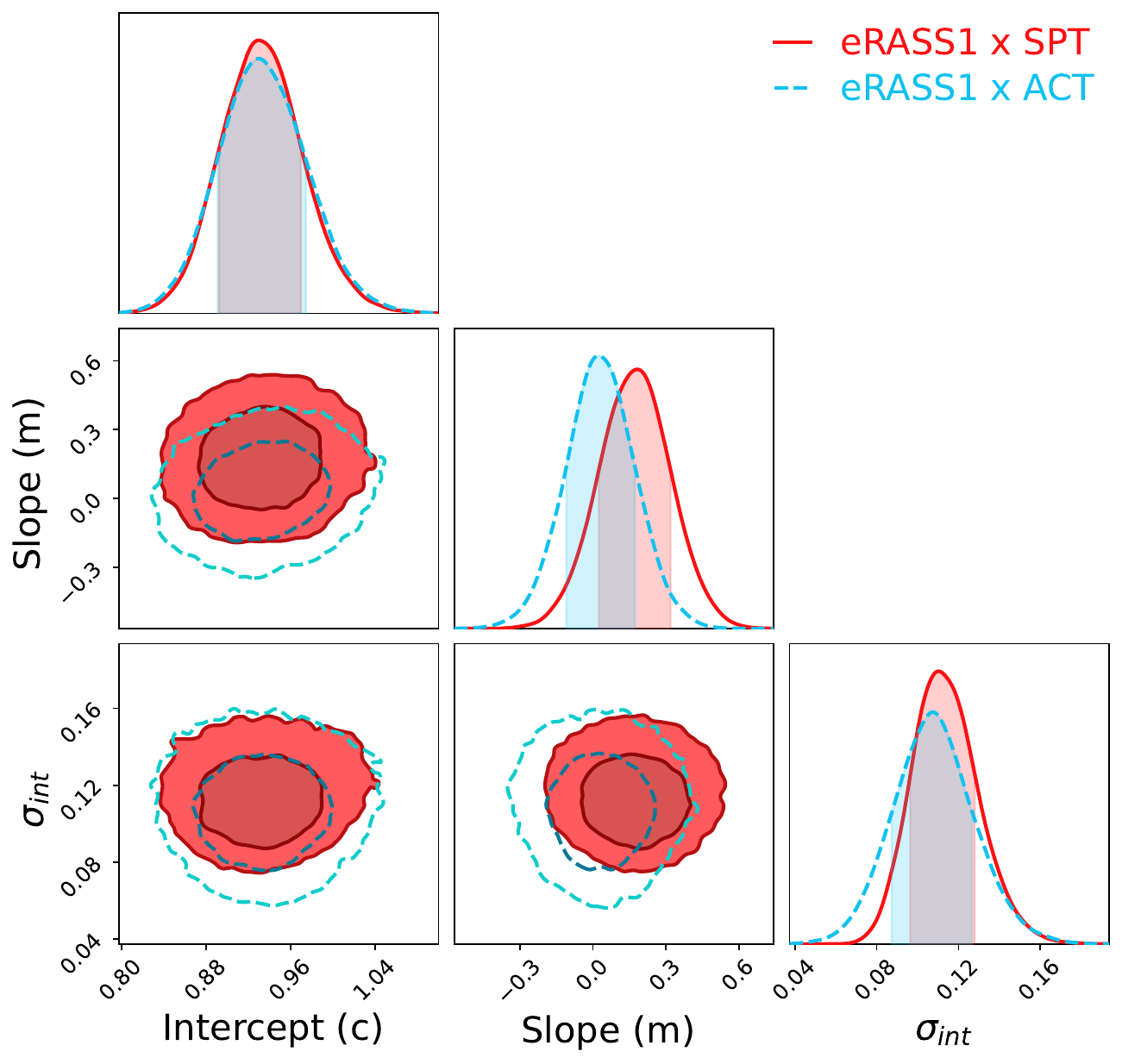}
    \caption{The posterior contours of the linear fit of the SZ to X-ray mass ratio as a function of axis ratio. There are three free parameters: intercept (c), slope (m) and intrinsic scatter ($\sigma_{\rm{int}}$). We neglect the uncertainties in $q$ for the fitting. We find a positive slope with a 86\,\% significance and 66\,\% significance for the SPT$\times$eRASS1 (red) and ACT$\times$eRASS1 (blue) samples respectively, i.e., at most marginal evidence for orientation bias in the SZ selection.}
    \label{fig:Mass_ratio_separate_and_joint_fit}
\end{figure}

We also investigate the dependence of SZ mass to X-ray mass ratio on the BCG shape. Figure \ref{fig:Mass_ratio_vs_q} shows the ratio of SZ mass estimate ($M^{\rm SZ}_{500}$) to X-ray gas mass ($M^{\rm Xray}_{500}$) vs. BCG axis ratio, $q$. If the SZ signal is significantly biased by orientation (and if the 2D BCG shape is an accurate proxy for orientation), then we would expect a positive slope of $\log(M^{\rm SZ}_{500}/M^{\rm Xray}_{500})$.
We fit a linear relation of $\log(M^{\rm SZ}_{500}/M^{\rm Xray}_{500})$ vs. $q$, and show the posterior contours in Figure \ref{fig:Mass_ratio_separate_and_joint_fit}.
Although both best-fit slopes are positive, the result is not statistically significant: for SPT, the measured slope is $0.17 \pm 0.14$ and for ACT, it is $0.03 \pm 0.14$. \footnote{We have not included the uncertainties in the axis ratio for the fit. However, including these should only increase the uncertainty on the measured slope, i.e., the overall trend will still be statistically insignificant keeping the conclusions unchanged.} 
Moreover, the effect is relatively small: the mean SZ to X-ray mass ratio at $q$=0.85 (median of round-BCG sample) is only $(10\pm9)$\% higher than that at $q$=0.6 (median of elliptical-BCG sample) for SPT and $(2\pm8)$\% for ACT. We note that the SPT result (i.e., the mean slope) is likely more robust, since in this comparison, only one SPT cluster was not matched to eROSITA, whereas for ACT, four clusters have no eROSITA match. By definition, the non-matched clusters have larger $M_{\rm SZ}/M_{\rm gas}$ ratios, i.e., they represent systems where the SZ signal is boosted.
Figure~\ref{fig:Mass_ratio_vs_q} shows that the BCGs of the non-matched ACT clusters tend to be rounder than typical, which suggests that these are systems aligned along the line-of-sight, with X-ray fluxes too low to be detected by eRASS1. With X-ray detections, these clusters would populate the top right of the plot, meaning that the slope measured here is likely to be biased low.

In comparison, \citet{Herbonnet-R-2019:Ellipticity} found a 20\,\% higher weak lensing mass than average for the round-BCG cluster sample and 20\,\% lower for the elliptical-BCG cluster sample, i.e., an overall effect of $40-50$\,\% between the two subsamples. This suggests that the effect of orientation bias is much larger on weak lensing observables than on the SZ signal. Therefore, while SZ-selected clusters are not truly mass-selected, they are close to being so. 

\subsection{Orientation effects in richness}

The richness of a cluster from an optical survey is a mass proxy that is independent of the SZ detection. Therefore, in the absence of projection effects, assuming that the SZ selection is close to the mass selection, one would expect the richness of the round-BCG cluster sample and that of the elliptical-BCG cluster sample to have statistically equivalent distribution of richness. However, we hypothesize that, since the richness of a cluster is the sum of the membership probabilities of all candidate member galaxies within the cluster radius, it is susceptible to the effects of cluster orientation. A prolate cluster aligned along the LOS would exhibit an increased number of member galaxies due to the LOS projections, while a cluster aligned along the plane-of-sky would show a reduced number of member galaxies. In this case, a certain fraction of the observed scatter in richness could be attributed to the effect of halo orientation. 
As a result, for a sample of observed clusters with a richness selection, clusters projected along the LOS with boosted richness values would be preferentially detected compared to those aligned perpendicular to the LOS, leading to a cluster sample that is over-represented by the LOS-oriented (round-BCG) clusters.

\begin{figure*}
    \centering
    \includegraphics[width=1.7\columnwidth]{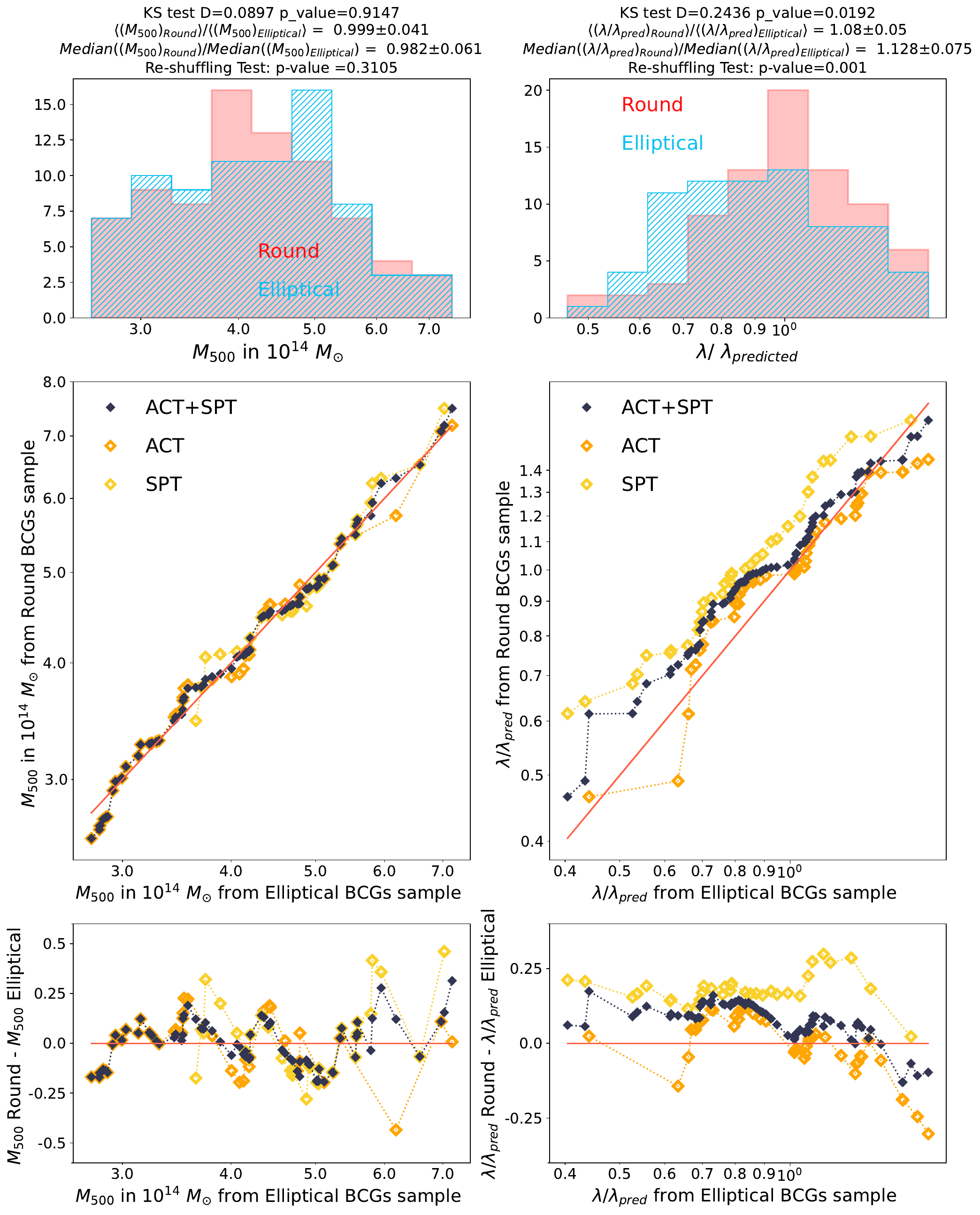}
    \caption{
    \textit{Top}: The distribution of SZ mass $M_{500}$ (left) and the distribution of relative richness (the ratio of the \textsc{redMaPPer} richness to the predicted richness using the scaling relation from \citet{Bleem-L-2020:SPTECS}, (right) for the combined ACT and SPT sample. \textit{Middle}: The Quantile-Quantile (QQ) plot for the cluster mass, $M_{500}$, of the round-BCG and the elliptical-BCG samples (left) and that for the relative richness (right). The SPT SZ clusters are plotted in yellow, the ACT SZ clusters in orange and the combined ACT+SPT sample in black. 
    \textit{Bottom}: The difference between the Round and the elliptical-BCG sample shown in the QQ plots. Note that the relative richness is systematically biased high for the round-BCG clusters (aligned along the LOS) compared to the elliptical-BCG clusters (aligned along the plane-of-sky). Above the top panels, we show a few summary statistics to compare the distributions of the round-BCG sample and the elliptical-BCG sample. We use the Kolmogorov–Smirnov (KS) statistics and a re-shuffling test to rule out the null hypothesis that the two sub-samples follow identical distributions. The KS 2-sample test and the shuffling test confirm the observed discrepancy in the richness distributions for the two samples with significance well above $3\,\sigma$.
    }
    \label{fig:QQ_PLOT_SPT_and_ACT}
\end{figure*}

With an SZ-selected sample matched to optical \textsc{redMaPPer} clusters one-to-one (see Section \ref{likelihood-matching}), we study the bias in the richness measurement at fixed mass and redshift. We compare the relative richness computed as a ratio of the observed richness to the richness computed from the scaling relation in \citet{Bleem-L-2020:SPTECS} to account for the dependence on mass. We use Quantile-Quantile (QQ) plots to compare the distributions of the round-BCG and the elliptical-BCG sub-samples (Section~\ref{Pair matching}), which are related to the scatter in relative richness at fixed mass.

The QQ plot for the relative richness for the SPT and the ACT data sets are shown in the right panels of Figure \ref{fig:QQ_PLOT_SPT_and_ACT}. The relative richness of the clusters in the round-BCG sample is systematically higher than that in the elliptical-BCG sample. 
The discrepancy is significant above 99.7\,\% ($3\,\sigma$) as reported by the p-values from the Kolmogorov–Smirnov 2-sample test (0.0006) and the re-shuffling test (0.0005). In the re-shuffling test, we generate a large number of sample pairs with randomly assigned labels (Elliptical or Round) to construct distributions, and we then compute the summary statistics (e.g., mean, median) for each randomly paired sample to derive the p-value between the observed round and elliptical-BCG sample distributions. We also show a similar set of plots in the left panels for the SZ mass instead of relative richness, from which we confirm that the mass distributions of the elliptical-BCG and the round-BCG sample are consistent with each other (see the p-values reported in Figure \ref{fig:QQ_PLOT_SPT_and_ACT}).

\subsection{Weak lensing method and observable} \label{Weak lensing method}
Given the expected correlation between the orientation of the galaxy cluster with respect to the LOS and its projected density, we investigate the excess surface density profiles ($\Delta\Sigma$) for the clusters stacked in the two bins of orientation, Round and Elliptical, using the weak-lensing method. 

Given a surface density profile $\Sigma(R)$ with $R$ being the radial distance measured from the BCG, the corresponding $\Delta\Sigma(R)$ can be expressed as
\begin{equation}
    \Delta \Sigma (R) = \frac{2}{R^2}\int_0^R \Sigma(R') R' \text{d}R' - \Sigma(R)\,,
    \label{eq:delta_sigma}
\end{equation}
where $\Delta\Sigma (R)$ is related to the tangential shear $\boldsymbol\gamma_{\rm t}(R)$ as
\begin{equation}
    \Delta \Sigma (R) = \Sigma_{\rm crit} \boldsymbol\gamma_{\rm t} (R) \,,
\end{equation}
and
\begin{equation}
    \Sigma_{\rm crit} = \frac{c^2}{4 \pi G} \frac{D_{\rm A,s}}{D_{\rm A,l} D_{\rm A, ls}}\,.
\end{equation}
Here, $D_{\rm A,s}$ represents the angular diameter distance to the source galaxy, $D_{\rm A,l}$ that to the lensing cluster, and $D_{\rm A,ls}$ that between the lensing cluster and the source galaxy.

In practice, we measure the reduced shear, $\textbf{g}$, rather than the true shear, $\boldsymbol{\gamma}$. The reduced shear is related to the true shear as:
\begin{equation}
    \textbf{g} = \frac{\boldsymbol{\gamma}}{1 - \Sigma/\Sigma_{\rm crit}}\,.
\end{equation}
However, the reduced shear can be approximated as the shear ($\textbf{g} \approx \boldsymbol{\gamma}$) as we move away from the cluster center. The convergence ($\Sigma (R) /\Sigma_{\rm crit}$), is negligible in most cases of weak lensing. We refer the reader to e.g., \citet{Bartelmann2017} for a more complete review of weak lensing.

Given these equations and the galaxy shapes measured by \textsc{metacalibration} (see Section \ref{Data}), our lensing estimator is,
\begin{equation}
    \label{eq:DeltaSigma_estimator}
    \Delta\tilde{\Sigma} (R) = \frac{\sum_{ij} s^{ij} \boldsymbol{{\rm e}}_{\rm t}^{ij} (R) }{\sum_{ij} s^{ij} \Sigma^{-1}_{\rm crit}(z^i_{\rm l},z^j_{\rm s,MC}) (\mathcal{R}+\mathcal{R}_{\rm s})^j}\,,
\end{equation}
where $i$ runs over the clusters and $j$ runs over the source galaxies.
Here, $\boldsymbol{{\rm e}}_{\rm t}$ is the tangential component of the measured ellipticity of source galaxies, $\Sigma^{-1}_{\rm crit}(z_{\rm l},z_{\rm s,MC})$ is the inverse of $\Sigma_{\rm crit}$ evaluated at the lens redshift with the source redshift randomly chosen from the probability distribution (pdf) given by \textsc{DNF} algorithm, $\mathcal{R}+\mathcal{R}_{\rm s}$ is the sum of the shear response and the selection response, and $s$ is the weight for each galaxy, 
\begin{equation}
    s^{ij} = \omega^j \Sigma^{-1}_{\rm crit}(z^i_{\rm l},z^j_{\rm s,mean})\,, 
\end{equation}
where $\omega$ denotes the square inverse of the shape measurement error for each source galaxy and $z_{\rm s,mean}$ is evaluated at the mean of the source redshift pdf from \textsc{DNF} \citep{DNF}. 
Note that according to \textsc{metacalibration} formalism, $\langle\boldsymbol{g}\rangle = \langle\boldsymbol{\textrm{e}}\rangle/\langle\mathcal{R}\rangle$.
To minimize the diluting effect from the foreground and the cluster member galaxies onto $\Delta\Sigma$, we apply a redshift cushion of $z_{\rm s,mean} > z_{\rm l} + 0.1$.
We refer readers to \citet{McClintock2019} for detailed validation of this estimator. 

Note that our estimator suffers from a few percent of multiplicative biases for the galaxy shape measurement and for the photometric redshift uncertainty leaking into the estimation of $\Sigma_{\rm crit}$ \citep{McClintock2019}. However, given that our goal is to compare $\Delta\Sigma$ of the round-BCG and elliptical-BCG cluster samples matched in SZ mass and redshift and that those biases affect the both samples equally, we do not apply those few percent level corrections.

Another important systematic bias in weak lensing is contamination due to the cluster member galaxies, which do not show any weak lensing effect, and thus dilutes our weak lensing signal.
It is generally called the ``boost factor'' ($\mathcal{B}$), which one must estimate and thereby correct the measured data. 
\citet{Varga-T-2019:DESY1}, based on the method suggested by \citet{Gruen-D-2014}, used the data from the DES Y1 to show that the boost factor could be unbiasedly obtained by decomposing the observed source redshift distribution around clusters, at each radial bin, into the background and the cluster contamination component:
\begin{equation}
    P(z|R) = f_{\rm cl}(R) \, P_{\rm cont}(z|R) + (1-f_{\rm cl}(R)) \, P_{\rm bg}(z)\,,
\end{equation}
where $f_{\rm cl}(R)$ represents the fraction of the cluster member contamination as a function of cluster-centric radius $R$ and $P_{\rm bg} (z)$ the background source redshift distribution evalulated at random points on the sky. 
As in \citet{Varga-T-2019:DESY1}, we assume a Gaussian distribution as a function of $z$ for the cluster member contamination, $P_{\rm cont} (z|R)$, with the mean and the standard deviation of the Gaussian being free parameters. 
Therefore, given $P(z|R)$ and $P_{\rm bg} (z)$ from the observations, we can constrain $f_{\rm cl}$ at each radial bin.
Accordingly, the boost factor is related to $f_{\rm cl}$ as 
\begin{equation}
    \mathcal{B}(R) = \frac{1}{1-f_{\rm cl}(R)}\,,
\end{equation}
and the corrected weak lensing profile becomes
\begin{equation}
    \Delta\tilde{\Sigma}_{\rm corr} = \mathcal{B}\Delta\tilde{\Sigma}\,.
\end{equation}

For the estimation of $\Delta \tilde {\Sigma}$ and $\mathcal{B}$, we use 12 logarithmically spaced radial bins ranging from 0.2 to $100~h^{-1} {\rm Mpc}$ in physical units. The uncertainties on the data points are estimated using 20,000 bootstrap samples of the clusters. We have checked that increasing the number of the bootstrap samples does not meaningfully change the covariance matrix. Note that we do estimate a separate boost-factor for the round-BCG and elliptical-BCG sample using the above procedure.

\subsection{Galaxy number density profile}
\label{sec:galden}
We measure the projected number density profiles of galaxies around each cluster sub-sample. From the DES Y6 GOLD galaxy catalog, we first select galaxies that are brighter than i-band magnitude of 23.1 with the uncertainty on the magnitude smaller than $0.1$. Furthermore, to ensure that we do not miss any galaxy due to the survey depth, we select the survey footprint which is deeper than $m_{\rm i}=23.1$, which removes $\sim$5\,\% of the sky coverage. 
Then for each cluster, following the estimator developed in \citet{Davis1983}, we calculate the correlation function, $\omega (R)$ between the cluster and the galaxy sample as 
\begin{equation}
    \omega(R) = \frac{N_{\rm DD}(R) - N_{\rm DR}(R)}{N_{\rm DR}(R)}\,,
\end{equation}
where $N_{\rm DD}(R)$ represents the number of pairs between the cluster and the galaxy sample separated by $R$ and $N_{\rm DR} (R)$ that between the cluster and the random points that follow the galaxy survey footprint. We have verified that including random points for the clusters using the estimator developed by 
\citet{LandySzalay93} does not change our data points significantly. 

To ensure that we select a similar galaxy population over the redshift range while including as many galaxies as possible per cluster, we further impose a cut on the absolute magnitude in $i$-band, $M_{\rm i}<-19.27$, which corresponds to the apparent magnitude cut $m_{\rm i}<23.1$ at $z=0.7$. 
We do not impose any cut on galaxies based on the photometric redshift estimation, since the correlation function itself measures the excess galaxy number density, and thus naturally
picks up galaxies that are correlated with the cluster position \citep[see e.g.,][for details]{Shin-T-2021:SZ}.
To this extent, when calculating the correlation function, we place all galaxies at the cluster redshift regardless of their photo-z when applying the absolute magnitude cut.
$\omega(R)$ is then converted to the mean-subtracted galaxy surface density as
\begin{equation}
    \Sigma_{\rm g} (R) = \langle \Sigma_{\rm g} \rangle \omega (R)\,,
\end{equation}
where $\langle \Sigma_{\rm g} \rangle$ represents the mean surface number density of the galaxies over the survey footprint. 
The final galaxy surface number density profile, averaged over our cluster samples, is 
\begin{equation}
    \bar{\Sigma}_{\rm g} (R) = \frac{\sum_i \Sigma_{{\rm g},i} (R) w_{{\rm tot},i}}{\sum_i w_{{\rm tot},i}}\,,
\end{equation}
where $i$ runs over the clusters and $w_{{\rm tot},i}$ is the sum of the lensing weights, $s \times (\mathcal{R}+\mathcal{R}_{\rm s})$, of the source galaxies for the $i$-th clusters (Eq.~\ref{eq:DeltaSigma_estimator}). Note that, to compare the galaxy density profiles to the lensing profiles unbiasedly, we include the total lensing weights to each cluster to account for the different lensing power of clusters.

\subsection{Results from the density profiles}
\label{sec:WL_result}

We measure the stacked lensing profiles and projected galaxy number density profiles of clusters in bins of ellipticity to compare the underlying matter distribution of the clusters with different LOS orientation. 

\subsubsection{Weak lensing profile results}
\begin{figure*}
    \centering
    \includegraphics[width=1.7\columnwidth]{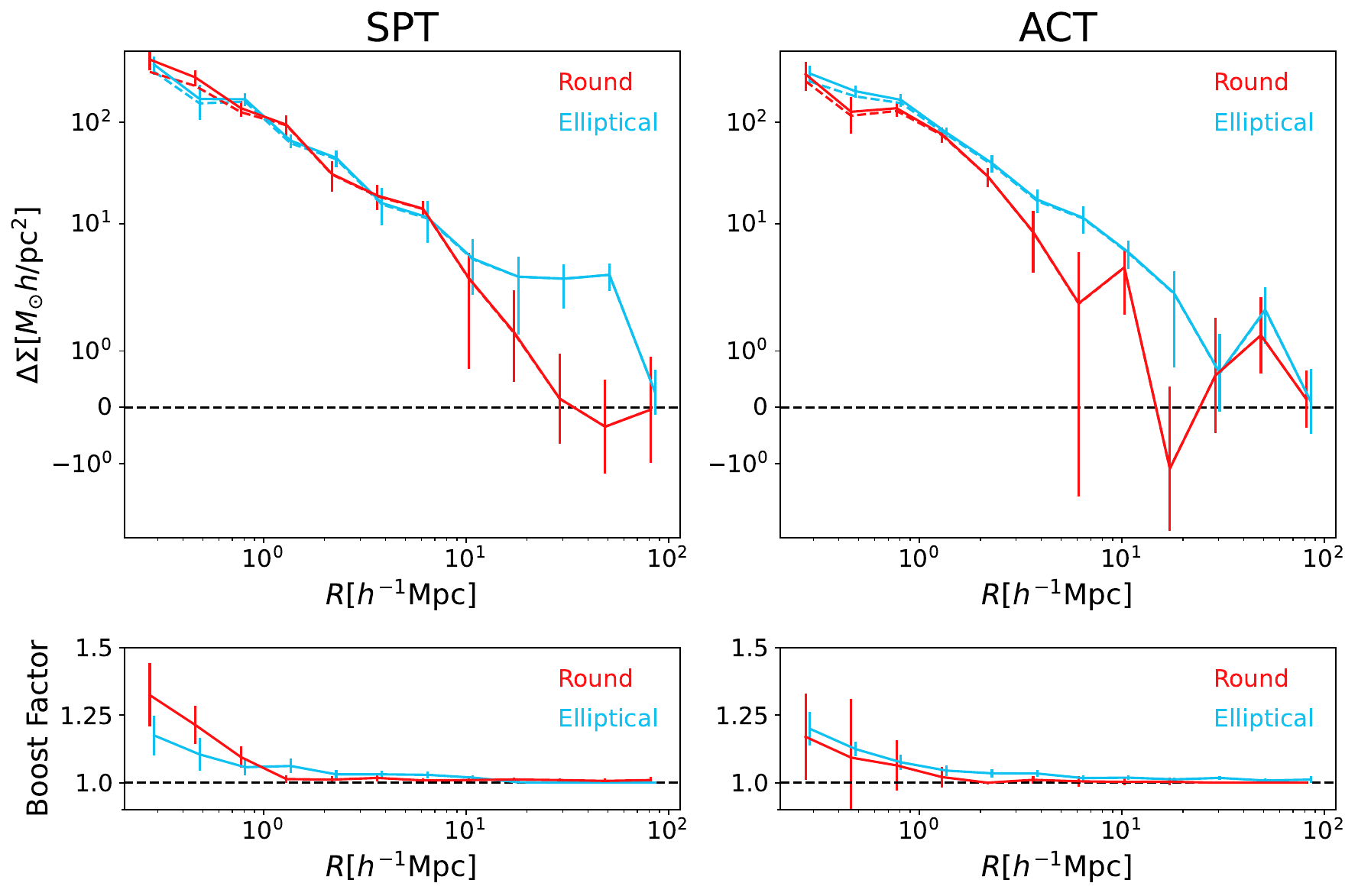}
    \caption{The stacked weak-lensing $\Delta\Sigma$ profiles for the SPT cluster sample (left) and for the ACT cluster sample (right). The red data points represent the round-BCG cluster sample and the blue data points represent the elliptical-BCG cluster sample with the split made at the 25th and the 75th percentiles. The solid (dashed) lines with corresponding colors represent the lensing profiles with the boost factor (not) corrected. We observe a significant discrepancy between the two samples at the 2-halo regime.}
    \label{fig:ACT_and_SPT_delta_sigma}
\end{figure*}

In Figure \ref{fig:ACT_and_SPT_delta_sigma} we show the stacked lensing profiles (with and without the boost factor correction) of the round- and the elliptical-BCG cluster sample for SPT and ACT. For both, the amplitudes of the profiles are similar at small scales and diverge at large scales, with the elliptical-BCG cluster sample having a larger projected density than the round-BCG cluster sample.  For SPT, this break occurs at $\sim 10~h^{-1} {\rm Mpc}$, with $\Delta \Sigma$ of the round-BCG sample being consistent with 0 at larger scales.  For ACT, the break occurs at smaller scales, around $\sim 3~h^{-1} {\rm Mpc}$.
The significance of the difference between the round- and the elliptical-BCG samples is $2.85\,\sigma$ for SPT above $10~h^{-1} {\rm Mpc}$  and $2.85\,\sigma$ for ACT above $3~h^{-1} {\rm Mpc}$ with a combined significance of $2.8\,\sigma$ when measured above $6~h^{-1} {\rm Mpc}$ for both SPT and ACT.

This result is counter-intuitive when compared with the predictions from the simulations. \citet{Osato-K-2018:Orientation}, in their Figure 2, show that the surface density profile of the halos aligned along the LOS is boosted at all scales compared with those aligned perpendicular to the LOS, showing a dumbbell-like residual in the ratio of the profiles. \citet{Zhang-Z-2023:Triaxiality} showed the same for $\Delta \Sigma$ profiles (presented in Figure \ref{fig:Mass_dependence_on_2halo}). Both studies find an excess in the density profile of clusters aligned along the LOS over the entire radial range in both the 1-halo regime (closer to the cluster center at $R<R_{200\rm m}$) and the 2-halo regime ($R \gtrsim R_{200\rm m}$).  In the 1-halo regime, the excess can be straightforwardly understood from the projection of the halo along its major axis, as well as of a significant fraction of the correlated large-scale structure, since filaments tend to feed into halos along the major axes.  Simulations predict that the anisotropy of the mass distribution continues to at least $100~h^{-1}{\rm Mpc}$ \citep[see e.g., Figure~6--8 of][]{Osato-K-2018:Orientation}, which explains the excess in the 2-halo regime; in simple terms, in this case the major filaments are viewed along their extent.
The expectation from the simulations is therefore that the weak-lensing profile of the round-BCG clusters should exceed that of the elliptical-BCG clusters on all scales, which is not what we find in Figure \ref{fig:ACT_and_SPT_delta_sigma}. To highlight the differences, we also show a ratio of the lensing profiles between the round-BCG and elliptical-BCG sample in Figure \ref{fig:ACT_and_SPT_ratio_plot}. The ratio is significantly less than 1 in the 2-halo regime.

\subsubsection{Galaxy number density profile results}

\begin{figure*}
    \centering
    \includegraphics[width=1.7\columnwidth]{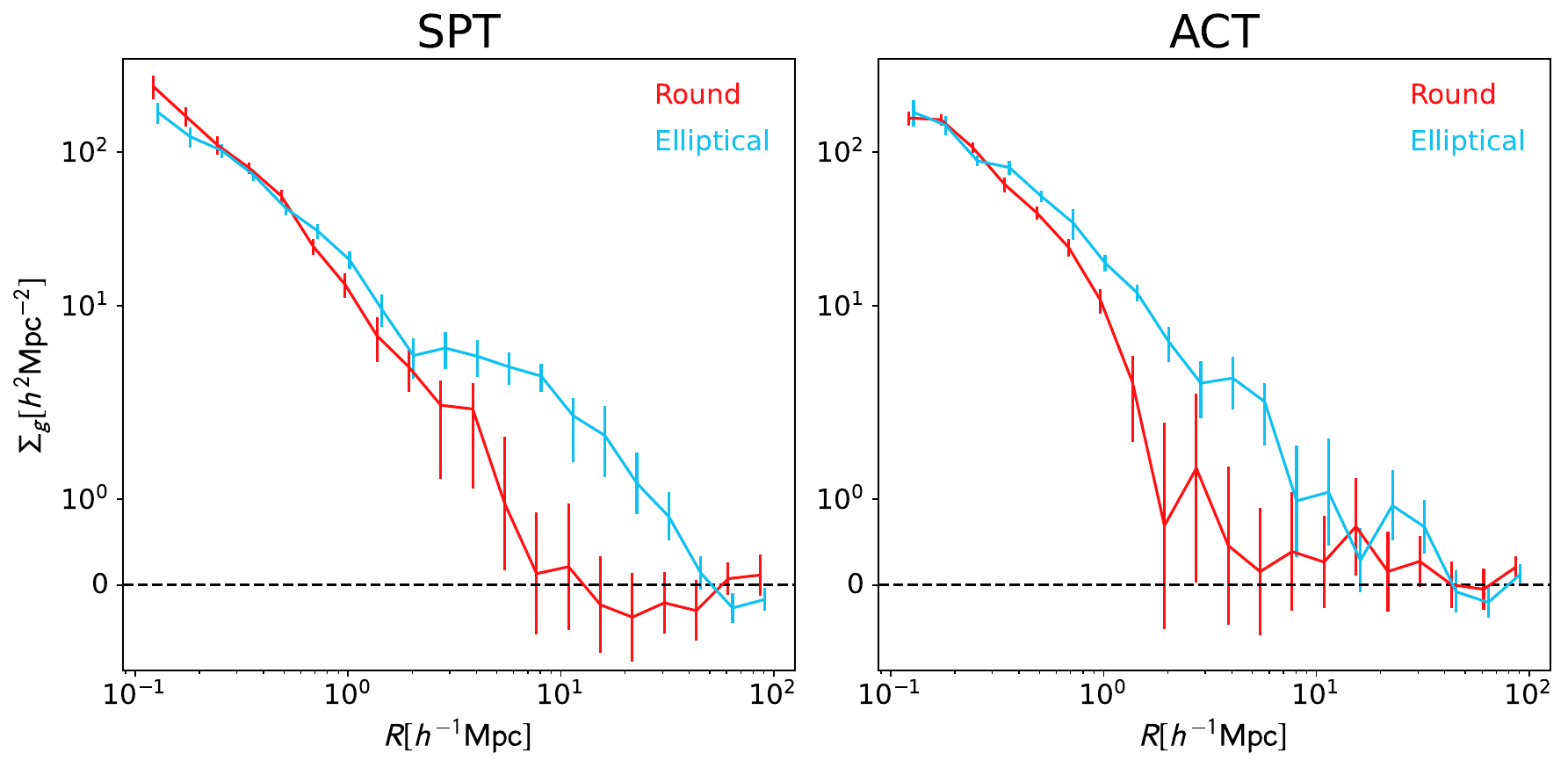}
    \caption{Surface number density profiles of galaxies ($\Sigma_{g}$) in the round-BCG (red) and elliptical-BCG (cyan) cluster samples (split at 25th-75th percentiles) for the SPT data set on the left and for the ACT data set on the right. Again, we observe a discrepancy between the two cluster samples at scales above $1~h^{-1} {\rm Mpc}$.}
    \label{fig:ACT_and_SPT_galaxy_density}
\end{figure*}

We use the galaxy density profiles as a second method to measure the matter density profiles around the cluster subsamples. 
The results are shown in Figure \ref{fig:ACT_and_SPT_galaxy_density}, which shows that the measured galaxy density profiles closely follow the trend of the weak-lensing profiles shown in Figure \ref{fig:ACT_and_SPT_delta_sigma}. At small scales ($< 1~h^{-1} {\rm Mpc}$), the amplitude of the galaxy density profiles is similar for the round-BCG and elliptical-BCG samples, whereas at larger scales the amplitude of the elliptical-BCG sample is significantly higher than that of the round-BCG sample, which is largely consistent with 0. The statistical significance of the difference is larger than in the weak-lensing profiles: above $1~h^{-1} {\rm Mpc}$, it is $4.5\sigma$ for SPT and $5.4\sigma$ for ACT.

\subsubsection{Consistency of profile trends between weak lensing and galaxy density}

\begin{figure}
    \centering
    \includegraphics[width=0.85\columnwidth]{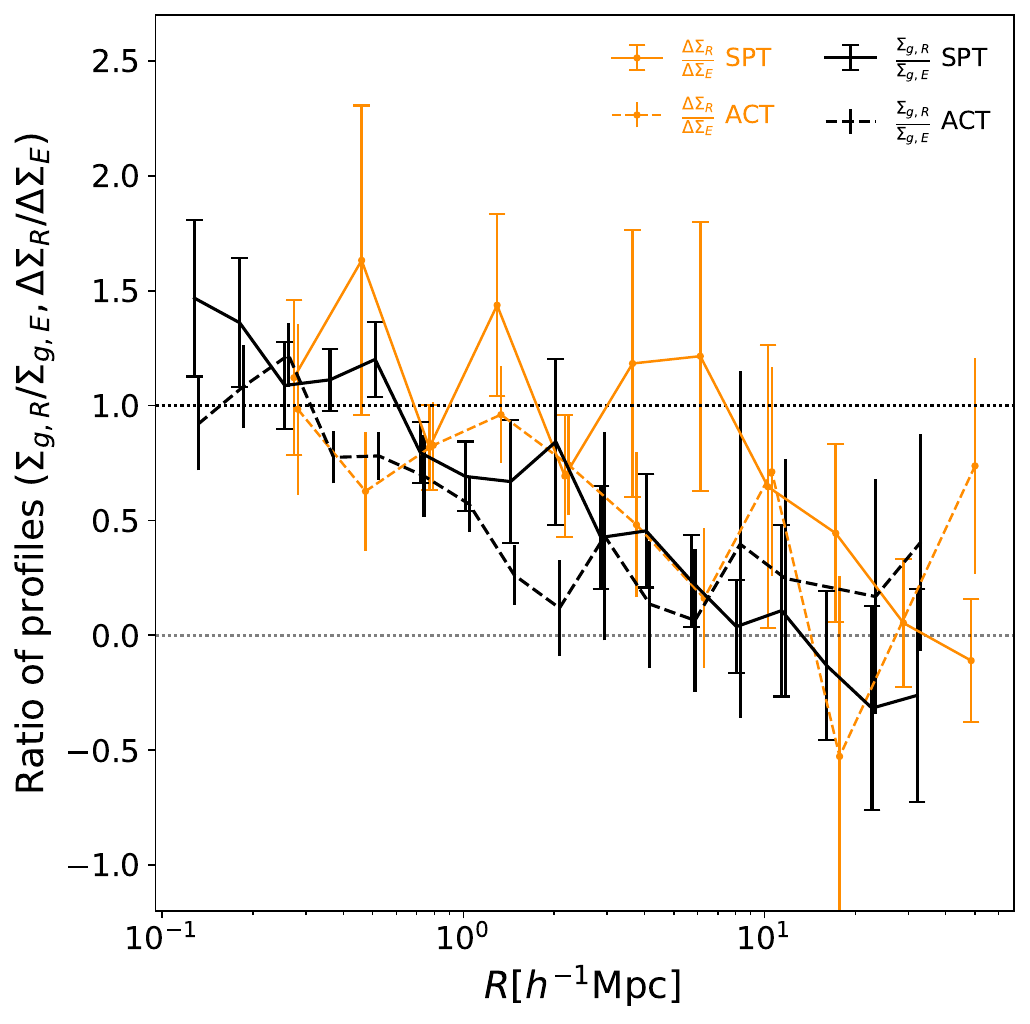}
    \caption{ Ratio of the round-BCG to elliptical-BCG lensing profiles in orange and of the surface number density profiles of galaxies in black. We show the profiles for SPT with solid lines and for ACT with dashed lines. We do not include data-points where the elliptical-BCG profile is consistent with zero within $1.5\,\sigma$.}
    \label{fig:ACT_and_SPT_ratio_plot}
\end{figure}

We show the ratio of both the weak-lensing and galaxy density profiles between the round-BCG and elliptical-BCG sample in Figure \ref{fig:ACT_and_SPT_ratio_plot}.
All four ratio profiles show a consistent picture of being $\sim 1$ on small scales ($\lesssim 1 h^{-1}$Mpc), and dropping to 0 on large scales ($\gtrsim 10 h^{-1}$Mpc).
For both the weak-lensing and galaxy density profiles, the profile bifurcation starts at a smaller scale for the ACT clusters than the SPT clusters. This similarity indicates that the galaxy distribution is a good proxy for the underlying dark matter distribution per our galaxy selection and also confirms that the trend seen in the weak-lensing profiles is most likely to be physical. 
It is also noteworthy that the bifurcation in the profiles between the elliptical-BCG and round-BCG sample occurs at a smaller scale ($\sim$$2~h^{-1}{\rm Mpc}$) for the galaxy density profile than that for the weak lensing profile ($\sim$$8~h^{-1}{\rm Mpc}$).  This scale difference between density profile and weak lensing profile is similar to the scale difference found by \citet{Wu-H-2022:Orientation} when comparing stacked profiles of clusters with orientation-induced selection biases to mass-selected samples. 

We note that we have tested where the profiles of an in-between sample fall.  Because of the pair-matched definition of the round-BCG and elliptical-BCG samples, the construction of such a sample is not straightforward.  For different definitions, we consistently find that the profiles of the in-between samples lie in between the round-BCG and elliptical-BCG samples, consistent with a systematic trend in $q$. However, in SPT the in-between sample is closer to the round-BCG sample, whereas in ACT it is closer to the elliptical-BCG sample.  This likely reflects intrinsic scatter in the BCG shape measurement (see also Figure~\ref{fig:shape_validation_SPT}), i.e., some leakage of the round-BCG / elliptical-BCG sample into the in-between sample, and vice-versa. Since the BCG classification differs between SPT and ACT, it is possible that the amount and direction of leakage is not the same between the two.

\section{Interpretation of the observed profiles}
\label{sect:toy-model}

Both the weak lensing profiles and galaxy density profiles trace the underlying matter distribution in galaxy clusters. While we find consistent results between the lensing and the galaxy density profiles, they are opposite to our initial expectation. In this section, we discuss possible explanations for our observations.

\subsection{Orientation bias in the SZ selection}

\begin{figure}
    \centering
    \includegraphics[width=0.85\columnwidth]{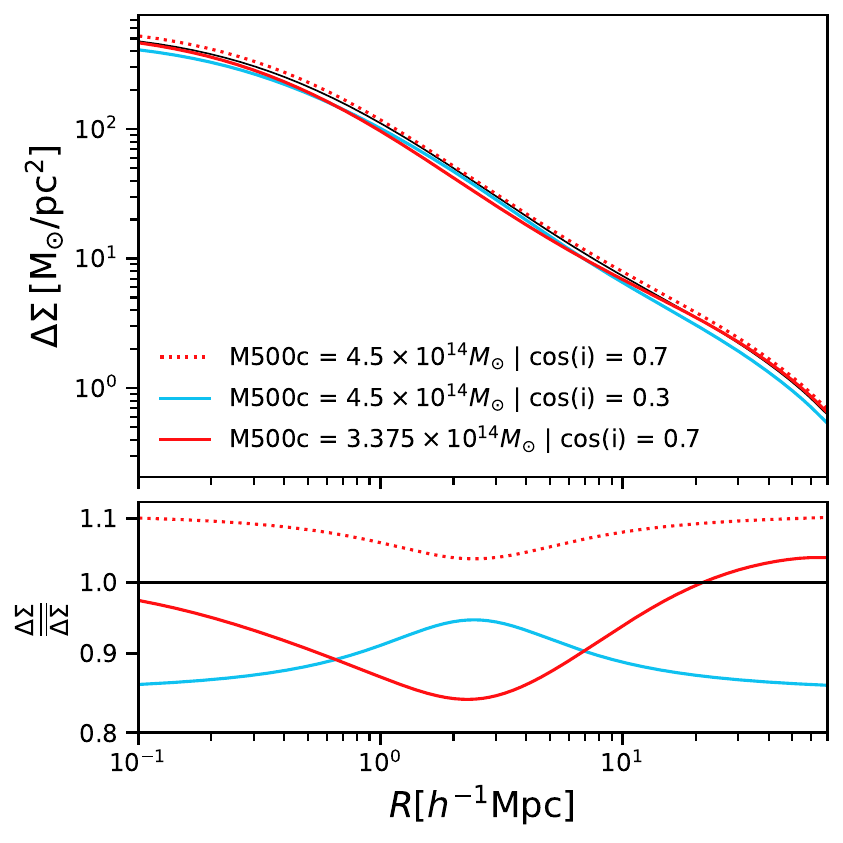}
    \caption{\textit{Top}: The theoretical (1-halo NFW + 2-halo given by the mass-bias relation of \citet{Tinker-J-2010:HaloBias}) $\Delta \Sigma$ profiles of two representative clusters, one aligned along the LOS (dotted red, cos(\textit{i}) = 0.7) and the other aligned to the plane-of-sky (solid blue, cos(\textit{i}) = 0.3) with the same mass, $M_{\rm 500\rm c} = 4.5 \times 10^{14} M_{\odot}$. Also plotted is the profile of a cluster aligned along the LOS (cos(\textit{i}) = 0.7) with a 25\,\% lower intrinsic mass, $M_{\rm 500\rm c} = 3.375 \times 10^{14} M_{\odot}$ to match the 1-halo amplitudes between the two halos (solid red). For reference the original spherical NFW halo is shown in black. 
    \textit{Bottom}: The ratio of each lensing profile to that of a same-mass cluster when orientation bias is not taken into account.}
    \label{fig:Mass_dependence_on_2halo}
\end{figure}

In the 1-halo regime, our expectation was that the lensing profiles of the round-BCG samples exceed those of the elliptical-BCG samples, owing to the increased projected surface density.  For X-ray-selected WtG clusters, \citet{Herbonnet-R-2019:Ellipticity} find that weak-lensing mass estimates of the round-BCG clusters are a factor of $\sim 1.5$ higher than those of elliptical-BCG clusters at fixed X-ray gas mass.  Instead, we find that the profile amplitudes of the round-BCG and the elliptical-BCG samples are very similar.  Partially, this
could be the result of orientation bias manifesting in the SZ selection: since the clusters from the two samples are matched in the {\it measured} SZ mass, if we see evidence for orientation bias in the SZ mass (Section \ref{Mass_ratio}), then it is plausible that the round-BCG sample has a lower true average mass than the elliptical-BCG sample, but that the profiles are boosted high due to projection effects.
The comparison with X-ray gas mass estimates (Section \ref{Mass_ratio} and Figure~\ref{fig:Mass_ratio_vs_q}) shows that the mass difference between the two samples cannot be very large, and that the round-BCG sample likely cannot be more than $\sim$10\% less massive than the elliptical-BCG sample.

For cluster-sized halos, the large-scale bias parameter $b$ which sets the amplitude of the 2-halo term is a strong function of halo mass \citep{Tinker-J-2010:HaloBias}.  If the round-BCG sample has a lower average mass than the elliptical-BCG sample, it also has a lower $b$, and the (spherically averaged) amplitude of the 2-halo profiles will be lower than for the elliptical-BCG sample.  However, this suppression of the 2-halo term of the round-BCG sample will be counteracted by the boosting due to the orientation along the LOS.  To estimate whether our results are simply due to a lower average halo mass of the round-BCG sample, we use the \textsc{Colossus} package \citep{Diemer-B-2018:Colossus} to compute the lensing profile of a spherical NFW halo with mass $M_{500\rm c} = 4.5 \times 10^{14} M_{\odot}$ (similar to the mean mass of the samples) and a 2-halo term given by the mass-bias relation of \citet{Tinker-J-2010:HaloBias}, shown as the black line in Figure~\ref{fig:Mass_dependence_on_2halo}.  \citet{Zhang-Z-2023:Triaxiality} provides an empirical model to correct halo weak lensing profiles for orientation bias; we use this model to re-scale the profile for inclination angles of $\cos(i)=0.7$ as proxy for round-BCG clusters, shown as the dotted red line in Figure~\ref{fig:Mass_dependence_on_2halo} and $\cos(i)=0.3$ as proxy for elliptical-BCG clusters, shown as the solid blue line in Figure~\ref{fig:Mass_dependence_on_2halo}.  We then reduce the mass of the $\cos(i)=0.7$ halo until the 1-halo amplitude roughly matches that of the $\cos(i)=0.3$ line, at approximately $3.375 \times 10^{14} M_{\odot}$ (solid red line in Figure~\ref{fig:Mass_dependence_on_2halo}).  
Thus, if the shape of the BCG perfectly traces inclination angle, this analysis suggests that the mass of the round-BCG sample is 25\,\% lower than that of the elliptical-BCG sample. However, such a large mass ratio does not seem plausible, given the results from Section \ref{Mass_ratio} and Figure~\ref{fig:Mass_ratio_vs_q}.  

Figure \ref{fig:Mass_dependence_on_2halo} also shows that if the true mass of the round-BCG sample were to be $\sim$25\,\% lower than that of the elliptical-BCG sample, the elliptical-BCG profile would indeed exceed the round-BCG profile in the transition region ($\sim0.7-7$~$h^{-1} {\rm Mpc}$).  At larger radii, however, the round-BCG profile would significantly exceed that of the elliptical-BCG profile, in contradiction with our observations (Figure~\ref{fig:ACT_and_SPT_delta_sigma}).

Hence, we conclude that our observations cannot be explained solely by orientation bias in conjunction with a lower average mass of the round-BCG sample. 

\subsection{Intrinsic BCG shape and assembly bias}
Our analysis so far assumes that the observed BCG shape is determined mainly by the BCG orientation angle along the LOS (along with random scatter); however, this assumption is only fully true if all BCGs have the same intrinsic (3-dimensional) shape.
From hydrodynamical simulations we expect that while the majority of BCGs are prolate with intrinsic axis ratios of $\sim 0.74 \pm 0.10$, there exists a small fraction of BCGs which are notably closer to being spherical \citep[see][especially their Figure 4]{Herbonnet-R-2022:Triaxiality}.  In our analysis, these clusters with intrinsically spherical BCGs would be included in the round-BCG sample, but would be missing from the elliptical-BCG sample.  Thus, if more spherical halos are preferentially located in sparser environments (i.e., they have lower $b$), they would reduce the average amplitude of the 2-halo term of the round-BCG sample. It is therefore possible that the different 2-halo-term amplitudes that we find are a manifestation of {\it assembly bias} \citep{Gao05,Wechsler2006}, in its broadest interpretation as a correlation between a property of the main halo and the amplitude of the large-scale bias $b$.

In the simplest scenario, the shape of the BCG follows the shape of the host halo, i.e., spherical BCGs would be hosted by spherical halos.  Simulations predict that more spherical halos live in denser environments and have higher $b$ \citep[][ \citeauthor{vanDaalen12} \citeyear{vanDaalen12}, note that the latter finds no difference or a possible reversal for cluster-scale halos]{Faltenbacher-A-2010}, i.e., at face value, the inclusion of spherical halos in the round-BCG sample should increase the amplitude of the 2-halo profile, not decrease it as observed.  However, the situation is more complex:  simulations also predict that halo ellipticity correlates strongly with concentration, in the sense that more spherical halos have higher concentrations \citep[][]{Lau-E-2021:Assembly}.  Although \citet{Mao18} caution that the correlation between two halo properties does not mean that they show similar trends in clustering amplitude, we can ask whether the round-BCG sample has a higher concentration, since assembly bias predicts that more concentrated cluster-sized halos are located in lower-density environments and have lower $b$ \citep[e.g.,][]{Gao05, Wechsler2006, Mao18}.

This is indeed plausible, since halo concentration is expected to correlate with the SZ signal at fixed mass \citep{Baxter-E-2024:ConcentrationSZ}, so more concentrated halos are more likely to up-scatter into the cluster sample during cluster selection.  
In this scenario, the round-BCG sample would be composed of clusters that are preferentially less massive, but more concentrated than those in the elliptical-BCG sample. 
Since halos with higher concentrations tend to be more spherical \citep{Lau-E-2021:Assembly}, the typical 3D ellipticity of these clusters is smaller, meaning that --- under the assumption that more spherical halos host more spherical BCGs --- a broader range of inclination angles $\cos(i)$ suffices for these clusters to be included in the round-BCG selection.  Vice versa, for a cluster to be included in the elliptical-BCG selection, the BCG not only needs to be nearly in the plane of the sky, but also to be intrinsically (very) elliptical; and highly elliptical halos tend to have lower concentrations \citep{Lau-E-2021:Assembly}.

\subsection{Concentration measurements} \label{sect:concentration}

\begin{figure*}[ht]
    \centering
    % Left subfigure
    \includegraphics[width=0.45\textwidth]{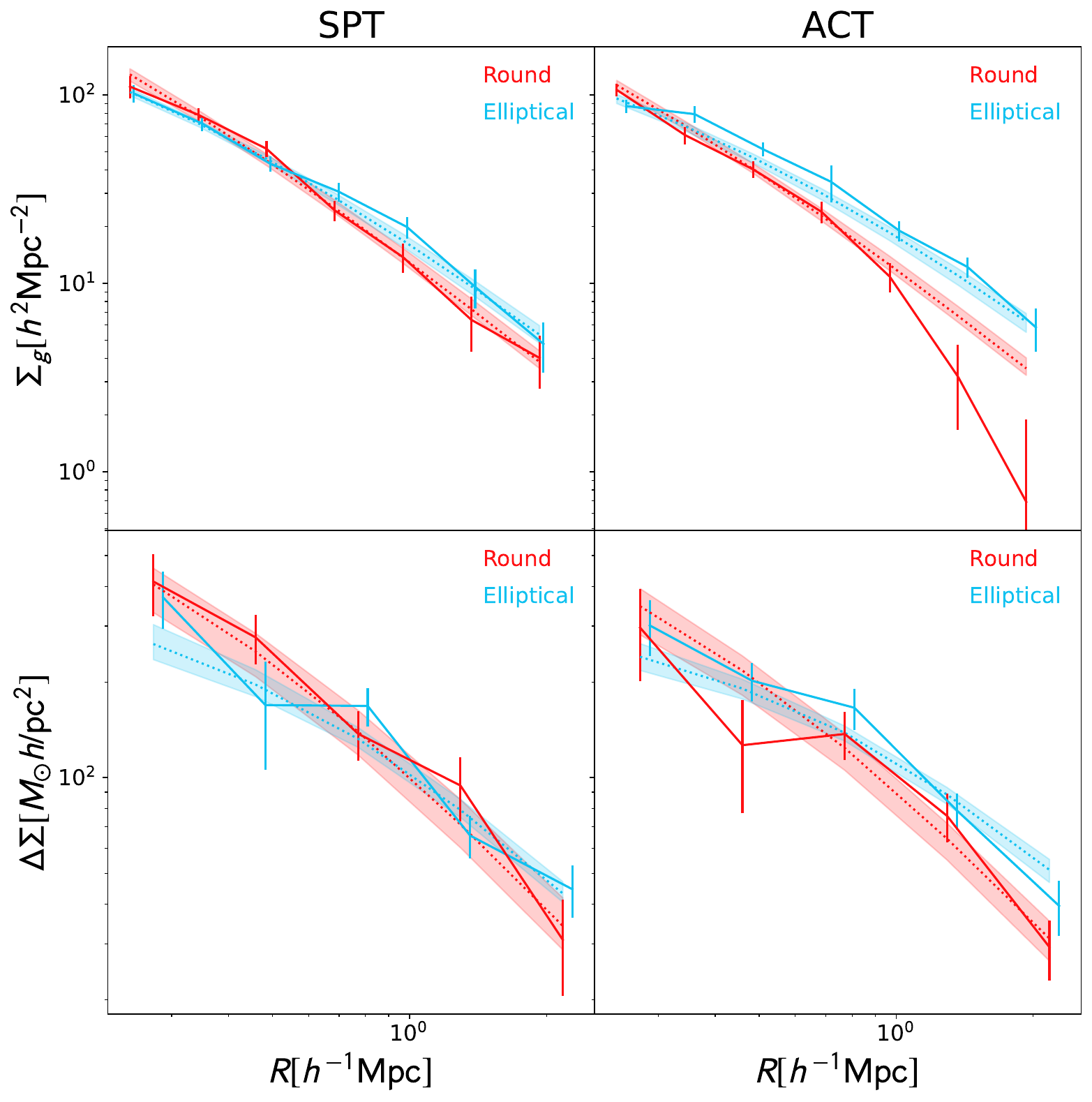}
    % Right subfigure
    \includegraphics[width=0.45\textwidth]{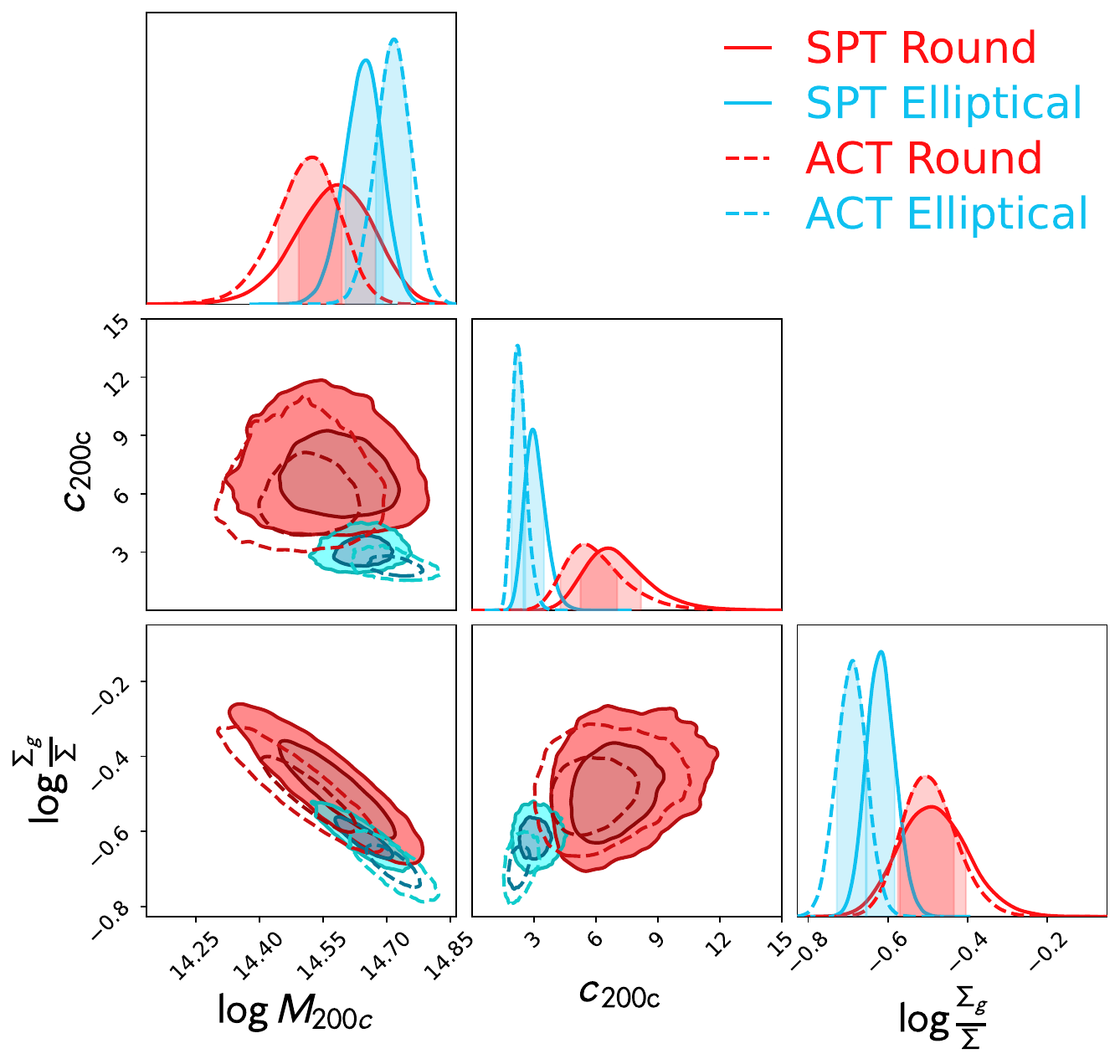}
    \caption{\textit{Left}: Jointly fit surface number density and weak-lensing profiles for SPT dataset on the left and for the ACT dataset on the right.\textit{Top row}: The surface number density profiles shown for the round-BCG sample in red and for the elliptical-BCG sample in blue. We show the data with the error-bars using solid lines, the best-fit model using dotted lines and the $1\,\sigma$ confidence bands with the shaded regions. \textit{Bottom row}: The weak-lensing profiles for the round and elliptical-BCG samples. \textit{Right}: The posterior contours from the joint fitting of surface number density profiles and the weak-lensing $\Delta\Sigma$ profiles for the SPT (solid contours) and ACT dataset (dashed contours). The three free parameters are: mass $M_{\rm 200c}$, concentration ($c_{\rm 200c}$), and multiplicative-factor ($\Sigma_{g}/\Sigma$). The blue contours represent the elliptical-BCG sample and the red contours represent the round-BCG sample.}
    \label{fig:ACT_and_SPT_joint_fitting}
\end{figure*}

Since the observations are projected quantities, it is not possible to directly measure the 3D concentration for mass distributions that are not spherically symmetric, i.e., in the presence of triaxiality and orientation bias.  In addition, even for the assumption of a spherically symmetric mass distribution, the weak-lensing profiles are quite noisy, given the relatively small number of clusters and shallow lensing data, and do not suffice to robustly constrain a two-parameter model (NFW with mass and concentration as free parameters).  However, the galaxy density profiles have higher signal-to-noise, and \citet{Shin-T-2021:SZ} have shown that, for massive SZ clusters from ACT ($\langle M_{\rm 500c} \rangle \gtrsim 3 \times 10^{14} M_{\odot}$), the mean galaxy density profile closely follows that of the total matter from weak lensing from small scales ($\sim$$0.2 h^{-1} {\rm Mpc}$) to large scales ($\sim$$10 h^{-1} {\rm Mpc}$). We therefore jointly fit the weak-lensing and galaxy density profiles as follows (note that the caveat about 2D and 3D concentration still applies). Assuming the lensing-weighted effective redshift of our cluster sample at $z\sim0.32$ ($0.35$) for SPT (ACT), our halo model includes an NFW profile with $M_{\rm 200c}$ and $c_{\rm 200c}$ as free parameters. We assume that the shapes of the density profiles are consistent between the galaxy density and the total matter density. To account for the difference in the overall amplitudes between these two probes, we introduce a multiplicative factor to adjust the amplitude of the galaxy density profiles with respect to that of the total matter, $f = \Sigma_{\rm g}/\Sigma_{\rm WL}$, where WL denotes weak lensing.  We fit this model over the 1-halo radial range only, i.e., $0.2 - 2.5~h^{-1}{\rm Mpc}$, where the outer limit is equivalent to that used for the weak-lensing mass estimates in the SPT cluster cosmology analysis \citep{Bocquet2024a}, but the inner limit is closer to the cluster center to gain more information on concentration (we neglect miscentering since we assume that the visually identified BCGs are at the true cluster centers).
Using this model, we perform a joint MCMC analysis of the galaxy density profiles and the weak lensing profiles, assuming no cross-correlation between them. We use the publicly available \textsc{Emcee} package \citep{Foreman-Mackey-D-2013:emcee}. 

We show the posteriors and the best-fit profiles in Figure \ref{fig:ACT_and_SPT_joint_fitting}.  The NFW model provides a good description for three of the four profiles, but it is not able to provide an adequate description of the ACT galaxy density profile (SPT round-BCG: $p=0.46$, SPT elliptical-BCG: $p=0.20$, ACT round-BCG: $p=0.01$, ACT elliptical-BCG: $p=0.17$ for the joint fit), which is very steep at large radii.  However, when fitting an Einasto profile, the data are not quite informative enough to robustly constrain all three parameters; tentatively, the steepness of the round-BCG ACT profile is absorbed by the parameter $\alpha$ that describes how quickly the slope changes with radius.  We therefore proceed with the NFW fitting results, noting this caveat for the ACT round-BCG subsample.

The NFW fits indeed show that the round-BCG sample is significantly more concentrated than the elliptical-BCG sample.  The best-fit concentrations are $7.1 \pm 1.6\,$/$\,6.1 \pm 1.6$ for the round-BCG SPT/ACT sample, and $3.9 \pm 0.5\,$/$\,2.3 \pm 0.3$ for the elliptical-BCG SPT/ACT sample, i.e., the ratio of measured concentrations is $1.8 \pm 0.5$ for SPT and $2.7 \pm 0.8$ for ACT.  Projection effects have been shown to also bias concentration measurements high: 
\citet{Osato-K-2018:Orientation} find a ratio of 1.2 -- 1.6 between $\cos{i} = 0.6 - 1.0$ and $\cos{i} = 0.0 - 0.4$ bins (see their Figure 4, comparing the first and second bins to the last and second-to-last bins).  The ratio here is therefore somewhat larger than expected from the simulations, especially when considering that the measured BCG shape is not a perfect tracer of $\cos{i}$, which agrees with the hypothesis that the round-BCG sample has higher intrinsic concentration.

\begin{figure*}
    \centering
    \includegraphics[width=1.7\columnwidth]{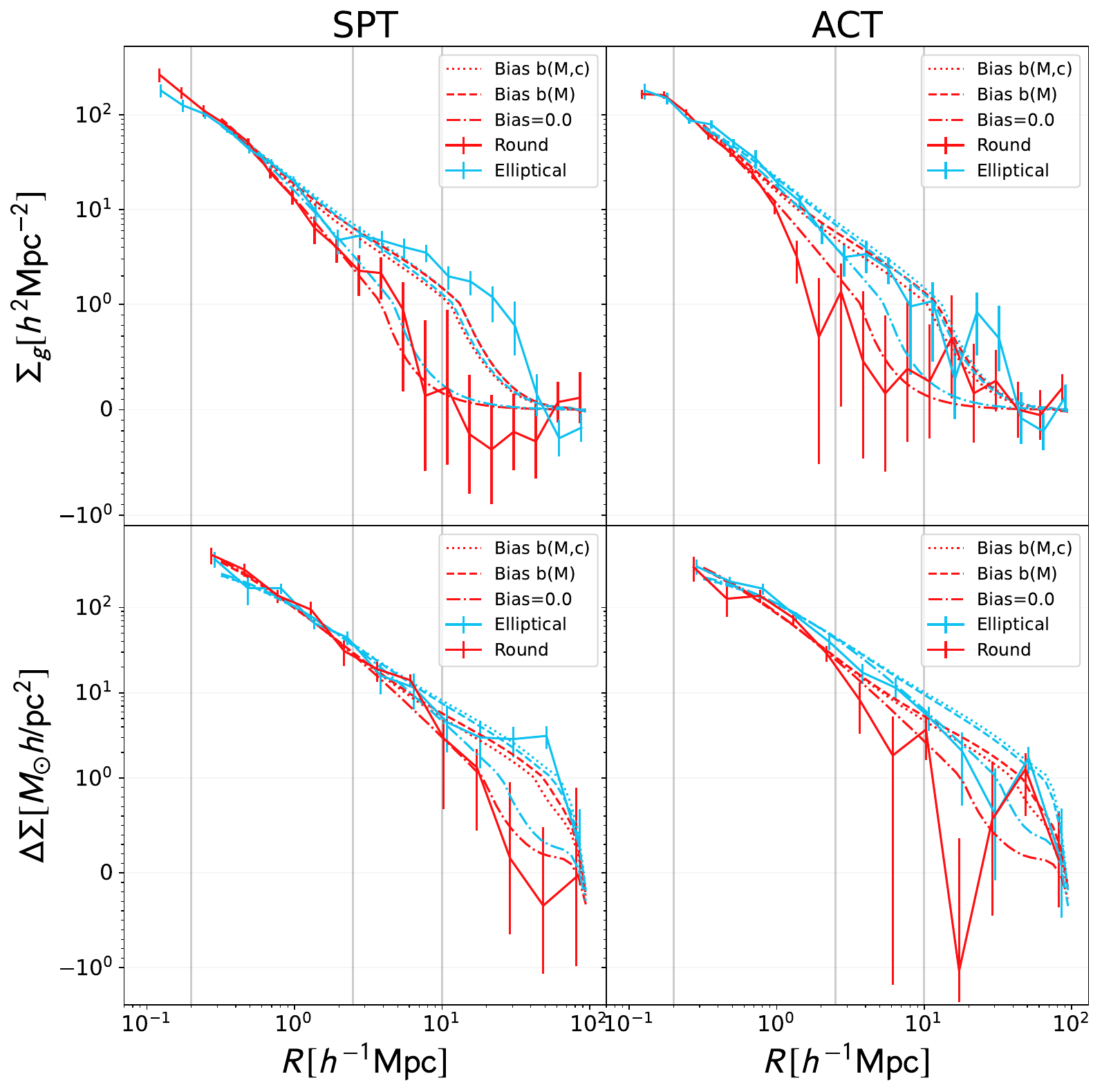}
    \caption{\textit{Top}: Surface number density profiles of galaxies ($\Sigma_{g}$) in the round-BCG (red) and elliptical-BCG (cyan) cluster samples (split at 25th-75th percentiles) for the SPT data set on the left and for the ACT data set on the right. \textit{Bottom}: The stacked weak-lensing $\Delta\Sigma$ profiles for the SPT cluster sample (left) and for the ACT cluster sample (right). With predictions from the best-fit NFW model in the 1-halo regime (marked by the gray vertical lines at $0.2~h^{-1} {\rm Mpc}$ and $2.5~h^{-1} {\rm Mpc}$), we compute a 2-halo bias (b(M)) based on \citet{Tinker-J-2010:HaloBias} and a bias b(M,c) as predicted from \citet{Wechsler2006}. We plot the 1-halo+2-halo model predictions from the various bias models, b(M,c), b(M) and a bias of zero (b=0) for each Round and Elliptical sample with dotted, dashed and dash-dotted lines respectively.}
    \label{fig:ACT_and_SPT_2halo_bias_prediction}
\end{figure*}

In the assembly bias scenario, cluster-mass halos with higher concentrations are less clustered, i.e., have a lower $b$.  We use the predictions from \citet{Wechsler2006} to translate these concentration measurements into a re-scaling of $b$ for the sub-samples, and in Figure~\ref{fig:ACT_and_SPT_2halo_bias_prediction} we show the predicted lensing and clustering profiles.  We also show the canonical profiles, as well as profiles for $b=0$. We see that the observed profile split is significantly larger than the change in profile amplitude due to the change in concentration.  The difference in concentration between the two samples is therefore not enough to explain the observations.  One should note, however, that the $b(M,c)$ predictions from \citet{Wechsler2006} are spherically symmetric (i.e., stacked along random sightlines), and do not account for the anisotropy of the large-scale structure around massive halos \citep{Osato-K-2018:Orientation} coupled with the orientation bias that we believe to be at play.  

Remarkably, Figure~\ref{fig:ACT_and_SPT_2halo_bias_prediction} suggests that the round-BCG profiles are best described by $b=0$, i.e., the profiles are consistent with NFW halos out to large radii, without evidence for a 2-halo term.  Simulations that split halo samples by concentration do not find such a low bias for the high-$c$ quartiles.  We note that \citet{Ramakrishnan2019} have argued that the anisotropy of the local web (on scales of a few Mpc) is a much stronger predictor of assembly bias, and could indeed cause differences in $b$ of factors of a few.  If that is the case, it suggests that the shape of the BCG is strongly coupled to the local tidal anisotropy.  Such a correlation should be studied in simulations, and future datasets with comprehensive spectroscopy of cluster outskirts should re-visit the work here to measure the local tidal anisotropy.  

Figure~\ref{fig:ACT_and_SPT_joint_fitting} also includes the posteriors for the best-fit mass, but similarly to concentration, these are difficult to interpret in the presence of projection effects.  
%Note that we show the model mass within 1.5~Mpc, which is less sensitive to biases in the amplitude of the weak-lensing profiles than $M_{200c}$.  
If the round-BCG and elliptical-BCG samples were intrinsically the same mass, but affected by projection effects, the mass estimate from the spherical model should be biased high for the round-BCG sample, and biased low for the elliptical-BCG sample.  Figure~\ref{fig:ACT_and_SPT_joint_fitting} shows that this is not the case: the mass estimates for the round-BCG samples are lower, or on par, with those of the elliptical-BCG samples.  Qualitatively, this result does not change when we 
fit for the mass using only the weak-lensing profiles, using the same fitting procedure as used for the SPT cluster cosmology analysis \citep[i.e., fixed concentration, linearly spaced bins from $0.5 - 2.3~h^{-1}{\rm Mpc}$; ][]{Grandis21,Bocquet2024a}: for SPT, we measure $M_{\rm 200c} = (4.9 \pm 1.0 \ /\ 4.4 \pm 0.3 ) \times 10^{14} h^{-1} M_{\odot}$ for the round- / elliptical-BCG subsamples, and for ACT, we measure 
$(3.6 \pm 0.6\ /\ 4.5 \pm 0.5) \times 10^{14} h^{-1} M_{\odot}$.  
Broadly speaking therefore, the mass estimates from the 1-halo term are consistent between the round-BCG and elliptical-BCG subsamples, which is significantly different from the result of \citet{Herbonnet-R-2019:Ellipticity}, but in agreement with our qualitative analysis in Sect.~\ref{sect:toy-model}.
We note that in the alternative scenario that there is no orientation bias, the much higher amplitude of the 2-halo term in the elliptical-BCG samples would bias the mass estimates from their 1-halo term high, so the measured results do not allow any definitive conclusion on the relative mass ratios of the two subsamples.  

Finally, the posteriors shown in Figure~ \ref{fig:ACT_and_SPT_joint_fitting} indicate that the ratio of projected galaxy density to projected matter density is larger for the round-BCG clusters, which is a reaffirmation of our earlier result that the measured galaxy richness of the round-BCG clusters is higher than that of the elliptical-BCG clusters.

\subsection{Proxies for cluster assembly history}

As the name suggests, assembly bias was thought to originate from different halo assembly histories; however, in simulations, the large-scale bias of cluster-sized halos does not correlate with proxies for assembly histories (e.g., the half-mass scale, i.e., the time where a halo reaches half of its current mass), although those proxies correlate with concentration, which itself does correlate with the large-scale bias \citep{Mao18}.  Observationally, assembly bias at the cluster mass scale has been difficult to detect, but intriguingly, \citet{Lin_22} were able to detect it by matching clusters to halos in a constrained simulation and using the halo assembly histories in the simulation to divide the observed clusters into an early-forming and a late-forming sample.  They found that the early-forming clusters have a higher concentration and a lower clustering amplitude, i.e., in this case, the halo formation history does predict both properties.   We therefore here also investigate whether the round-BCG and elliptical-BCG sample differ in observational proxies that are thought to relate to the cluster assembly history, namely the blue galaxy fraction, the magnitude gap, and the X-ray morphology.

\subsubsection{Blue galaxy fraction}

The galaxy population in clusters is notably different from that in the field, in that the majority of cluster members have ceased star formation.  The dominant mechanism to quench star formation in in-falling star-forming galaxies is thought to be strangulation, where ram pressure of the ICM removes the hot gas reservoir of the galaxy, leading to an exponential decline of star formation activity \citep[e.g.,][]{Balogh99,Poggianti06,vonderLinden10, Wetzel2013}.  Infalling blue galaxies thus turn red on a timescale of $\sim 1$~Gyr \citep{Adhikari-S-2021:Splashback}.  Therefore, the fraction of blue galaxies is an indicator of the recent mass accretion rate, in the sense that older clusters with a lower mass accretion rate exhibit a redder galaxy population than their younger counterparts. If the round-BCG sample is indeed older (as suggested by the higher concentration), it should therefore have a lower blue galaxy fraction than the elliptical-BCG sample.

We split the galaxies into a red and blue galaxy population in the $g-i$ vs. $i$ color-magnitude space. For each cluster redshift bin of $dz=0.05$, we first identify the red sequence by fitting a straight line with scatter to the population of \textsc{redMaPPer} member galaxies. We then define the red galaxies as the ones that are redder than the bluest 1\% of the \textsc{redMaPPer} member galaxies, and the blue galaxies as all galaxies bluer than that. The fraction of each color is then calculated by dividing the density profile with a certain color by the total density profile (Figure~\ref{fig:ACT_and_SPT_galaxy_density}).

\begin{figure}
    \centering    \includegraphics[width=0.85\columnwidth]{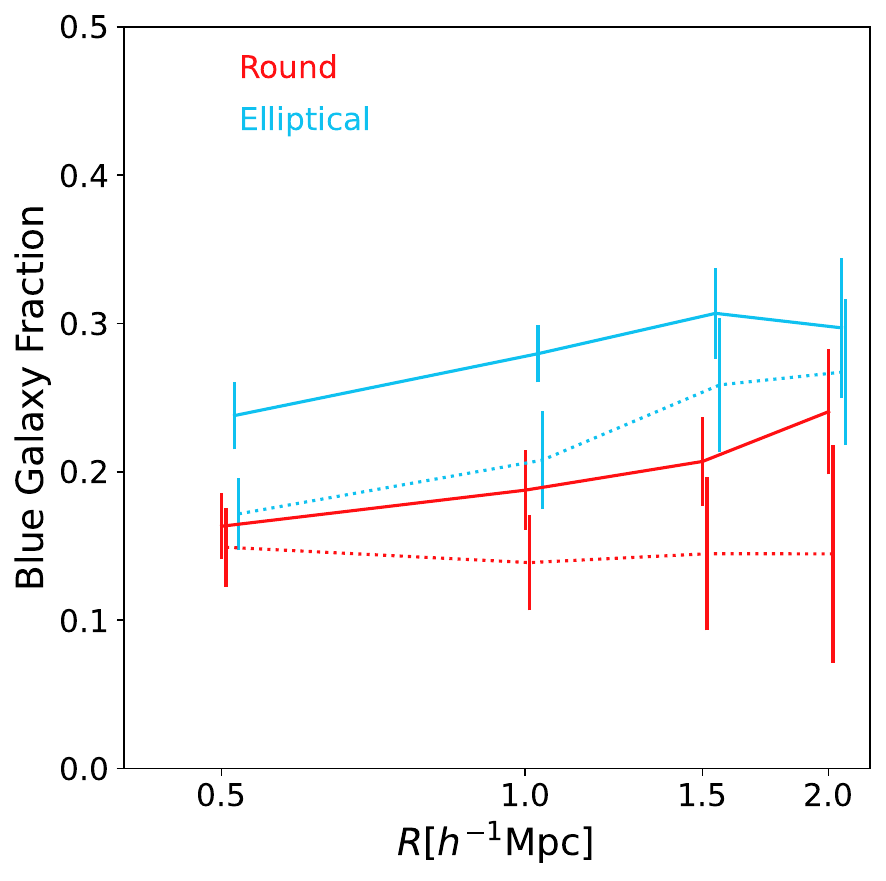}
    \caption{The radial cumulative distribution of the blue galaxy fraction for the SPT (solid lines) and ACT (dotted lines) clusters split by their BCG ellipticity at the 25th and 75th percentiles (shown in red for the round-BCG sample and blue for the elliptical-BCG sample).}
    \label{fig:Galaxy_fraction}
\end{figure}

In Figure \ref{fig:Galaxy_fraction}, we show the cumulative fraction of blue galaxies for three apertures around the cluster centers.  There are statistically significant differences within every radial aperture. The blue galaxy fraction of the elliptical-BCG sample is indeed significantly higher than that of the round-BCG sample in support of the age hypothesis. The significance of this difference within $1.5~h^{-1}{\rm Mpc}$, for SPT is $2.3\,\sigma$, while for ACT it is $1.7\,\sigma$.

It is worth noting that since we do not have spectroscopic information, the measurement here extends over all galaxies that contribute to the statistical overdensity along the LOS, i.e., we effectively measure the blue fraction in a cylinder.  The measurement therefore includes galaxies projected along the LOS, and we expect a larger contribution from surrounding structure to the clusters aligned along the LOS, i.e., the round-BCG sample.  Since the blue galaxy fraction is higher in the surrounding structures than the cluster itself, the true blue galaxy fraction of the round-BCG clusters might be even lower than measured here, which would increase the significance of the detection.    
Overall, therefore, we find good evidence for a lower blue fraction, and lower mass accretion rate, of the round-BCG sample in SPT.

\begin{figure*}
    \centering
    \includegraphics[width=1.7\columnwidth]{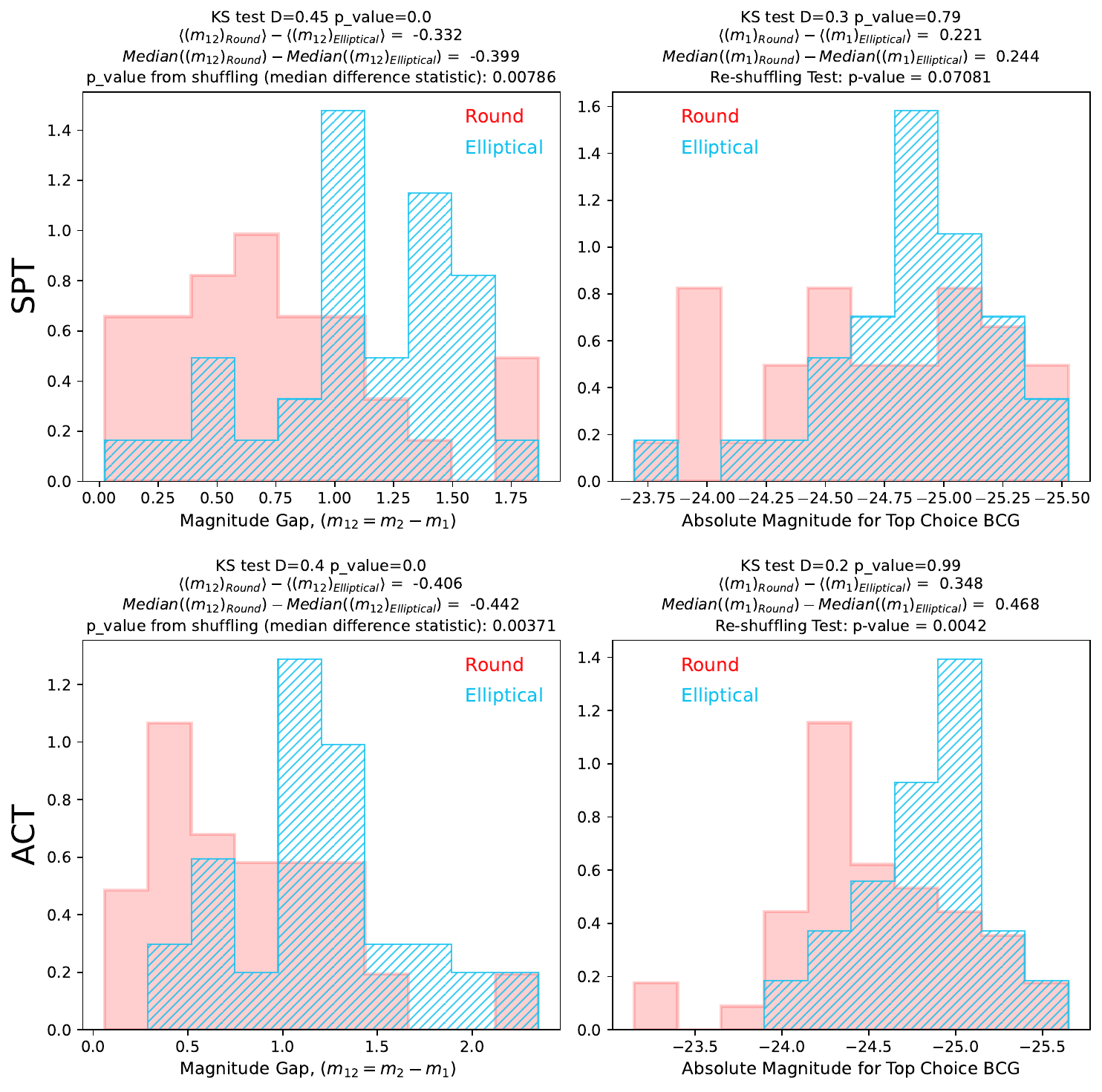} 
    \caption{Magnitude gap (left) between the visually inspected BCG and the brightest of rest of the cluster galaxies for the SPT sample (top) and for the ACT sample (bottom). Red-filled histogram represents the distribution of magnitude gaps for the round-BCG cluster sample and the blue-hatched histogram represents the same distribution for the elliptical-BCG cluster sample. The KS statistic represents the differences in the two samples. The distribution of the absolute magnitudes for the BCGs in the two samples is shown on the right for the SPT sample (top) and for the ACT sample (bottom). The magnitude gaps and absolute magnitudes are computed using the ``MODEL\_MAG'' magnitudes (calculated using a model fit to the galaxy luminosity profile) in the $z$-band.}
    \label{fig:SPT_and_ACT_magnitude_gap}
\end{figure*}

\subsubsection{Magnitude gap}\label{Magnitude gap}

The magnitude gap $m_{12} = m_2 - m_1$ between the magnitudes $m_1$ of the BCG and $m_2$ of the second brightest galaxy is expected to be correlated with the halo assembly history, since the timescale of the central galaxies in major merger events to merge is of the timescale of 
a few ($>2$) Gyr \citep{Ragone-Figueroa-C-2020:Alignment}.  Indeed, observations have found a clear correlation between large magnitude gaps and halo properties indicative of a relaxed dynamical state, including concentration, and presence of a cool core \citep[][]{Smith-G-BCGAssemblyBias:2010}. Therefore, the age hypothesis predicts that the round-BCG clusters should have greater magnitude gaps than the elliptical-BCG clusters.

We measure $m_{12}$ from the $z$-band composite model magnitudes (cModel) that were used in the construction of the DES \textsc{redMaPPer} cluster catalog \citep{Rykoff-E-2016:RedmapperDES}.  We take the brightest of all \textsc{redMaPPer} member galaxies with a \textsc{redMaPPer} membership probability greater than 0.8 to be the second-brightest cluster member.  In the left panels of Figure \ref{fig:SPT_and_ACT_magnitude_gap}, we show the magnitude gap ($m_{12}$) distribution for the round-BCG and elliptical-BCG samples.  We do find statistically significant differences in the distributions in both SPT and ACT; however, we find that the elliptical-BCG sample tends to have larger $m_{12}$, opposite to the prediction of the age hypothesis.  There are, however, again some caveats to this measurement: in particular, bright galaxies in surrounding filaments can be mistaken as cluster members, including as the BCG and the second-brightest galaxy, when the cluster is viewed along the LOS, which could effectively ``fill in'' the magnitude gap of the round-BCG cluster.  In addition, the probability of a galaxy to have been selected as the BCG (both by \textsc{redMaPPer} and in our visual inspection) takes the brightnesses of other member galaxies into account, and thus the magnitude gap is not independent of the BCG classification.  

In the right-hand panels of Figure~\ref{fig:SPT_and_ACT_magnitude_gap}, we investigate whether the difference in magnitude gaps is due to variations in the brightness of the BCGs themselves by comparing the distributions of absolute magnitudes $M_z$.  We see that at the bright end, the distributions agree remarkably well, but the magnitude distribution of round BCGs extends to fainter magnitudes.  
Since BCG stellar mass has been shown to correlate with halo mass \citep{Golden-Marx25}, this could indicate a greater contamination of the round-BCG sample by low-mass systems. 
Indeed, Figure~\ref{fig:Mass_ratio_vs_q} shows that clusters with BCGs fainter than $M_z > -24$ tend to have larger $M_{\rm SZ}/M_{\rm gas}$ ratios, i.e., are less massive at fixed SZ signal. 
Such contamination would indeed lower the average halo bias and the weak-lensing amplitude of the round-BCG sample, as observed.  For optically selected samples, contamination by filaments of lower-mass systems that are aligned along the line-of-sight is expected \citep{Sunayama-T-2020:Projection}; however, the SZ selection used here should be largely robust to such contamination due to the steeper mass-observable relation and lower intrinsic scatter.  A substantial caveat here is that we show the magnitudes as measured in the catalog, whereas detailed studies of the light and mass content of BCGs require custom image analyses \citep{Huang-H-2016:Orientation, Huang2022, Golden-Marx25}.  Future studies should re-visit this question, in particular when measurements of the BCG outer stellar mass \citep[which have been shown to be accurate tracers of halo mass, as well as good BCG identifiers; ][]{Kwiecien2024} become available from deeper imaging data, such as from the Rubin Observatory.

\begin{figure*}
    \centering
    \includegraphics[width=1.7\columnwidth]{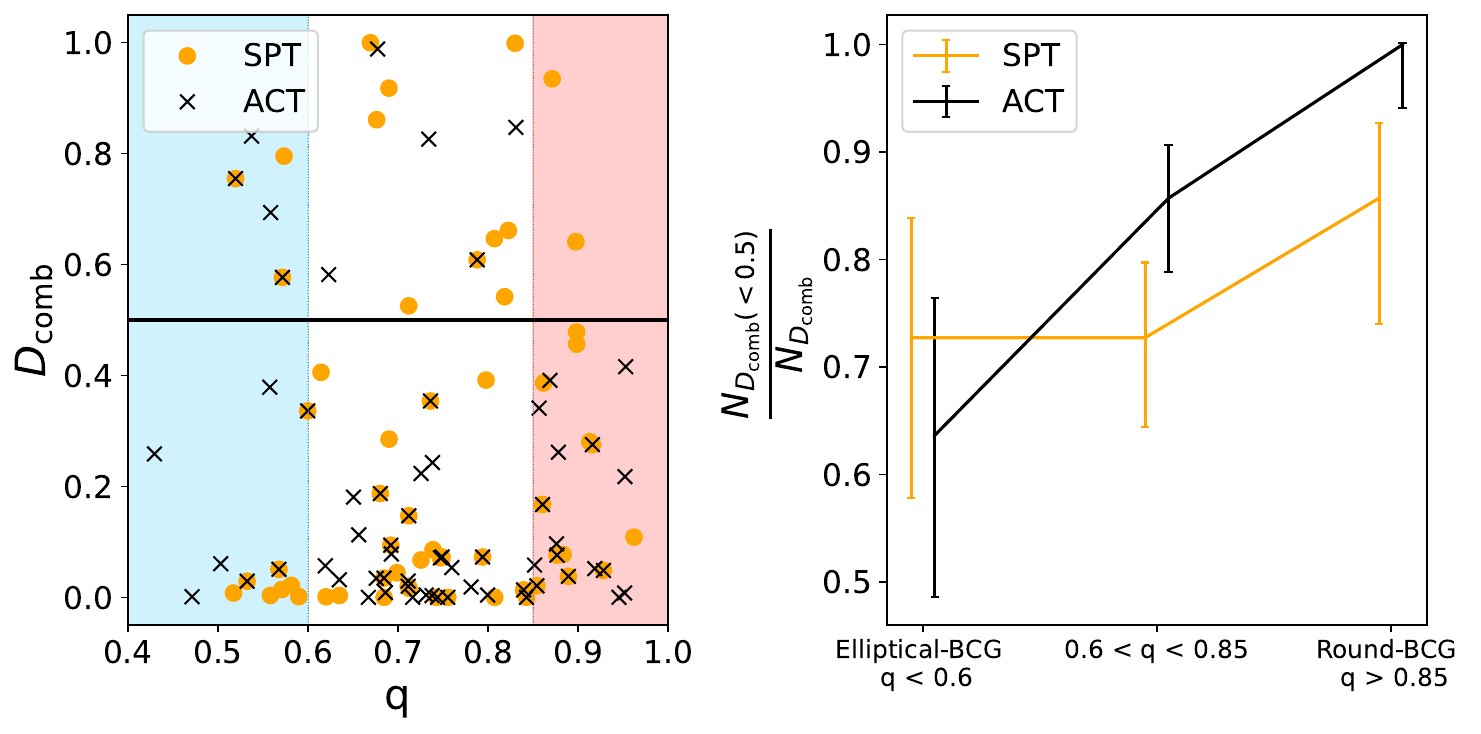} 
    \caption{\textit{Left}: The distribution of combined disturbance score ($D_{\rm comb}$) from \citet{Sanders-J-2025:XrayMorph} for eRASS1xSPT (yellow points) and eRASS1xACT clusters (black crosses) in relation to the BCG axis ratio (q). The shaded regions represent the clusters that have a elliptical-BCG (blue) or a round-BCG (red). \textit{Right}: The fraction of clusters with the combined disturbance score below 0.5 for each bin: Elliptical, in-between and round.}
    \label{fig:ACT_and_SPT_D_COMB}
\end{figure*}

\subsubsection{X-ray morphology}
Cluster morphology as seen in X-rays can be a powerful probe of the dynamical state of a cluster and therefore its formation history \citep[e.g., ][]{Rasia13,Darragh-Ford23}.  Given that nearly all of the clusters in this work are also in the eRASS1 catalog, we test whether the round-BCG and elliptical-BCG subsamples show significant differences in the eROSITA combined disturbance score $D_{\rm comb}$, which encompasses several X-ray morphological properties (concentration, ellipticity, sloshing, centroid offset, and power ratios) \citep{Sanders-J-2025:XrayMorph}.

In the left panel of Figure \ref{fig:ACT_and_SPT_D_COMB}, we show the distribution of $D_{\rm comb}$ for clusters in the eRASS1xSPT and eRASS1xACT samples. 
A lower $D_{\rm comb}$ score reflects a more relaxed X-ray cluster morphology; \citet{Sanders-J-2025:XrayMorph} noted that eROSITA finds a large fraction of relaxed clusters, and we see that the same is true for the SZ-selected samples here, at least at this mass scale.  In all three ranges of BCG ellipticity (elliptical, medium, round), clusters occupy the full range of $D_{\rm comb}$, with the notable exception of the round-BCG ACT sample, where no cluster would be classified as disturbed.
In the right panel of Figure \ref{fig:ACT_and_SPT_D_COMB}, we plot the fraction of $D_{\rm comb} < 0.5$ clusters.  In ACT, we see a notable trend of an increasing fraction of relaxed clusters with BCG axis ratio.  The SPT clusters are consistent with such a trend, but do not favor a trend on their own.  The signficance of the difference  between the round-BCG and elliptical-BCG samples is $2.6\,\sigma$ for ACT and SPT. Therefore, the X-ray morphology of the ACT sample is consistent with the hypothesis that the round-BCG sample is preferentially older / more relaxed than the elliptical-BCG sample, but the SPT sample is inconclusive.

\section{Discussion} 
\label{Discussion}
In this work, we measure the dependence of optical cluster observables (richness, weak-lensing, and galaxy density profiles) on the orientation of the halo relative to the LOS, as traced by the observed (2D) shape of the BCG. While we find a bias in richness as expected from orientation bias (round-BCG clusters have higher richnesses), we find that the 2-halo term in both weak-lensing and galaxy number density profiles is substantially lower in the round-BCG sample than the elliptical-BCG sample, which is opposite to the expectation from orientation bias. We investigate whether SZ selection biases and assembly bias, in addition to orientation bias, could explain this finding, but a clear explanation is missing.

\subsection{Masses of the round-BCG and elliptical-BCG subsamples}

Since halo bias is a strong function of mass for cluster-sized halos, we investigate whether the lower 2-halo term of the round-BCG sample is due to orientation bias in the SZ selection, and that the round-BCG sample has a lower true mass than the elliptical-BCG sample.  Importantly, the results from fitting spherically symmetric mass distribution the weak lensing profiles (Sect.~\ref{sect:concentration}) cannot be directly interpreted in terms of halo mass since weak-lensing profiles are very susceptible to orientation and projection effects.
Instead, we construct a toy model to estimate what the mass ratio of the two samples needs to be so that the lensing profile amplitudes are the same at small scales; we find this ratio to be $\sim 25$\% (Figure \ref{fig:Mass_dependence_on_2halo}). Notably, in this toy model, even a mass difference of 25\% would not be enough to explain the reversal of the profiles on large scales.
Moreover, such a large mass difference is not supported by the comparison of SZ mass estimates to X-ray gas mass estimates $M_{\rm gas}$, which are expected to be much less susceptible to orientation bias since X-ray emissivity scales with the square of the density; we find at most mild evidence that the SZ-to-$M_{\rm gas}$ ratio increases with BCG axis ratio (Figure~\ref{fig:Mass_ratio_vs_q}). The SZ-to-$M_{\rm gas}$ comparison suggests that the round-BCG sample cannot be more than $5-10$\% less massive than the elliptical-BCG sample.  

We note that the comparison of BCG absolute magnitudes shows a largely identical distribution at the bright end, but a tail towards fainter BCGs in the round-BCG sample, which could be indicative of a higher contamination of low-mass systems (Figure~\ref{fig:SPT_and_ACT_magnitude_gap}).  However, we caution that this comparison is based on catalog-based magnitudes, and a more robust analysis should be done with dedicated measurements of the light profiles and stellar mass content of BCGs \citep{Kwiecien2024}.

We also note that the results on the weak-lensing profile are markedly different from those for the {\it Weighing the Giants} (WtG) sample \citep{Herbonnet-R-2019:Ellipticity}, which found a weak-lensing mass ratio ($\sim 1.5$) between round-BCG and elliptical-BCG clusters.  WtG is an X-ray-selected cluster sample, and therefore expected to be even less affected by orientation bias than the SZ-selected sample here.  The WtG weak-lensing measurements were done on targeted observations, and thus did not probe the 2-halo term.  The mass ratio in \citet{Herbonnet-R-2019:Ellipticity} is from individual cluster weak lensing mass measurements within $r_{500}$ as measured from a low-scatter proxy ($M_{\rm gas}$), which is not directly comparable to the stacked measurements here based on matching the subsamples in the survey SZ observable.  In the future, higher signal-to-noise in both the weak-lensing and X-ray observables will make a more direct comparison possible.

\subsection{Assembly bias}

We hypothesize that the observed BCG shape depends not only on the inclination angle, but also on other halo properties which correlate with the amplitude of the large-scale bias, i.e., a manifestation of assembly bias in the broadest sense.  
Since halo concentration is the most established observable proxy for assembly bias, we jointly fit the weak-lensing and galaxy number density profiles with (spherical) NFW profiles within the 1-halo term, and indeed find that the concentrations of both the SPT and ACT round-BCG samples are substantially higher than those of the elliptical-BCG samples (Sect.~\ref{sect:concentration}).  
However, when propagating the expected difference in 2-halo amplitudes due to concentration, the split is much smaller than observed.  The difference in concentration alone therefore cannot explain our observations.  We note that the local tidal anisotropy has been proposed as a much stronger indicator of assembly bias \citep{Ramakrishnan2019}, and future work should investigate a potential link to BCG shape.

Simulations suggest that the SZ signals of more concentrated clusters are higher \citep{Baxter-E-2024:ConcentrationSZ}, implying that the SZ selection is biased towards clusters with higher concentrations, in addition to the orientation bias previously discussed.  The round-BCG sample therefore likely includes lower-mass clusters whose SZ signal is biased due to the orientation bias, concentration, or both.

Although simulations do not find direct measures of halo assembly history to predict assembly bias \citep{Mao18}, there is observational evidence that ``older'' clusters live in less dense environments \citep{Lin_22}.  Hydrodynamical simulations predict that the ICM temperatures (and therefore the SZ signals) of clusters with low current accretion rates are higher than on average \citep{Chen-H-2019}; implying that older clusters are overrepresented in SZ-selected samples.  Since halo accretion history is correlated with halo ellipticity \citep{Lau-E-2021:Assembly}, older halos are expected to be rounder and, to first approximation, have rounder BCGs.  With this line of argument, we would expect the round-BCG sample to be older than the elliptical-BCG sample.   
Indeed, we find a lower blue galaxy fraction for the round-BCG samples in both SPT and ACT (Figure~\ref{fig:Galaxy_fraction}), suggesting that they indeed have lower infall rates than elliptical-BCG clusters. 
On the other hand, the round-BCG sample has smaller magnitude gaps than the elliptical-BCG sample on average, which seemingly contradicts the age hypothesis.  
In X-rays, the ACT sample does indeed show a trend between BCG shape and a relaxed X-ray morphology; in particular, all of the round-BCG ACT clusters are relaxed.  In SPT, there is no statistically significant trend, however.  The observational evidence for the round-BCG clusters being older is therefore inconclusive, unless future spectroscopic studies can show that the magnitude gap of round-BCG clusters is often ``filled in'' by projection effects.

\subsection{Implications for Cosmology}

These results have important implications for cosmology: (1) We have shown that round-BCG clusters have higher richnesses and are thus overrepresented in optically selected cluster samples, as expected for clusters aligned along the LOS.  (2) Since round-BCG clusters have a surprisingly low large-scale galaxy bias, caution needs to be used when utilizing the 2-halo regime for the weak-lensing mass calibration in cluster count cosmology.  In the DES Y1 cluster cosmology analysis \citep{Abbott-T-2020:Y1cluster}, the amplitude $b$ of the 2-halo term was a fixed function of halo mass; \citet{To-2021a} introduced a nuisance parameter $b_{\rm sel}$ to re-scale $b$ and showed that in simulations, projection effects cause $b_{\rm sel}$ to be $\sim 1.5$ times larger than for a random halo selection.  Intriguingly, in data, \citet{To-2021b} measure a lower value of $b_{\rm sel} = 1.15^{+0.11}_{-0.09}$, possibly related to the overrepresentation round-BCG clusters, which have lower 2-halo amplitude.  (3) Our results suggest that there are selection biases present in SZ-selected clusters that favor the inclusion of more concentrated, older halos. These effects have been predicted in simulations before; but it is somewhat surprising that the observed BCG shape should be a sensitive tracer of halo assembly history --- this connection warrants further study, e.g., on hydrodynamical simulations.  It is important to note that cluster cosmology analyses that use hierarchical population models are largely robust against the details of the causes of selection biases, as long as the model includes a covariance term between SZ mass and weak-lensing mass estimates, as is the case for the SPT cluster cosmology analyses \citep{Bocquet-S-2019:SPTSZ,Bocquet2024a}.  

The cluster sample studied here is relatively small, and it is imperative to perform similar analyses on larger cluster samples in order to more completely study selection biases, as well as the correlation between BCG ellipticity and the amplitude of the 2-halo term, for clusters selected in SZ, optical and X-rays.  Fortunately, with current and upcoming cluster surveys such as SPT-3G, SO, DES, LSST, Euclid and eROSITA, the number of known clusters is increasing rapidly.  However, a major bottleneck in our work has been the robust identification of the BCG. The deeper imaging provided by Euclid and LSST, along with 
targeted spectrocopic datasets to confirm the BCG redshifts, will substantially facilitate automated, robust BCG identification by detecting the faint stellar envelopes indicative of true central galaxies, and enabling measurements of the outer stellar mass \citep{Huang2022,Kwiecien2024}.  

\section{Summary and Conclusion}

In this work, we show that SZ-selected cluster samples that are identical in SZ-mass estimate and redshift, but are split by observed BCG ellipticity, have statistically different optical properties:
\begin{itemize}
\item Round-BCG clusters exhibit higher richness than elliptical-BCG clusters, in agreement with the expectation from orientation bias.
Since the richnesses used here are those from the DES \textsc{redMaPPer} catalog, an immediate corollary is that round-BCG clusters are overrepresented in the DES selection (compared to a mass selection with random scatter). This selection bias is expected from simulations \citep{Wu-H-2022:Orientation,Zhang-Z-2023:Triaxiality}, and this result is %(to our knowledge) 
the first direct evidence seen in data.

\item In the projected density profiles probed by galaxy number densities and lensing, the 2-halo term is lower for round-BCG clusters than for elliptical-BCG clusters (whereas the 1-halo terms are consistent).  This result is opposite to the expectation from orientation bias but is in the same direction as the effect expected from assembly bias; however, our estimate shows that assembly bias alone is too weak to explain the observed difference.

\item Round-BCG clusters have higher (projected) concentrations than elliptical-BCG clusters.  This could be caused by a combination of orientation bias (enhanced 2D concentration) and older cluster age (enhanced 3D concentration).

\item For cluster age proxies, we find that round-BCG clusters have a lower blue galaxy fraction and more relaxed X-ray morphology (indicating older age).  However, we have found that round-BCG clusters have smaller magnitude gaps (indicating the opposite).

\end{itemize}

Orientation bias and assembly bias alone are not enough to explain our observations, in particular the puzzlingly low amplitude of the weak-lensing and galaxy density profiles at large radii.  Future work should reinvestigate this analysis with larger cluster samples; moreover, upcoming survey data will be able to shed light on the details of the selected cluster samples, e.g., with the use of the BCG outer stellar mass as additional mass proxy and BCG identifier, with individual weak-lensing mass estimates, with better $M_{\rm gas}$ measurements, and with spectroscopy of candidate cluster members in the clusters and their outskirts.  

\section*{Acknowledgements}

\addcontentsline{toc}{section}{Acknowledgements}
The authors thank Jessie Golden-Marx, Matthew Hilton, Ravi Sheth, and Shenming Fu for helpful discussions, and Prakruth Adari, Hsin Fan, Jiyun Di, Alden Beck, Xiangyu Huang, and Leonardo Pierre for their contributions in visual inspection of BCGs. 

The authors acknowledge the usage of \textsc{NumPy} \citep{Harris-C-2020:Numpy}, \textsc{SciPy} \citep{Virtanen-P-2020:Scipy}, \textsc{Astropy} \citep{Price-Whelan-A-2018:Astropy}, \textsc{Matplotlib} \citep{Hunter-J-2007:matplotlib},\textsc{emcee} \citep{Foreman-Mackey-D-2013:emcee}, \textsc{corner} \citep{Foreman-Mackey-D-2016:corner}, \textsc{scikit-learn} \citep{Pedregosa-F-2011:Sklearn} and \textsc{ChainConsumer} \citep{Hinton-S-2016:ChainConsumer}. This work also used \textsc{TOPCAT} \citep{Taylor-M-2005:Topcat} for viewing and handling of tabular data.

RS, ThS, AvdL, RH and BL are supported by the US Department of
Energy under awards DE-SC0018053, DE-SC0023387 and DE-SC0025309. RS thanks the LSST-DA Data Science Fellowship Program, which is funded by LSST-DA, the Brinson Foundation, the WoodNext Foundation, and the Research Corporation for Science Advancement Foundation; his participation in the program has benefited this work.

The South Pole Telescope program is supported by the National Science Foundation (NSF) through awards OPP-1852617 and OPP-2332483. Partial support is also provided by the Kavli Institute of Cosmological Physics at the University of Chicago. PISCO observations were supported
by US NSF grant AST-0126090.

Funding for the DES Projects has been provided by the U.S. Department of Energy, the U.S. National Science Foundation, the Ministry of Science and Education of Spain, 
the Science and Technology Facilities Council of the United Kingdom, the Higher Education Funding Council for England, the National Center for Supercomputing Applications at the University of Illinois at Urbana-Champaign, the Kavli Institute of Cosmological Physics at the University of Chicago, the Center for Cosmology and Astro-Particle Physics at the Ohio State University,
the Mitchell Institute for Fundamental Physics and Astronomy at Texas A\&M University, Financiadora de Estudos e Projetos, Funda{\c c}{\~a}o Carlos Chagas Filho de Amparo {\`a} Pesquisa do Estado do Rio de Janeiro, Conselho Nacional de Desenvolvimento Cient{\'i}fico e Tecnol{\'o}gico and the Minist{\'e}rio da Ci{\^e}ncia, Tecnologia e Inova{\c c}{\~a}o, the Deutsche Forschungsgemeinschaft and the Collaborating Institutions in the Dark Energy Survey. 

The Collaborating Institutions are Argonne National Laboratory, the University of California at Santa Cruz, the University of Cambridge, Centro de Investigaciones Energ{\'e}ticas, Medioambientales y Tecnol{\'o}gicas-Madrid, the University of Chicago, University College London, the DES-Brazil Consortium, the University of Edinburgh, the Eidgen{\"o}ssische Technische Hochschule (ETH) Z{\"u}rich, Fermi National Accelerator Laboratory, the University of Illinois at Urbana-Champaign, the Institut de Ci{\`e}ncies de l'Espai (IEEC/CSIC), the Institut de F{\'i}sica d'Altes Energies, Lawrence Berkeley National Laboratory, the Ludwig-Maximilians Universit{\"a}t M{\"u}nchen and the associated Excellence Cluster Universe, the University of Michigan, NSF NOIRLab, the University of Nottingham, The Ohio State University, the University of Pennsylvania, the University of Portsmouth, SLAC National Accelerator Laboratory, Stanford University, the University of Sussex, Texas A\&M University, and the OzDES Membership Consortium.

Based in part on observations at NSF Cerro Tololo Inter-American Observatory at NSF NOIRLab (NOIRLab Prop. ID 2012B-0001; PI: J. Frieman), which is managed by the Association of Universities for Research in Astronomy (AURA) under a cooperative agreement with the National Science Foundation.

The DES data management system is supported by the National Science Foundation under Grant Numbers AST-1138766 and AST-1536171. Data access is enabled by Jetstream2 and OSN at Indiana University through allocation PHY240006: Dark Energy Survey from the Advanced Cyberinfrastructure Coordination Ecosystem: Services and Support (ACCESS) program, which is supported by U.S. National Science Foundation grants 2138259, 2138286, 2138307, 2137603, and 2138296. The DES participants from Spanish institutions are partially supported by MICINN under grants PID2021-123012, PID2021-128989 PID2022-141079, SEV-2016-0588, CEX2020-001058-M and CEX2020-001007-S, some of which include ERDF funds from the European Union. IFAE is partially funded by the CERCA program of the Generalitat de Catalunya.

We  acknowledge support from the Brazilian Instituto Nacional de Ci\^encia e Tecnologia (INCT) do e-Universo (CNPq grant 465376/2014-2).

This document was prepared by the DES Collaboration using the resources of the Fermi National Accelerator Laboratory (Fermilab), a U.S. Department of Energy, Office of Science, Office of High Energy Physics HEP User Facility. Fermilab is managed by Fermi Forward Discovery Group, LLC, acting under Contract No. 89243024CSC000002.

\appendix

\section{Validation of BCG shapes} \label{Shape validation}
\begin{figure*}
    \centering
    \includegraphics[width=1.7\columnwidth]{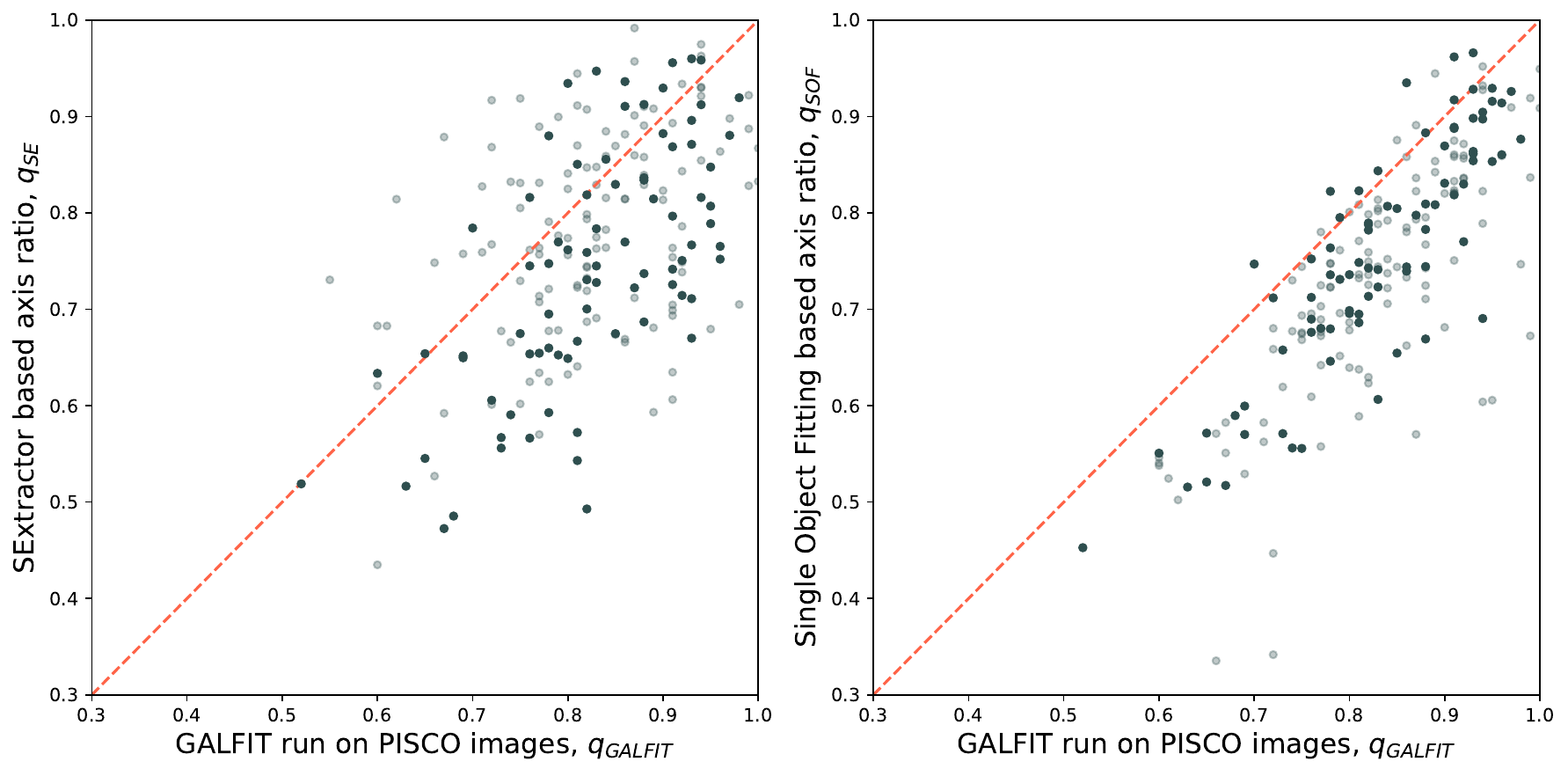}
    \caption{The comparison of axis ratios derived from different shape measurements. On the left the SExtractor-derived axis ratios are compared to the \galfit{} modelled shapes from PISCO imaging for the clusters. Black points highlight the clusters that have undergone the visual inspection process in Appendix \ref{BCG-VI} and thus the BCGs are both well-identified and devoid of blending and other blatant shape measurement issues. On the right the same plot has been constructed for the Single Object Fitting (SOF) method of shape measurement. Due to their lower scatter, we use SOF shapes in our main analysis.}
    \label{fig:shape_validation_SPT}
\end{figure*}

We use the deeper PISCO images available for a subset of 300 SPT clusters to inspect the BCG shapes and model them with \galfit{}. Several runs of \galfit{} were made with updated masks for removing contaminations and for exclusion of the brightest galaxy cores as they skewed the fits considerably. The procedure is detailed in \citet{Herbonnet-R-2019:Ellipticity}. The shapes derived from \galfit{} were compared to the shapes provided in the DES Y3 catalogs (Single Object Fitting (SOF) and SExtractor) as shown in Figure \ref{fig:shape_validation_SPT}. While the scatter in the shapes is critical to our analysis, the overall bias is not as relevant as we split our shapes into quantiles of extremes. The plots show a tighter correlation between SOF and \galfit{} than SExtractor shapes; therefore, we select SOF shapes for our analyses.

\section{Visual Inspection of BCG} \label{BCG-VI}

As mentioned in Section \ref{BCG selection}, it is important to identify the BCG for each cluster as its shape is the proxy for the orientation of the cluster. The \textsc{redMaPPer} cluster finder algorithm identifies the 5 most-likely BCG candidates and reports their associated probabilities. To minimize the impact of mis-centering, which is known to affect 20-25\,\% of the \textsc{redMaPPer} clusters, we visually inspect the BCGs using the optical imaging from DES. To accomplish this, we required multiple-stages of inspections per cluster. We had a group of 10 annotators at each stage, perusing through a google form, populated with the cutouts and the corresponding questionnaire. The responses were recorded in a separate google sheet, blinded from view until the respective inspection round was completed. With 266 clusters to be visually identified by at least 3 annotators over two rounds of inspection, each annotator spent a total of 20 hours on average. Given the large amount of time spent over the inspection, coordinating the efforts was a significant challenge. We used a quick round, to sift through and separate the easily identifiable cases and followed it with a more detailed round, to pick apart the more difficult cases. In the following sub-sections, each round of inspection is described in detail.

\subsection{Visual inspection of BCGs - Quick Round}

\begin{figure*}
    \centering
    \includegraphics[height=1.7\columnwidth]{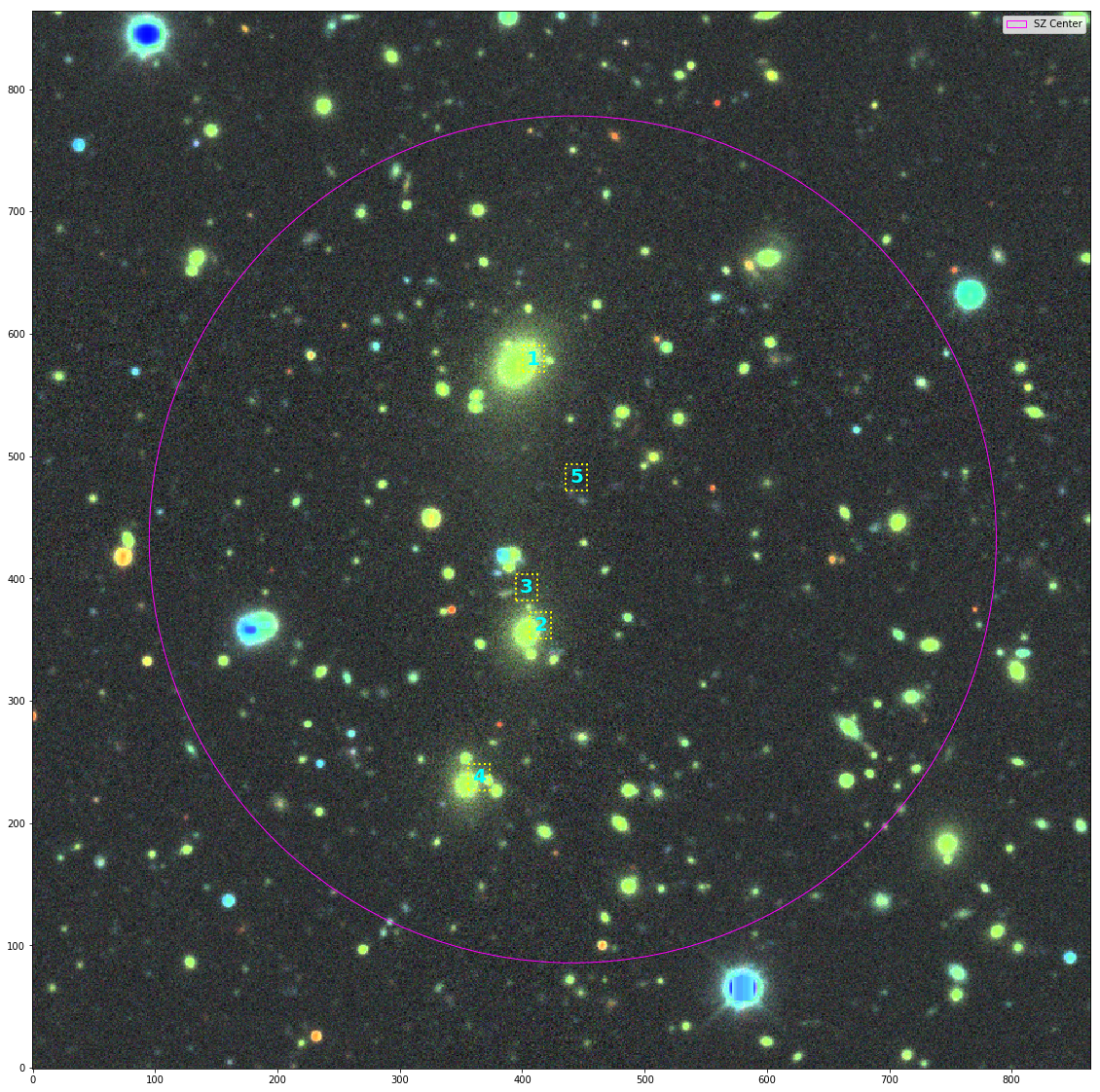}
    \caption{1 Mpc box-cutout from the DES images hosted at \url{https://des.ncsa.illinois.edu/releases} for the SPT cluster, SPT-CLJ0516-5430, at redshift of 0.295. The SZ ring in purple is plotted with a radial size equal to the core-radius of 1.5'. Each \textsc{redMaPPer} identified BCG is numbered (cyan) in the order of its reported probability. BCG 1 is identified as the highest likely candidate both visually and by the \textsc{redMaPPer} algorithm.}
    \label{fig:Large_416771661_zrg.png}
\end{figure*}

\begin{figure*}
    \centering
    \includegraphics[width=1.7\columnwidth]{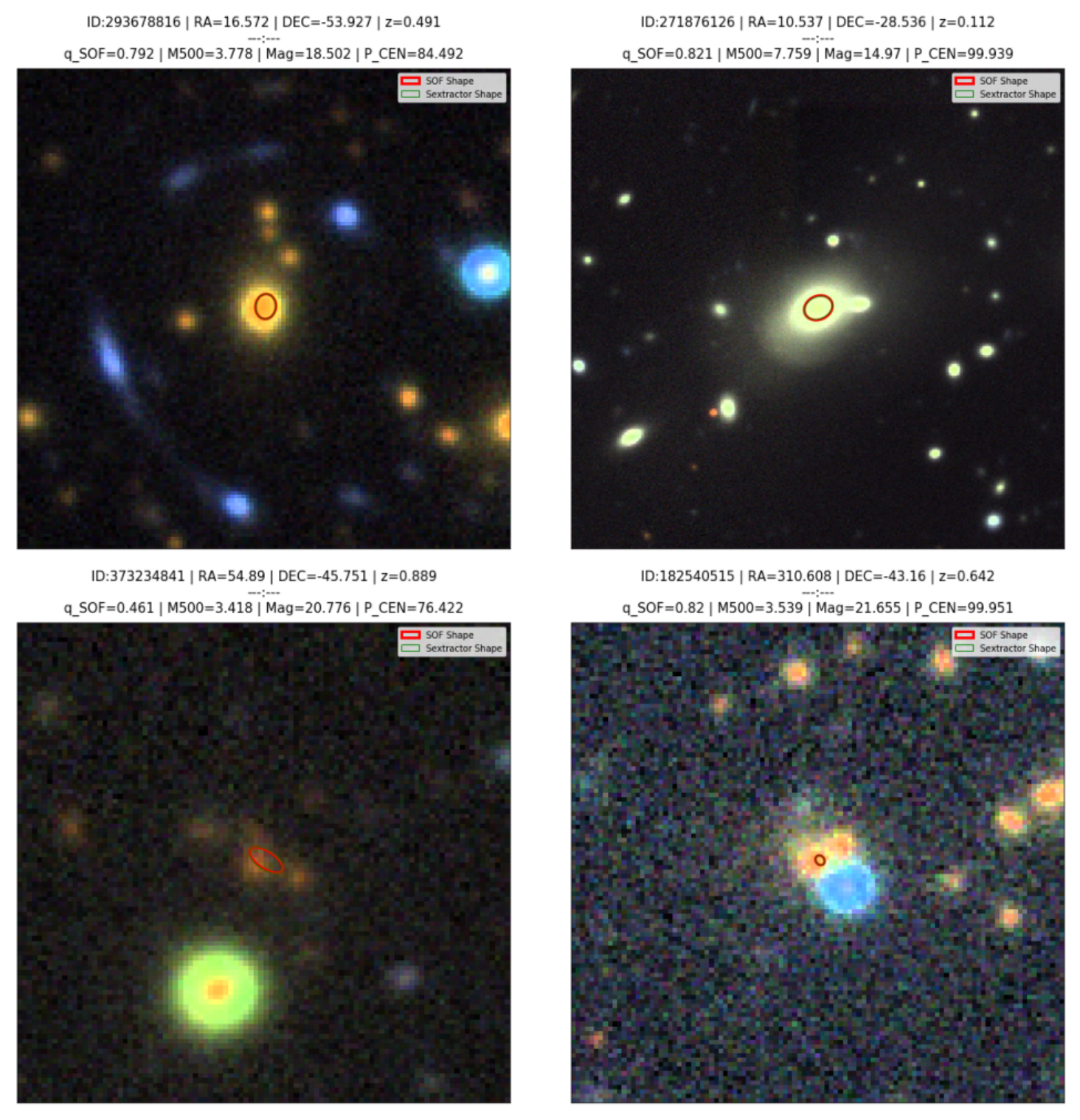}
    \caption{A gallery of BCGs that were visually inspected. We overlayed the shape measurement, red and green ellipse for SOF and Sextractor shapes respectively, and compared it to the extended diffuse envelope of the BCG. In the top row are the BCGs that passed the inspection while the bottom row shows the BCGs whose shape measurement was skewed due to multiple blended objects.}
    \label{fig:BCG_gallery}
\end{figure*}

We asked for the ``easiest-to-identify'' BCGs to be filtered out in the first round of visual inspection. We also designed flags for identifying the common issues such as blending, improper color-scaling, or being overly complicated and time consuming to inspect. This made the first round a preliminary, but rapid analysis of all the BCGs in the SPTxDES sample in the redshift range of [0.15,0.7] and with $\xi \geq 5.0$ (266 BCGs). It provided us a preliminary statistical summary of the classification and inspired the restructuring of the subsequent rounds of inspection.

For this round, we queried cutouts, centered at the SZ centroid, of two box sizes: 1. 1 Mpc and 2. 500 kpc sourced from the Dark Energy Survey using their public database at \url{https://des.ncsa.illinois.edu/desaccess/}. We chose the g,r and i-band filters to make our RGB images. Also, we applied appropriate scaling to each image to visualize the BCGs better than bright foreground objects, for example stars, etc. We also over-plot the SZ centroid on the image using equally-spaced concentric rings with fixed angular size. The rings were used to aid with visualizing the SZ signal. While the SZ centroid was used as a proxy for the cluster center, we also followed closely the proximity of the BCG to the red-sequence population during the inspection. 

Lastly, we queried 5 images of box size 200 kpc centered around each of the 5 BCG candidates. Thus, for every galaxy cluster, there were in total 7 images presented to 3 randomly assigned annotators. We requested a rating out of 5 for every BCG candidate. A net sum of score between the 5 candidate BCG choices was not requested to be a constant, so that only the singularly top-scoring BCGs were accepted out at this stage. If two or more BCGs had comparable scores, they were passed over to the Round 2 of visual inspection.

We set up several conditions to pick out the top scored BCGs. Each BCG had to score an aggregate of 12 out of 15 by the 3 annotators. Additionally, the minimum score by any annotator should not be below 4, discounting the cases such as (2,5,5) or (3,4,5). Furthermore, if any of the blended,improper-scaling or overly-complicated flags were set for the BCG by any of the annotators, it was sent over to the next round of inspection. With these strict conditions applied, we found 114 clusters of 266 clusters ``easy-to-identify'' and ready to proceed into the science sample. Out of these, 108 clusters have the same BCG as the top-ranked BCG from \textsc{redMaPPer}. 

\subsection{Visual inspection of BCGs - Detailed Round}

We were left behind with a net total of 152 clusters for this round. Since, in the results of the quick round, we found several cases of poor-scaling due to foreground bright objects such as stars, we applied a new re-scaling scheme using the $arcsinh$ stretch as prescribed in \citet{Lupton-R-2004:Image-processing}. We also introduced a redshift dependence on the parameters of the scaling function to make the high redshift clusters (z $> 0.7$) brighter and easier to inspect. Also, two separate sets of RGB images were constructed using the filters, (\textit{g,r,i}) and (\textit{r,i,z}). This gave us more flexibility while inspecting the high redshift clusters and to identify star-forming blue-BCGs in the most massive clusters. We also introduced a larger cutout with box size of 2 Mpc to accomodate clusters with larger offsets between the BCG candidates and the SZ centroid. A single SZ ring was plotted at an angular size of the maximum of 1 arcminute or the core radius as provided by SPT. Keeping a single outer SZ ring had the advantage of minimizing the bias of repeatedly choosing the BCG closest to the SZ center as the best candidate as shown in Figure \ref{fig:Large_416771661_zrg.png}. Lastly, we also provided PISCO images and Hubble images from the Hubble Legacy Archive (HLA) whenever available. We decided to use a larger range for the scoring system (out of 10) and to implement a relative score, where the BCGs could only be awarded a net sum of 10 by each annotator. This allowed us to compare the BCGs more closely. 

We assigned a minimum cut on the mean of the scores as 8 out of 10 and a maximum of 2 on the standard deviation. Combined, this would result in a list of BCGs with the highest scores and the least variance between the annotators. Like before, we again restricted the flags on the blending and bad scaling to none and allowed the shape to be verified as good by at least two out of the three annotators. We share some examples of BCGs flagged as good versus bad in Figure \ref{fig:BCG_gallery}. We found another 61 BCGs out of the 152 that satisfied all our constraints. Of these, 58 BCGs were the same as the top-ranked BCG from \textsc{redMaPPer}. These were appended to the Round 1 sample to give us a net total of 166 clusters. Therefore, we have 62\,\% of the SPT clusters that are accepted after visual inspection and are used for the analyses of orientation effects. Despite the cost of having a reduced signal-to-noise ratio for the observed trends, it is critical to remove the clusters with incorrect BCGs to reduce the chances of contamination from the mis-identification of the projected orientation that can lead to the direct dilution of the observed trends, if any. We reduce the ACT data set using a combination of feature engineering and a quick visual inspection.

\section{Feature Engineering with SPT and revisiting the ACT Data set} \label{feature-engg-details}

\begin{figure*}
    \centering
    \includegraphics[height=1.7\columnwidth]{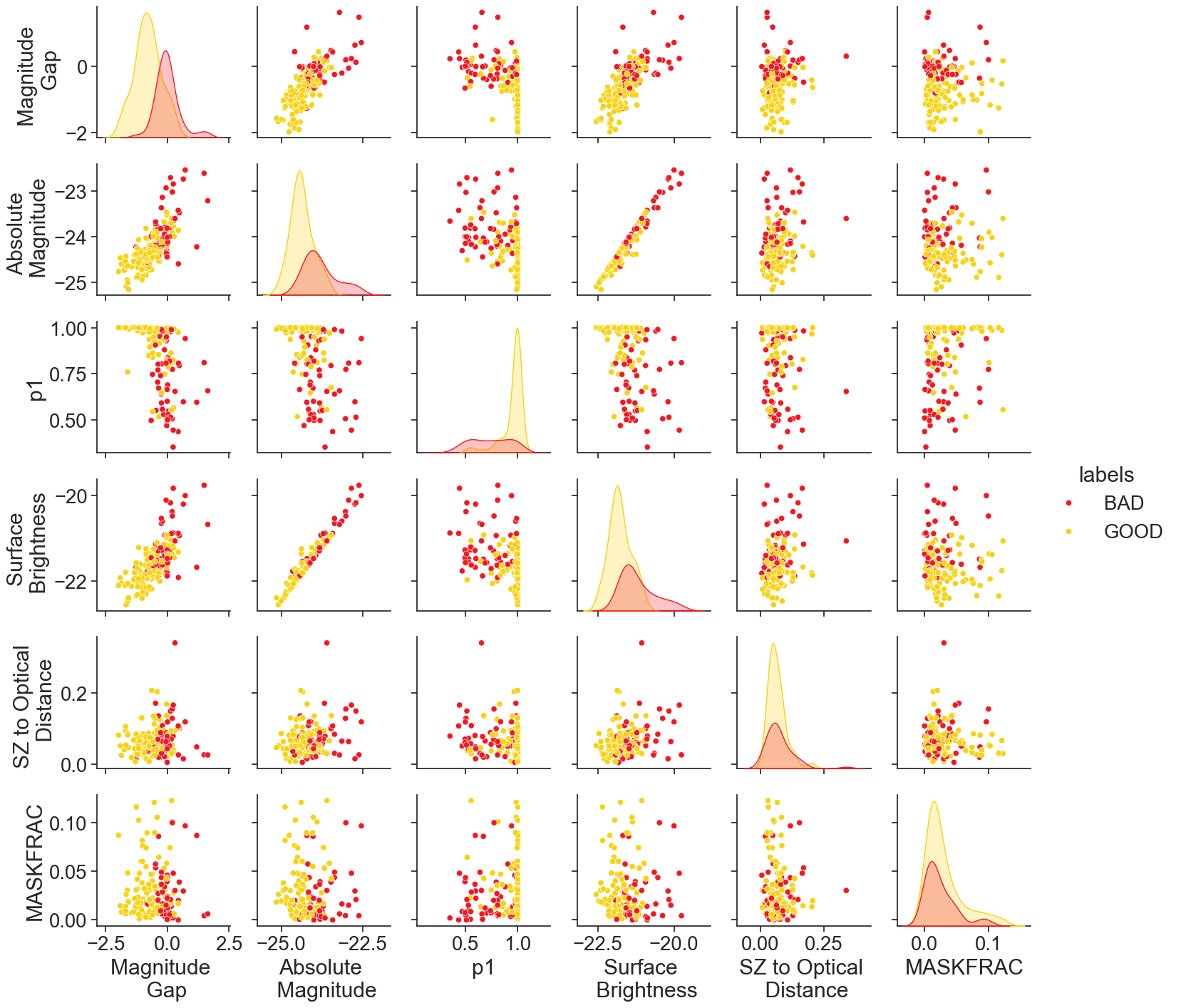}
    \caption{Giant scatter plot of all the BCG properties for the SPT data set. The yellow points mark the BCGs labelled as ``GOOD'' in the visual inspection process and the red points mark the BCGs labelled as ``BAD''. The purity fraction of the GOOD BCGs in the sample is $66.4\,\%$.}
    \label{fig:Feat_engg_before}
\end{figure*}

We investigated a host of features of BCG in the SPT data set, such as \textsc{redMaPPer} probabilities, absolute magnitude, magnitude gaps, surface brightness, etc. to reproduce the expensive visual classification on the ACT data set. BCG mis-identifications are quite critical as the key to identifying the halo orientation lies in the shape of the BCG and an incorrect BCG/shape directly affects our analysis. It is therefore vital to pick our BCGs carefully. 

\begin{figure*}
    \centering
    \includegraphics[height=1.7\columnwidth]{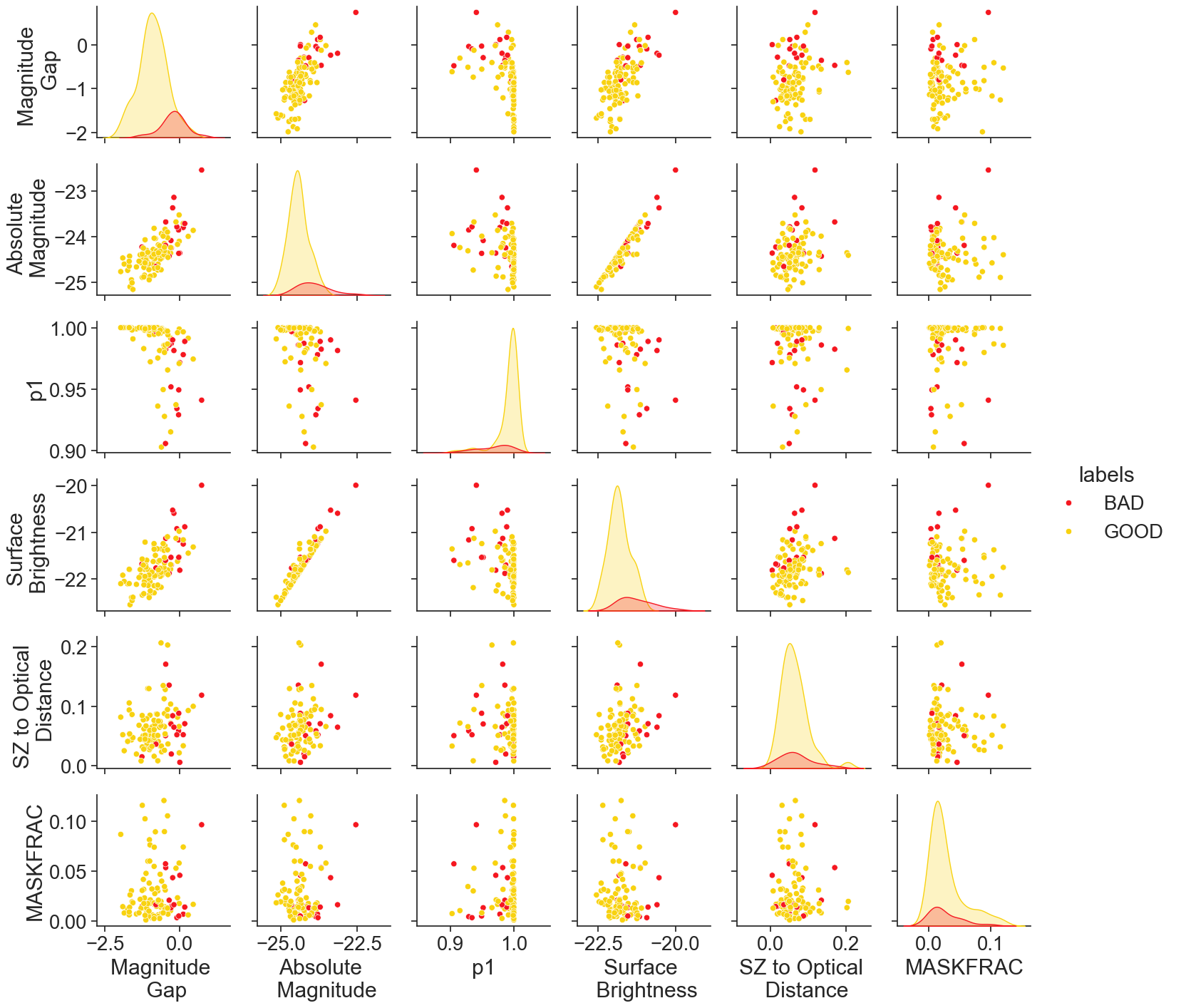}
    \caption{Same as Figure \ref{fig:Feat_engg_before}, but after selecting on ``p1'' > 0.9, the probability of first BCG as provided by \textsc{redMaPPer} and on ``Mask Fraction'' < 0.15. We achieve a ``GOOD'' purity fraction of 84.6\,\% while taking a hit to the completeness fraction at 82.2\,\%.}
    \label{fig:Feat_engg_after}
\end{figure*}

The giant scatter plots, Figure \ref{fig:Feat_engg_before} and Figure \ref{fig:Feat_engg_after}, compare all the BCG properties and their distributions for the visually inspected and accepted top-ranked \textsc{redMaPPer} BCGs (labelled as ``GOOD'') versus the rejected BCGs (labelled as ``BAD''). We do not use the BCGs that have a bad shape measurement (blended/bad scaling from foreground bright objects) from the full sample as we want the classifier to be only sensitive to the BCGs with a decent shape measurement. 

From the scatter plot in Figure \ref{fig:Feat_engg_before}, we can see that most BCGs which were accepted (yellow data-points) had a high \textsc{redMaPPer} probability, which is as expected. Interestingly, the absolute magnitude of the BCG and the magnitude gap were also good identifiers of BCGs, with the accepted ``GOOD'' BCGs (shown in yellow) tending towards a larger magnitude gap and brighter magnitudes. We also observed the magnitude and the magnitude gap as a clear identifier of the dominant BCG in a cluster during the visual inspection process. However, since magnitude gaps are also significant in the characterization of the assembly history of a cluster, we construct our main sample without the use of this property. Our objective was to find the simplest collection of cuts that can retain most of the ``GOOD'' sample and reject the majority of the ``BAD'' sample. 

Consequently, we placed strategic cuts on the \textsc{redMaPPer} probability to maximize the sample of ``GOOD'' BCGs in our final sample. With ``P1'' > 0.9 that is a minimum of 90\,\% chance given by \textsc{redMaPPer} and `Mask fraction' < 0.15, we found a good compromise of 84.6\,\% purity with only around 18\,\% of our ``GOOD'' BCGs removed in the process i.e., 82\,\% completeness. With a significantly high completeness of our sample, we can reproduce our results from SPT visual inspection in the ACT data set. Thus, we ported over the following cuts that result in the subset of our ACTxDES clusters with a well-identified BCG and well-measured shape, keeping 263 clusters of the 416 clusters in the sample: 
\begin{itemize}
    \item ``P1'' > 0.9
    \item AND ``MASKFRAC'' < 0.15 ; P1 is the probability assigned to the BCG and MASKFRAC is the masked fraction per cluster as reported by \textsc{redMaPPer} 
\end{itemize} 

While we did not make a feature-cut on informative features like the magnitude gaps for the main analysis, we tried the use of various machine learning classifiers like RF (Random Forest), kNNs (k nearest neighbours), GP (Gaussian Process), etc. to classify BCGs. We show the results and discuss the training process in Appendix \ref{ML}.

\section{Validation of Visual Inspection and Feature Engineering} \label{Validation-of-VI}

\begin{figure}
    \centering
    \includegraphics[width=0.85\columnwidth]{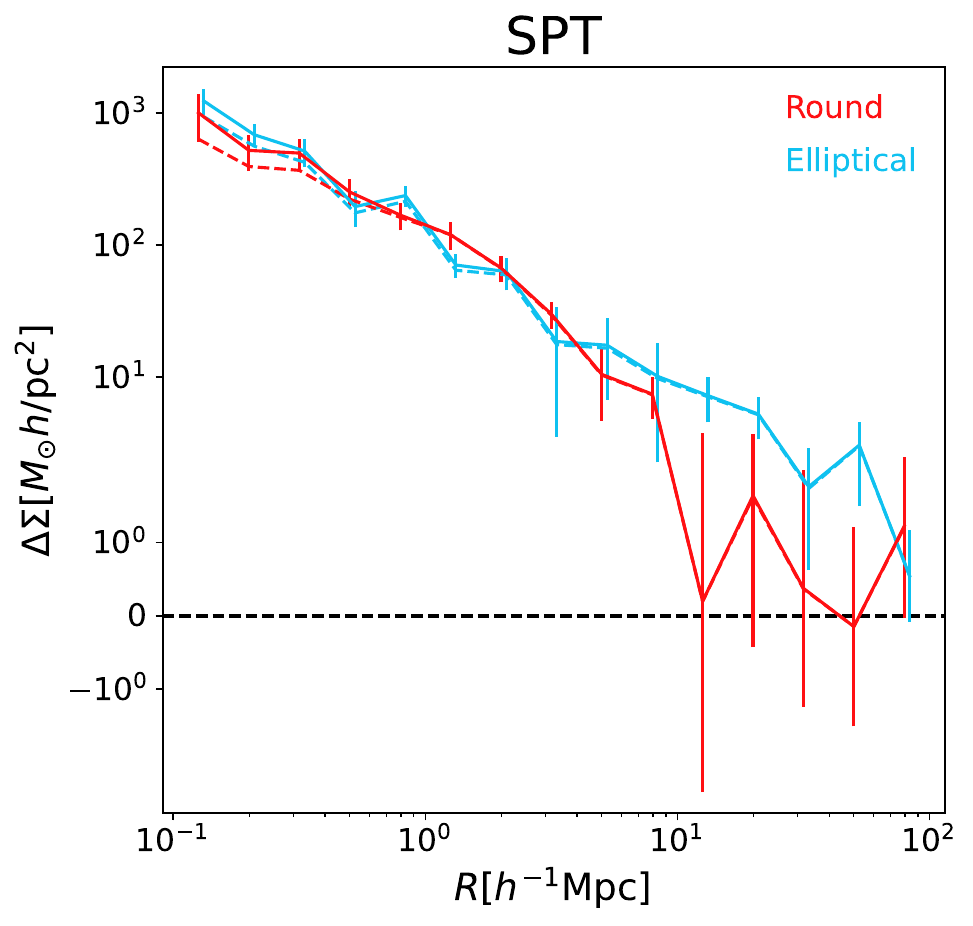}
    \caption{The stacked $\Delta\Sigma$ for the SPT cluster sample with feature engineering replacing the visual inspection results. As before, the red data points represent the round-BCG cluster sample and the blue data points represent the elliptical-BCG cluster sample and the dashed line representing the sample constructed without the boost factor. We observe a statistically significant ($3\,\sigma$ above $10~h^{-1}{\rm Mpc}$) discrepancy between the two samples which is almost as significant as for the visually inspected sample.}
    \label{fig:Without_VI_SPT_delta_sigma}
\end{figure}

\begin{figure}
    \centering
    \includegraphics[width=0.85\columnwidth]{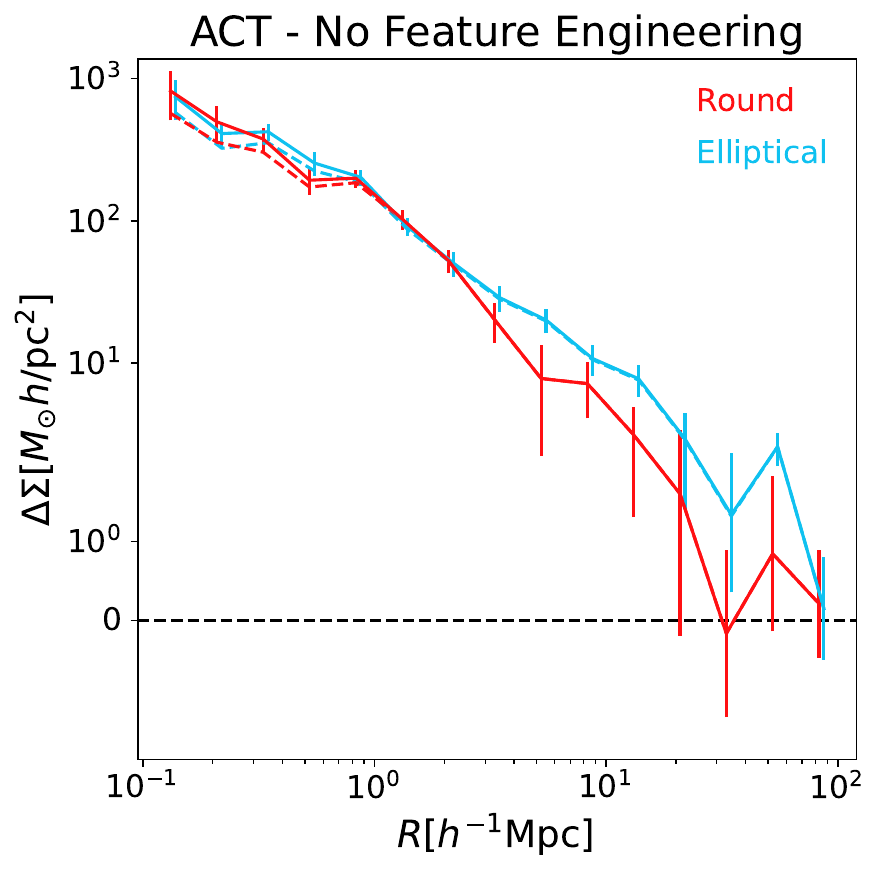}
    \caption{The stacked $\Delta\Sigma$ for the ACT cluster sample with no feature engineering cuts. The red data points represent the round-BCG cluster sample and the blue data points represent the elliptical-BCG cluster sample and the dashed line representing the sample constructed without the boost factor. We observe a consistent split of the elliptical and round-BCG profiles at 2-halo regime agreeing with the sample after feature engineering. We conclude that feature engineering likely does not impart an additional selection effect to the ACT sample.}
    \label{fig:No_feature_engineering_ACT_delta_sigma}
\end{figure}

A valid question to look-out for, is the need for visual inspection/feature engineering and the selection effects, if any, imparted on the sample in the due process. We mimic the feature engineering analysis from the ACT data set, without making any omissions as dictated by the visual inspection on the SPT data set. With only the visual check of the shape measurement (discarding BCGs where there is a bright foreground object biasing the shape measurement) included as part of the selection. The fraction of BCGs affected for the two data sets is mentioned in Section \ref{BCG selection}. 

These moderate cuts result in the stacked lensing profiles of Figure \ref{fig:Without_VI_SPT_delta_sigma}. The SPT profiles are consistent with the lensing results from Figure \ref{fig:ACT_and_SPT_delta_sigma}, where the full-fledged visual inspection process is applied. This is the first result where both ACT and SPT sample are treated equally in reference to the data cuts applied on them. Furthermore, we plot the lensing profiles for the ACT sample when feature engineering cuts are discarded in Figure \ref{fig:No_feature_engineering_ACT_delta_sigma}. Again we find the lensing profiles for ACT sample consistent with Figure \ref{fig:ACT_and_SPT_delta_sigma}. These results together show that the visual inspection and feature engineering likely did not cause an additional selection effect in the data sets. Also, this sets a precedent to use the feature engineering method in larger data sets as a competitive but significantly less time-consuming alternative to visual inspection for the filtering of mis-identified BCGs from the sample.

\section{Machine Learning approach to Feature Engineering} \label{ML}

We trained a variety of machine learning algorithms on the visually inspected sample from SPT which was relatively limited in size. However, there were several informative features that made it feasible. We split our main SPT visually classified sample into a training data set (70\,\%) and testing/validation data set (30\,\%). This split was made with random sampling to avoid any selection effects at this stage. The test data set was never employed in the training phase and is “unseen” data for the model.

\begin{figure*}
    \centering
    \includegraphics[width=1.9\columnwidth]{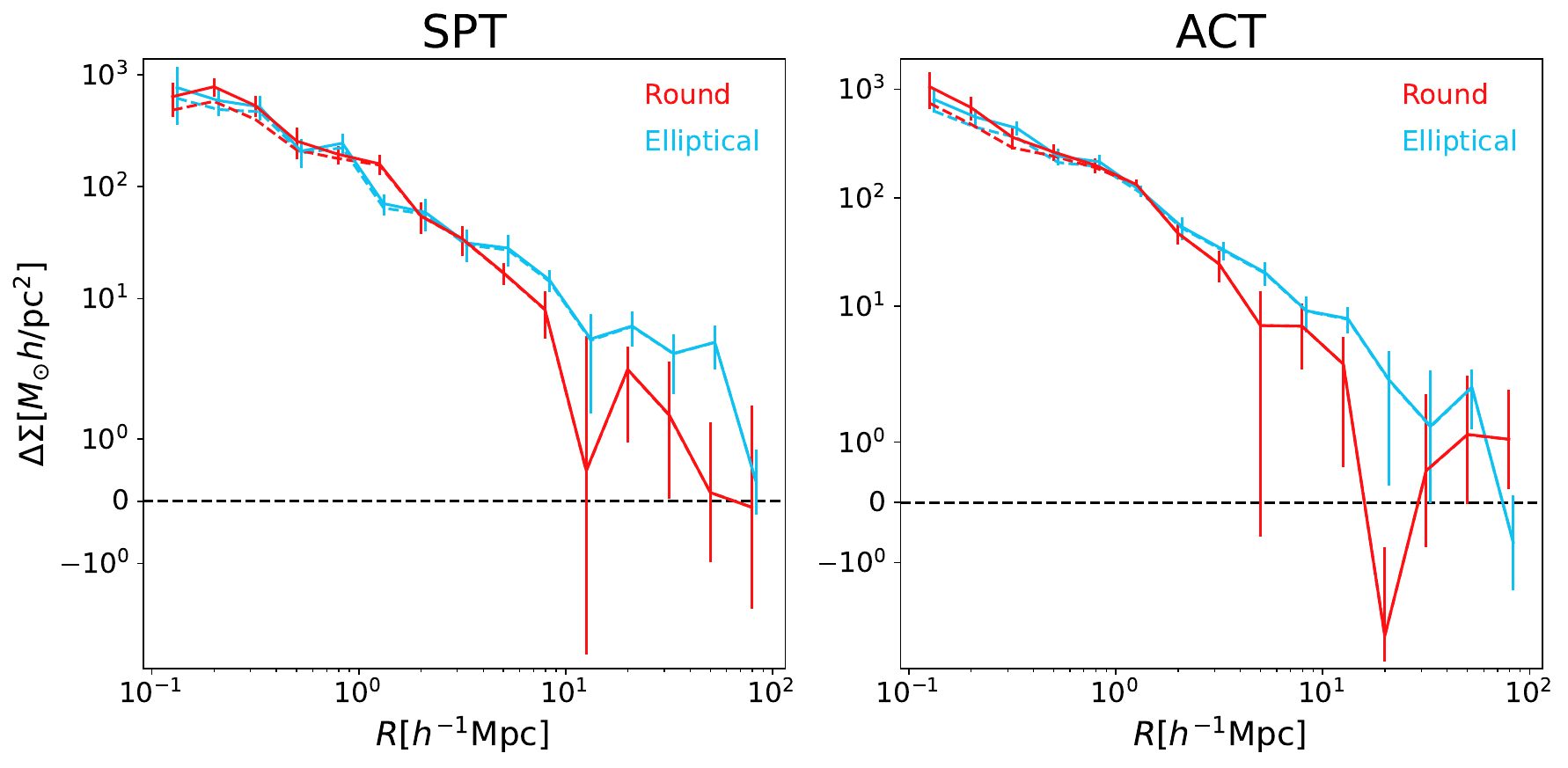}
    \caption{The stacked $\Delta\Sigma$ for the SPT cluster sample (left) and for the ACT cluster sample (right) when classified using the Random Forest classifier. As before, the solid line representing the boost-factor corrected profiles and the dashed line without the use of the boost-factor correction. We observe a statistically significant ($2.3\,\sigma$ for SPT and $2.6\,\sigma$ for ACT above $10~h^{-1}{\rm Mpc}$) discrepancy between the two sub-samples at large scales.}
    \label{fig:RF_ACT_and_SPT_delta_sigma}
\end{figure*}

While training the Random forest classifier, we identified the Magnitude Gap, p1 (\textsc{redMaPPer} probability of top BCG) and Absolute Magnitude to be the three most informative features in the order of most to least informative. Limiting ourselves to just these three features we trained a host of machine learning algorithms: Random Forest, k Nearest Neighbours, Gaussian process classifier and the Multi-layered Perceptron (MLP) network. We found in each case the networks performed well on the testing data set, classifying the BCGs with a purity of 90-95\,\% and a completeness or 85-90\,\%. To test for overfitting, we compared the figure of merit between the training and test samples. Using the precision score as the figure of merit, we found the training sample to give a figure of merit between 0.82-0.9 and for the test sample between 0.8-0.86. We concluded that there was no significant overfitting in the classifiers.
They performed as well as the feature engineering sample created just using the \textsc{redMaPPer} probabilities. As can be seen in Figure \ref{fig:RF_ACT_and_SPT_delta_sigma}, the lensing profiles are quite similar to our main sample. The difference between the round-BCG and elliptical-BCG sample at large scales for SPT was $2.3\,\sigma$  and $2.6\,\sigma$ for the ACT data set. We find similar results for all the other machine learning algorithms. This is a successful test on the capabilities of machine learning algorithms in classifying BCGs and will be pursued as a promising avenue for future analysis.

\bibliographystyle{apj}
\bibliography{lensing_low_1}

%%%%%%%%%%%%%%%%%%%% AFFILIATIONS %%%%%%%%%%%%%%%%
\section*{Affiliations}
%\small
$^{1}$Department of Physics and Astronomy, Stony Brook University, Stony Brook, NY 11794, USA\\
$^{2}$University Observatory, Faculty of Physics, Ludwig-Maximilians-Universit{\"a}t, Scheinerstr. 1, 81679, Munich, Germany\\
$^{3}$High Energy Physics Division, Argonne National Laboratory, Argonne, IL, USA 60439\\
$^{4}$Kavli Institute for Cosmological Physics, University of Chicago, Chicago, IL, USA 60637\\
$^{5}$Department of Physics, Southern Methodist University, Dallas, TX 75205, USA\\
$^{6}$State Key Laboratory of Dark Matter Physics, \& Tsung-Dao Lee Institute, Shanghai Jiao Tong University, Shanghai 200240, China\\
$^{7}$Institute of Space Sciences (ICE, CSIC),  Campus UAB, Carrer de Can Magrans, s/n,  08193 Barcelona, Spain\\
$^{8}$Department of Astrophysical Sciences, Princeton University, Peyton Hall, Princeton, NJ 08544, USA\\
$^{9}$Department of Physics, University of Cincinnati, Cincinnati, OH 45221, USA\\
$^{10}$Physics Department, 2320 Chamberlin Hall, University of Wisconsin-Madison, 1150 University Avenue Madison, WI  53706-1390\\
$^{11}$Argonne National Laboratory, 9700 South Cass Avenue, Lemont, IL 60439, USA\\
$^{12}$Department of Physics and Astronomy, University of Pennsylvania, Philadelphia, PA 19104, USA\\
$^{13}$Department of Physics, Carnegie Mellon University, Pittsburgh, Pennsylvania 15312, USA\\
$^{14}$NSF AI Planning Institute for Physics of the Future, Carnegie Mellon University, Pittsburgh, PA 15213, USA\\
$^{15}$Instituto de Astrofisica de Canarias, E-38205 La Laguna, Tenerife, Spain\\
$^{16}$Laborat\'orio Interinstitucional de e-Astronomia - LIneA, Av. Pastor Martin Luther King Jr, 126 Del Castilho, Nova Am\'erica Offices, Torre 3000/sala 817 CEP: 20765-000, Brazil\\
$^{17}$Universidad de La Laguna, Dpto. Astrofísica, E-38206 La Laguna, Tenerife, Spain\\
$^{18}$Center for Astrophysical Surveys, National Center for Supercomputing Applications, 1205 West Clark St., Urbana, IL 61801, USA\\
$^{19}$Department of Astronomy, University of Illinois at Urbana-Champaign, 1002 W. Green Street, Urbana, IL 61801, USA\\
$^{20}$Department of Astronomy and Astrophysics, University of Chicago, Chicago, IL 60637, USA\\
$^{21}$Department of Physics, Duke University Durham, NC 27708, USA\\
$^{22}$NASA Goddard Space Flight Center, 8800 Greenbelt Rd, Greenbelt, MD 20771, USA\\
$^{23}$Centro de Investigaciones Energ\'eticas, Medioambientales y Tecnol\'ogicas (CIEMAT), Madrid, Spain\\
$^{24}$Lawrence Berkeley National Laboratory, 1 Cyclotron Road, Berkeley, CA 94720, USA\\
$^{25}$Fermi National Accelerator Laboratory, P. O. Box 500, Batavia, IL 60510, USA\\
$^{26}$Universit\'e Grenoble Alpes, CNRS, LPSC-IN2P3, 38000 Grenoble, France\\
$^{27}$Department of Physics and Astronomy, University of Waterloo, 200 University Ave W, Waterloo, ON N2L 3G1, Canada\\
$^{28}$California Institute of Technology, 1200 East California Blvd, MC 249-17, Pasadena, CA 91125, USA\\
$^{29}$SLAC National Accelerator Laboratory, Menlo Park, CA 94025, USA\\
$^{30}$Universit\"at Innsbruck, Institut f\"ur Astro- und Teilchenphysik, Technikerstr. 25/8, 6020 Innsbruck, Austria\\
$^{31}$School of Physics and Astronomy, Cardiff University, CF24 3AA, UK\\
$^{32}$Department of Applied Mathematics and Theoretical Physics, University of Cambridge, Cambridge CB3 0WA, UK\\
$^{33}$Kavli Institute for Particle Astrophysics \& Cosmology, P. O. Box 2450, Stanford University, Stanford, CA 94305, USA\\
$^{34}$MIT Kavli Institute for Astrophysics and Space Research, 70 Vassar Street, Cambridge, MA \\
$^{35}$Instituto de F\'isica Gleb Wataghin, Universidade Estadual de Campinas, 13083-859, Campinas, SP, Brazil\\
$^{36}$Nordita, KTH Royal Institute of Technology and Stockholm University, Hannes Alfv\'ens v\"ag 12, SE-10691 Stockholm, Sweden\\
$^{37}$University of Copenhagen, Dark Cosmology Centre, Juliane Maries Vej 30, 2100 Copenhagen O, Denmark\\
$^{38}$Department of Physics, University of Genova and INFN, Via Dodecaneso 33, 16146, Genova, Italy\\
$^{39}$School of Physics, The University of Melbourne, Parkville VIC 2010 Australia\\
$^{40}$Jodrell Bank Center for Astrophysics, School of Physics and Astronomy, University of Manchester, Oxford Road, Manchester, M13 9PL, UK\\
$^{41}$Departament de F\'{\i}sica, Universitat Aut\`{o}noma de Barcelona (UAB), 08193 Bellaterra, Barcelona, Spain\\
$^{42}$Institut de F\'{\i}sica d'Altes Energies (IFAE), The Barcelona Institute of Science and Technology, Campus UAB, 08193 Bellaterra (Barcelona) Spain\\
$^{43}$Department of Physics, University of Arkansas, 825 W Dickson Street, Fayetteville, AR\\
$^{44}$Brookhaven National Laboratory, Bldg 510, Upton, NY 11973, USA\\
$^{45}$Department of Physics, Faculty of Science, Chulalongkorn University, 254 Phyathai Road Patumwan, Bangkok 10330, Thailand\\
$^{46}$Vera C. Rubin Observatory Project Office, 950 N Cherry Ave, Tucson, AZ 85719, USA\\
$^{47}$NSF NOIRLab, 950 N. Cherry Ave., Tucson, AZ 85719, USA\\
$^{48}$Center for Astrophysics, Harvard \& Smithsonian, 60 Garden Street, MS 42, Cambridge, MA 02138, USA\\
$^{49}$Institut de Recherche en Astrophysique et Plan\'etologie (IRAP), Universit\'e de Toulouse, CNRS, UPS, CNES, 14 Av. Edouard Belin, 31400 Toulouse, France\\
$^{50}$INAF-Osservatorio Astronomico di Trieste, via G. B. Tiepolo 11, I-34143 Trieste, Italy\\
$^{51}$Physik-Institut, University of Zürich, Winterthurerstrasse 190, CH-8057 Zürich, Switzerland\\
$^{52}$Institute of Cosmology and Gravitation, University of Portsmouth, Portsmouth, PO1 3FX, UK\\
$^{53}$Department of Physics, Northeastern University, Boston, MA 02115, USA\\
$^{54}$Department of Physics \& Astronomy, University College London, Gower Street, London, WC1E 6BT, UK\\
$^{55}$School of Mathematics and Physics, University of Queensland,  Brisbane, QLD 4072, Australia\\
$^{56}$Astronomy Unit, Department of Physics, University of Trieste, via Tiepolo 11, I-34131 Trieste, Italy\\
$^{57}$Institute for Fundamental Physics of the Universe, Via Beirut 2, 34014 Trieste, Italy\\
$^{58}$Hamburger Sternwarte, Universit\"{a}t Hamburg, Gojenbergsweg 112, 21029 Hamburg, Germany\\
$^{59}$Department of Physics, IIT Hyderabad, Kandi, Telangana 502285, India\\
$^{60}$Instituto de Fisica Teorica UAM/CSIC, Universidad Autonoma de Madrid, 28049 Madrid, Spain\\
$^{61}$Santa Cruz Institute for Particle Physics, Santa Cruz, CA 95064, USA\\
$^{62}$Jet Propulsion Laboratory, California Institute of Technology, 4800 Oak Grove Dr., Pasadena, CA 91109, USA\\
$^{63}$George P. and Cynthia Woods Mitchell Institute for Fundamental Physics and Astronomy, and Department of Physics and Astronomy, Texas A\&M University, College Station, TX 77843,  USA\\
$^{64}$Aix Marseille Univ, CNRS/IN2P3, CPPM, Marseille, France\\
$^{65}$Instituci\'o Catalana de Recerca i Estudis Avan\c{c}ats, E-08010 Barcelona, Spain\\
$^{66}$Observat\'orio Nacional, Rua Gal. Jos\'e Cristino 77, Rio de Janeiro, RJ - 20921-400, Brazil\\
$^{67}$Department of Physics and Astronomy, Pevensey Building, University of Sussex, Brighton, BN1 9QH, UK\\
$^{68}$Computer Science and Mathematics Division, Oak Ridge National Laboratory, Oak Ridge, TN 37831\\
$^{69}$Berkeley Center for Cosmological Physics, Department of Physics, University of California, Berkeley, CA 94720, US\\
$^{70}$Max Planck Institute for Extraterrestrial Physics, Giessenbachstrasse, 85748 Garching, Germany\\
%%%%%%%%%%%%%%%%%%%%%%%%%%%%%%%%%%%%%%%%%%%%%%%%%%
\end{document}